\documentclass[a4paper,10pt]{article}
\usepackage{jheppub}
\pdfoutput=1
\usepackage{amsmath,amssymb,amsfonts, bm, natbib}
\usepackage{epsfig}
\usepackage{graphicx}
\usepackage{slashed}
\usepackage{physics}
\usepackage{multirow}
\usepackage{caption}
\usepackage{subcaption}
\usepackage{slashed} 
\usepackage{mathrsfs}
\usepackage{soul}
\usepackage{soul}
\usepackage{braket}
\usepackage{hyperref}
\usepackage{calc}
\usepackage{xcolor}
\usepackage{tablefootnote}

\usepackage[font=small]{caption}

\def\beqn{\begin{eqnarray}}
\def\ba{\begin{array}{c}}
\def\bat{\begin{array}{cc}}
\def\bat{\begin{array}{cc}}
\def\ea{\end{array}}
\def\bat{\begin{array}{cc}}
\def\batt{\begin{array}{ccc}}
\def\eeqn{\end{eqnarray}}

\newcommand{\bea}{\begin{eqnarray}}
\newcommand{\eea}{\end{eqnarray}}
\newcommand{\beq}{\begin{equation}}
\newcommand{\eeq}{\end{equation}}
\newcommand{\ec}{\end{center}}
\newcommand{\bc}{\begin{center}}

\newcommand{\pdir}{p\kern -5.2pt\raise 0.2ex\hbox {/}}

\newcommand{\vdir}{v\kern -5.75pt\raise 0.15ex\hbox {/}}
\newcommand{\kdir}{k\kern -5.75pt\raise 0.15ex\hbox {/}}
\newcommand{\epsdir}{\epsilon\kern -5.0pt\raise 0.15ex\hbox {/}}
\newcommand{\bvdir}{\bar{v}\kern -5.75pt\raise 0.15ex\hbox {/}}
\newcommand{\Ddir}{D\kern -7.75pt\raise 0.20ex\hbox {/}}
\newcommand{\Adir}{A\kern -7.75pt\raise 0.20ex\hbox {/}}
\newcommand{\ldir}{l\kern -5.0pt\raise 0.2ex\hbox{/}}
\newcommand{\varepsdir}{\varepsilon\kern -5.5pt\raise 0.15ex\hbox{/}}

\makeatother

\definecolor{niceblue}{rgb}{0.15,0.15,0.6}
\definecolor{nicegreen}{rgb}{0.1,0.5,0.1}
\definecolor{Red}{rgb}{1.,0.,0.}

\definecolor{Green}{rgb}{0.2,.7,0.2}

\makeatletter
\newcommand*{\rom}[1]{\expandafter\@slowromancap\romannumeral #1@}
\makeatother

\definecolor{gray}{rgb}{0.4,0.4,0.4}
\definecolor{darkblue}{rgb}{0.0,0.0,0.3}
\hypersetup{colorlinks,breaklinks,
            linkcolor=darkblue,urlcolor=darkblue,
            anchorcolor=darkblue,citecolor=darkblue}

\title{\Large On lepton flavour universality in  semileptonic {\bf $B_c \to \eta_c, J/\psi$} decays }
 \author[a]{Domagoj Leljak,} 
\author[a]{Bla\v zenka Meli\'c,}
\author[b]{Monalisa Patra}

 \affiliation[a]{Institut Rudjer Bo\v skovi\'c, Division of Theoretical Physics, Bijeni\v cka 54, HR-10000, Croatia}
  \affiliation[b]{Jo\v{z}ef Stefan Institute, Jamova 39, P. O. Box 3000, 1001 Ljubljana, Slovenia}
 
\abstract{
We discuss  $B_c \to \eta_c$ and $B_c \to J/\psi$ semileptonic decays within the Standard Model (SM) and beyond. The relevant transition form factors,  being the main source of theoretical uncertainties,  are calculated in the sum rule approach and are provided in a full $q^2$ range. We calculate the semileptonic branching fractions and find for the ratios, $R_{\eta_c}|_{\rm SM}= 0.32 \pm 0.02$, $R_{J/\psi}|_{\rm SM} = 0.23 \pm 0.01$.  Both predictions are in agreement with other existing calculations and support the current tension in $R_{J/\psi}$ at 2$\sigma$ level with the experiment.  To extend the potential of testing the SM in the semileptonic $B_c$ decays, we consider the forward-backward asymmetry and polarization observables. We also study the 4-fold differential distributions of $B_c \rightarrow J/\psi (J/\psi \to \tilde{\ell}^-\tilde{\ell}^+ ) \ell^- \bar{\nu}_\ell$, where $\tilde{\ell} = e, \mu$, in the presence of different new physics scenarios and  find that the new physics effects can significantly modify the angular observables and can also produce effects which do not exist in the SM. Using the constraints on the new physics couplings  from the recent combined analysis of BaBar, Belle and LHCb data on semileptonic $B \to D^{(\ast)}$ decays, where the effects of new physics could be visible, we find that these different new physics scenarios are not able to simultaneously explain the current experimental value of $R_{J/\psi}$.
}

\begin{document}

\maketitle

%\tableofcontents

\section{Introduction}
In September 2017 the LHCb collaboration announced the first measurement of testing the lepton flavor universality using charmed-beauty meson semileptonic decays to $J/\psi \tau^+ \nu_{\mu}$ and $J/\psi \mu^+ \nu_{\mu}$~\cite{Aaij:2017tyk}. The result for the measurement of the ratio of the branching fractions is
\begin{eqnarray}
R_{J/\psi}|_{\rm exp} = \frac{BR(B_c^+ \to J/\psi \tau^+ \nu_{\tau})}{BR(B_c^+ \to J/\psi \mu^+ \nu_{\mu})}  = 0.71 \pm 0.17 \pm 0.18,
\label{eq:expRJpsi}
\end{eqnarray}
and is more than 2$\sigma$ away from the Standard model (SM) prediction. Currently there are many model dependent calculations of $R_{J/\psi}$~\cite{Wen-Fei:2013uea,Kiselev:2002vz,preparation,Huang:2007kb,Scora:1995ty,AbdElHady:1999xh,Nobes:2000pm,Ebert:2003cn,Hernandez:2006gt,Qiao:2012vt,Wang:2008xt,Anisimov:1998uk,Ivanov:2006ni} within the SM and they give the results in the range (without including model uncertainties)
%The Standard model (SM) prediction of most models (without including model uncertainties) is given in the range    
\begin{eqnarray}
\label{eq:RJSM}
R_{J/\psi}|_{\rm SM} = 0.24 - 0.30,
\end{eqnarray}
%and is more than 2$\sigma$ away from the reported measurement. 
However, the $R_{J/\psi}$ measurement is challenging. Due to the presence of invisible $\nu$'s, both decays are observed only through 3 muons, two of them coming from $J/\psi$ decays and being perfectly identified.  The third muon makes a difference and enables distinguishing the semileptonic $B_c$ decays to $\tau$ and to $\mu$ from the background. Therefore it is still premature to speak about the new physics effects in these decays, although one can consider this probability having in mind that BABAR, Belle and  LHCb have also found other intriguing anomalies in the semileptonic decays of $B$ mesons, known as $R_D$ and $R_{D^\ast}$~\cite{Lees:2013uzd,Sato:2016svk,Aaij:2015yra,Amhis:2016xyh}. These experimental collaborations have revealed a significant deviation of 2.3$\sigma$, and 3.5$\sigma$ of the ratios $R_D$ and $R_{D^\ast}$ from the SM predictions. Also some deviations in $b \to s$ semileptonic decays are still  present~\cite{Aaij:2017vbb}.

Moreover, calculations of these semileptonic heavy-meson decays involve theoretical uncertainties coming from imprecise determination of the hadronic transition form factors describing the hadronic effect in the transition from the initial to the final meson state. 

The calculation of $B_c$ form factors are difficult and leads to big uncertainties. If we summarize values of $B_c$ into $S$-wave charmonia form factors at $q^2 = 0$ calculated in different models (perturbative QCD (pQCD)~\cite{Wen-Fei:2013uea}, three-point QCD sum rules (3ptQCDSR)~\cite{Kiselev:2002vz,preparation}, light cone sum rules (LCSR)~\cite{Huang:2007kb}, relativistic quark model (RQM)~\cite{Scora:1995ty,AbdElHady:1999xh,Nobes:2000pm,Ebert:2003cn}, nonrelativistic quark models (NRQM)~\cite{Hernandez:2006gt,Qiao:2012vt}, light-font quark model (LFQM)~\cite{Wang:2008xt}, constituent quark model (CQM))~\cite{Anisimov:1998uk}, relativistic quark model (RCQM)~\cite{Ivanov:2006ni}) in the literature we obtain: 
\begin{eqnarray}
f_+(0) = f_0(0) =  0.20 - 1.43, \
\end{eqnarray}
for $B_c \to \eta_c$ form factors, and
\begin{eqnarray}
V(0) = 0.17 - 1.63  , \nonumber \\
A_1(0) = 0.21 - 1.19 , \nonumber \\
A_2(0) = 0.23 - 1.27 , \nonumber \\
A_0 (0) = 0.12 - 1.09 ,
\end{eqnarray}
for $B_c \to J/\psi$ form factors. It is obvious that with such a large range of estimated form factors it is impossible to make any reliable prediction for $R_{\eta_c}$ and $R_{J/\psi}$ ratios. Moreover, in many estimations of form factors, the theoretical errors were not given or they are not under control. Although some of the uncertainties cancel in the ratio, the model predictions of $R_{J/\psi}$ calculated in different approaches and taking the theoretical uncertainties into account vary in a huge range~\cite{Watanabe:2017mip,Dutta:2017xmj,Cohen:2018dgz,Berns:2018vpl,Murphy:2018sqg,Tran:2018kuv}
%\cite{Cohen:2018dgz,Cohen:2018vhw,Berns:2018vpl,Murphy:2018sqg,Huang:2007kb,preparation,Kiselev:2002vz,Wen-Fei:2013uea,Ivanov:2006ni,Tran:2018kuv,Issadykov:2017wlb,Ebert:2003cn,Scora:1995ty,Nobes:2000pm,AbdElHady:1999xh,Wang:2008xt,Colangelo:1999zn,Anisimov:1998uk,Hernandez:2006gt,Qiao:2012vt,Dutta:2017xmj,Watanabe:2017mip}{\bf [delete references which make no sense or leave them and correct the eq. below?]}:
\begin{eqnarray}
R_{J/\psi}|_{\rm SM} = 0.17-0.41. % {\color{red} 0.03-0.41}
\end{eqnarray}
%and we would like to focus first on the sum rule calculation of %transition form factors.
The lattice QCD calculation for $B_c \to J/\psi$ form factors $V(q^2)$ and $A_1(q^2)$ are available now from the preliminary results of the HPQCD collaboration, at several points for $V(q^2)$ and $A_1(q^2)$ ~\cite{Colquhoun:2016osw}. Earlier, the same collaboration has also produced results for $B_c \to \eta_c$ form factors, which were reported on in the same proceedings.

In this paper we will address the calculation of the form factors  for $B_c \to S$-wave charmonia in the full $q^2$ range using the LCSR-inspired  approach. The LCSR method was proven to be a reliable method for calculating transition form factors of many heavy-to light decays, such as $B_{(s)},D_{(s)} \to \pi,\rho, K, K^{\ast}, \eta, \eta'$~\cite{BallBraun1,Duplancic:2008tk,BallZwicky2004,Duplancic:2008ix,Duplancic:2015zna} and even for $\Lambda_b \to \Lambda_c$ decays~\cite{Khodjamirian:2011jp,Bell:2013tfa}.
We will compare our results with the existing QCD lattice points for $f_+(q^2)$ and $f_0(q^2)$ from $B_c \to \eta_c$ and $V(q^2)$ and $A_1(q^2)$ from $B_c \to J/\psi$ and will show the nice agreement, specially having in mind that the lattice results are still preliminary and do not include systematical errors. Following ~\cite{Cohen:2018dgz,Berns:2018vpl} we have assigned $20\%$ uncertainty to the lattice QCD results.

We also cite the results of our recent calculation derived by using the 3ptQCDSR method~\cite{preparation}\footnote{A very brief discussion on the 3ptQCDSR calculation is provided in Appendix~\ref{sec:app1}.} and show that the form factors from two sum rule approaches, although calculated by using different quark-hadron duality assumptions, appear to be consistent and precise enough to enable precise determination of the ratios $R_{\eta_c}$ and $R_{J/\psi}$ in the SM. Recently, there also appeared a model-independent estimation of the SM bounds on $R_{\eta_c}$ and  $R_{J/\psi}$~\cite{Cohen:2018dgz,Berns:2018vpl,Murphy:2018sqg}. Such analysis rely on the data, available lattice results and the heavy-quark spin-symmetry (HQSS) relations for the form factors at the zero recoil and predict the $R$-ratios consistent with Eq.(\ref{eq:RJSM}) and our calculation, clearly at $2 \sigma$ discrepancy with the experiment. The HQSS and nonrelativistic QCD (NRQCD) relations used in these approaches will be carefully examined in Sec.~\ref{sec:HQrelations}.

The possible new physics (NP) effects in the semileptonic $B_c \to \eta_c(J/\psi )\ell \nu_\ell$ decays have been recently considered in the context of either specific models, such as the leptoquark models~\cite{Chauhan:2017uil}, left-right symmetric models, $R$-parity violating supersymmetric models, etc.~\cite{He:2017bft,Wei:2018vmk,Biswas:2018jun} or in a model independent approach based on the most general effective Hamiltonian~\cite{Watanabe:2017mip,Dutta:2017xmj,Tran:2018kuv,Dutta:2017wpq,Huang:2018nnq,Li:2018lxi}. To account for possible NP effects in $B_c \to \eta_c$ and $B_c \to J/\psi$ semileptonic decays we consider here the effective Hamiltonian approach consisting of all possible four-Fermi operators. The constraints on contributions of these NP operators and the corresponding Wilson coefficients are obtained from the experimental results of $R_D, R_{D^\ast}$, polarizations of $\tau$ and $D^{\ast}$ in $B\to D^{(\ast)} l \nu$ decays, as well as on the $B_c$ lifetime. There are various studies~\cite{Alonso:2016gym,Greljo:2018tzh,Blanke:2018yud,Feruglio:2018fxo,Alok:2017qsi,Alok:2018uft} performing a global fit on these NP operators considering the presence of only one or two NP operators simultaneously. We have taken the latest constraints on the Wilson coefficients from Ref.~\cite{Blanke:2018yud} and analysed the effects of these NP operators on various observables such as the ratio of the branching fractions, the forward-backward asymmetry, the convexity parameter and the longitudinal as well as the transverse polarization components of $\tau$ in the final state. We have also preformed the first study of the full 4-fold differential decay rate $B_c \to J/\psi \,(  J/\psi \to \mu^+\mu^-, e^+ e^-) l \nu_l$, where the leptons from the $J/\psi$ decay are of opposite helicities. The 4-fold decay distribution in this case is proportional to three angles and the momentum transfer $q^2$.  The three distinct angles gives the freedom to construct additional asymmetries which are sensitive to the real as well as the imaginary part of the new physics couplings. 

The structure of the paper is as follows. We compute the form-factors in the context of our sum rule model in Sec.~\ref{sec:LCSR_ff} and present the  results in the whole $q^2$ range.
We compare the results with those existing in the literature and with available preliminary lattice results. The discussion of the heavy-quark symmetry limit of form factors at the zero recoil is given in Sec. ~\ref{sec:HQrelations}. The general effective Hamiltonian of the $b\rightarrow c \ell \nu_\ell$ transition is introduced in Sec.~\ref{sec:diff_dist}, and we obtain the semileptonic decay distributions for $B_c \to \eta_c,J/\psi$ in the presence of NP operators  using the helicity technique.  We compare predictions for different physical observables in the SM and in the presence of NP. A detailed comparison of predictions of $R_{\eta_c,J/\psi}$ in the SM, with the form factors calculated in our model,  with the predictions from other approaches is also provided.  In Sec.~\ref{sec:diff_dist_decay}, we extend the calculation of $B_c \to J/\psi l \nu_l$ to the $J/\psi$ decay into a pair of muons or electrons, and  discuss the full 4-fold distribution. A set of new observables is considered and the results are compared for the SM case and beyond. Finally we conclude in the last section, Sec.~\ref{sec:conclusion}.
 
%%%%%%%%%%%%%%%%%%%%%%%%%%%%%%%%%%%%%%%%%%%%%
%%%%%%%%%%%%%%%%%%%%%%%%%%%%%%%%%%%%%%%%%%%%%

\section{Sum rule calculations for the form factors }\label{sec:LCSR_ff}

We will perform the estimation of the $B_c \to \eta_c$ and $B_c \to J/\psi$ form factors using the LCSR-inspired method. 
We will follow the standard QCD sum rule method, by interpolating the $B_c$ meson with an appropriate quark current and describing the S-wave charmonia by the distribution amplitudes (DAs) of an increasing twist. 

The method of the LCSR is very well know and we will just briefly outline the
procedure here in order to properly define all ingredients necessary for calculating the form
factors. In the calculation we will use the following approximations: the twist-2 light-cone distribution amplitudes will be calculated in the NRQCD model~\cite{Bodwin:2007fz}, and the Gegenbauer polynomials expanded at the scale $\mu$. The twist-3 and twist-4 DAs will be taken in their asymptotic form. Moreover, the Wandzura-Wilczek approximation will be applied, where the three-particle DA are neglected and therefore the twist-3 and twist-4 DAs are expressed in terms of the twist-2 distributions. The effects of the final state masses, $m_{\eta_c}$ and $m_{J/\psi}$ are  included \cite{BharuchaStraubZwicky}. 

The calculation of the form factors for $B_c \to \eta_c$ proceeds in a similar way as those for $B \to \pi, K.$ \cite{Duplancic:2008tk,Duplancic:2008ix,Duplancic:2015zna,Braun:1999uj,Ball:2001fp, Ball:2004ye,Khodjamirian:2011ub}, while $B_c \to J/\psi$ form factor calculation closely follows the derivation of the form factors of $B \to K^*$~\cite{Ball:1998kk,Ball:2004ye,BallZwicky2004,Altmannshofer:2008dz,BharuchaStraubZwicky}. We have checked that with appropriate changes in the expressions, all our results agree with previous calculations. 

\subsection{Definitions}\label{sec:2}

The form factors of the $B_c \to \eta_c$ decay are defined as 
\begin{eqnarray}\label{eq:etaC_ff}
\langle \eta_c(p)| \bar{c} \gamma_\mu b | B_c(p_{B_c}) \rangle &=& \left [ ( p + p_{B_c})_\mu - \frac{m_{B_c}^2 - m_{\eta_c}^2}{q^2} q_\mu \right]f_+(q^2) + \left [ \frac{m_{B_c}^2 - m_{\eta_c}^2}{q^2} q_\mu\right ] f_0 (q^2)\,,
\nonumber \\
\langle \eta_c(p)| \bar{c} \sigma_{\mu\nu} q^{\nu} b | B_c(p_{B_c}) &=& \frac{i}{m_{B_c} + m_{\eta_c}}\left [q^2 ( p + p_{B_c})_\mu - (m_{B_c}^2 - m_{\eta_c}^2)q_\mu      \right]f_T(q^2) \,,
\end{eqnarray}
where $f_+(0)  =f_0(0)$  and $0\leq q^2\leq (m_{B_c}-m_{\eta_c})^2$. 
The scalar form factor $f_0(q^2)$ follows also from the conservation of the vector current as 
\begin{eqnarray}
\langle 0 |\bar{c} b| B_c(p_{B_c}) \rangle =  \frac{m_{B_c}^2 - m_{\eta_c}^2}{m_b(\mu) - m_c(\mu)} f_0 (q^2).
\end{eqnarray}
The decay $B_c\to J/\psi  \ell^+ \nu_l$ is described by the following form factors defined as \cite{BallZwicky2004} 
\begin{eqnarray}\label{eq:etaC_Jpsi1}
\lefteqn{
\langle J/\psi(p,\epsilon) | \bar c\gamma_\mu(1-\gamma_5) b | B_c(p_{B_c})\rangle  =  
-i \epsilon^\ast_\mu (m_{B_c}+m_{J/\psi})
A_1(q^2) + i (p_{B_c}+p)_\mu (\epsilon^\ast \cdot q)\,
\frac{A_2(q^2)}{m_{B_c}+m_{J/\psi}}}\hspace*{2.8cm}\nonumber\\
&& {}+  i q_\mu (\epsilon^\ast \cdot q) \,\frac{2  m_{J/\psi}}{q^2}\,
\left(A_3(q^2)-A_0(q^2)\right) +
\varepsilon_{\mu\nu\rho\sigma}\epsilon^{\ast \nu} p_{B_c}^\rho p^\sigma\,
\frac{2 V(q^2)}{m_{B_c}+m_{J/\psi}},
\label{eq:FF1}
\end{eqnarray}
where $\epsilon$ is the polarization vector of the $J/\psi$ meson, $q=p_{B_c}-p$ is the momentum transfer varying in the range $0\leq q^2\leq (m_{B_c}-m_{J/\psi})^2$ and 
\begin{eqnarray}
A_3(q^2) & = & \frac{m_{B_c}+m_{J/\psi}}{2  m_{J/\psi}}\, A_1(q^2) -
\frac{m_{B_c}- m_{J/\psi}}{2 m_{J/\psi}}\, A_2(q^2)\,,
\end{eqnarray}
satisfying the relation
\begin{eqnarray}
A_3(0) =  A_0(0). 
\end{eqnarray}
The form factor $A_0$ is the pseudoscalar form factor which can also be defined by applying the equation of motion to the derivative of the axial current:  
\begin{eqnarray}
\langle J/\psi |\bar c i\gamma_5 b | B_c\rangle = \frac{2 m_{J/\psi}}{(m_b(\mu)+m_c(\mu))}
(\epsilon^\ast \cdot q) A_0(q^2)
\end{eqnarray}
and, as it can be seen below, contributes to the $B_c \to J/\psi l \nu$ decay only if the lepton in the decay is considered to have a non-vanishing mass, which will be a case for the $\tau$ particle. The tensor form factors are usually defined as 
\begin{eqnarray}\label{eq:etaC_Jpsi2}
\lefteqn{\langle J/\psi(p,\epsilon) | \bar c \sigma_{\mu\nu} q^\nu (1+\gamma_5) b |
B(p_{B_c})\rangle = 2 i\varepsilon_{\mu\nu\rho\sigma} \epsilon^{\ast \nu}
p_{B_c}^\rho p^\sigma \, T_1(q^2) }   \nonumber\\
& & {} 
+ T_2(q^2) \left [ \epsilon^\ast_\mu (m_{B_c}^2- m_{J/\psi}^2) - (\epsilon^\ast \cdot q) \,(p_{B_c}+p)_\mu \right ] + T_3(q^2) (\epsilon^\ast \cdot q) \left [ q_\mu - \frac{q^2}{m_{B_c}^2- m_{J/\psi}^2}\, (p_{B_c}+p)_\mu
\right ],
\nonumber \\
\end{eqnarray}
and
\begin{eqnarray}
T_1(0) = T_2(0). 
\end{eqnarray}
However, as discussed in \cite{BallZwicky2004,BharuchaStraubZwicky}, in the standard QCD sum rule one has to consider the off-shell $p_{B_c}$ momentum ($p_{B_c}^2 \neq m_{B_c}$) and in order to avoid any ambiguity in the interpretation of $p_{B_c}^2$ appearing at different steps of calculation it is more appropriate to use the following matrix element as a definition of the tensor form factors;
\begin{eqnarray}
\langle J/\psi(p,\epsilon)|\bar c \sigma_{\mu\nu}\gamma_5 b | B(p_{B_c})\rangle & = &
A(q^2) \left\{\epsilon^\ast_\mu (p_{B_c}+p)_\nu - (p_{B_c}+p)_\mu \epsilon^\ast_\nu\right\} - B(q^2)
\left\{\epsilon^\ast_\mu q_\nu - q_\mu \epsilon^\ast_\nu\right\}\nonumber\\
& & {} -  2 C(q^2) \,\frac{\epsilon^\ast \cdot q}{m_{B_c}^2-m_{J/\psi}^2} \,
\left\{ p_\mu q_\nu - q_\mu  p_\nu \right\}.
\end{eqnarray}
where $A(q^2)$, $B(q^2)$ and $C(q^2)$ are related to $T_i(q^2)$  
defined in Eq.~(\ref{eq:etaC_Jpsi2}) as
\begin{equation}
T_1(q^2) = A(q^2),\quad T_2(q^2) = A(q^2)-\frac{q^2}{m_{B_c}^2-m_{J/\psi}^2}\, B(q^2),\quad
T_3(q^2) = B(q^2)+C(q^2)\,.
\end{equation}

The form factors are extracted from the
correlation function of the T-product of the weak current $j_{\Gamma}, \Gamma = V,A,S,P,T$ 
%, being eitherthe $V-A$ current or the tensor current 
and an interpolating current of the $B_c$ meson $j_{B_c} = m_b \bar c i \gamma_5 b$  among the vacuum and the external on-shell meson M ($M = J/\psi, \eta_c$), 
\begin{eqnarray}\label{eq:correlator}
\Pi(q^2,p_{B_c}^2) &=& 
i\int d^4x e^{iqx} \left \langle M(p) \left | T \left\{ j_{\Gamma}(x) j_{B_c}^\dagger(0) \right \} \right | 0 \right \rangle \,.
%\Pi^{\rm OPE}(q^2,p_{B_c}^2) &=& 
%i\int d^4x e^{-i p_{B_c} x} \left \langle J/\psi/\eta_c (p)  \left | T \left\{ j_{\Gamma}(0) %j_{B_c}^\dagger(x) \right \} \right | 0 \right \rangle,\\
%\Pi^{\rm OPE}(q^2,p_{B_c}^2, p_{D^{(*)}}^2) && 
%i^2\int\int d^4x d^4y e^{-i ( p_{B_c}x - p_{D^{(*)}}y) } \left \langle 0 \left | T \left\{ %j_{D^{(*)}}(y) j_{\Gamma}(0) j_{B_c}^\dagger(x) \right \} \right | 0 \right \rangle
%\nonumber 
\end{eqnarray}
Both $B_c \to M$ decays proceed through $b$-quark decays and we assume that in the region of the large $m_b^2 -q^2 \leq \mathcal{O}(m_b\Lambda_{QCD})$ and $m_b^2 -p_{B_c}^2 \leq \mathcal{O}(m_b\Lambda_{QCD})$ virtualities, the correlation function  Eq.~(\ref{eq:correlator}) are dominated by the light-like distances and the description in terms of the products of perturbatively calculable hard-scattering kernels with non-perturbative and universal light-cone distribution amplitude (LCDA), ordered by increasing
twist, is appropriate. 

By inserting the sum over states with $B_c$ quantum numbers and by using 
\begin{eqnarray}
\langle 0 |j_{B_c} | B_c(p_B) \rangle = \frac{ f_{B_c} m_{B_c}^2}{m_b(\mu) + m_c(\mu)}
\end{eqnarray}
for the ground state, with the use of hadronic dispersion relation in the virtuality $p_{B_c}^2$ of the $B_c$ 
channel, we can relate the correlation function Eq.~(\ref{eq:correlator}) to the $B_c \to M$ matrix elements and the form factors defined above. As usual, the quark-hadron duality is used to approximate heavier state contribution by introducing the effective threshold parameter $s_0^{B_c}$
and the ground state contribution of the $B_{c}$ meson is enhanced by the Borel transformation in the variable  $p_{B_c}^2 \to \sigma^2$.

The strategy which we use to fix the sum rule parameters, in particular the continuum threshold parameter $s_0^{B_c}$, is to use the lattice results for the decay constant of $B_c$, Eq.~(\ref{eq:latBc}) and fix the continuum threshold parameters by calculating the constant with the 2-point functions calculated in the LCSR. This is done by using the NLO expression and the pole $m_b,m_c$ masses. The $\overline{\rm MS}$ masses used in the paper are taken as $m_b (\overline{m_b}) = 4.18$ GeV and $m_c (\overline{m_c}) = 1.27$ GeV.  We have achieved the stability of the sum rules, i.e. that continuum and  higher-order corrections are suppressed and that also the mass of $B_c$ is correctly reproduced for $\mu = 3.9 \pm 0.3$ GeV. With the calculated $s_{0}^{B_c} = 46.8 \pm 0.8$ GeV$^2$ we  have also checked the stability of the sum rules  for $B_c \to M$ transitions. In both cases, the results are very stable on the variation of the Borel parameter, allowing $\sigma^2$ to vary between $70-90 \,{\rm GeV}^2$ with almost no change.
Other parameters used in the paper are taken from  the lattice results or from the NRQCD models described afterwards.

 The method of the LCSR was extensively used for calculating the heavy-to-light transition form factors. Here the situation is far more complicated since the final meson is a quarkonium state $\eta_c$ or $J/\psi$. Therefore, to properly account for the non-negligible large mass corrections ${\cal O}(2 m_c)$ in the correlator, one would have to do a systematic expansion of the correlator near the light cone including those corrections. This is a highly nontrivial task and according to our knowledge has not been done yet. In the future, to improve the whole picture, one has to do a revised consideration of the LCDA for charmonia, similar to what was done for heavy hadrons ($B$-mesons and $\Lambda_b$), by proving the factorization theorems and deriving the RG evolution kernels of LCDA by considering full mass corrections. But, such a calculation for charmonia is far more complicated since there is no help from HQET and heavy-quark symmetries, neither can one achieve fast convergence in the heavy-mass expansion. Such a calculation, if consistently doable for charmonia, is far beyond the scope of our paper.  Here we  assume that these potentially large intrinsic mass effects can be effectively described using proper phenomenological model of DAs. So, we will follow a simplified sum rule model where we treat charmonia of $B_c$-decays as light particles in the correlator (\ref{eq:correlator}) expansion near the light-cone and will closely follow the approach of the standard LCSR in what follows. On the other hand, to describe nonperturbative properties of charmonia we will use the NRQCD-inspired DAs  which exactly reproduce leading NR moments of charmonia at $\sim$ 1 GeV energies.  To resolve the right DA structure at the $\sim m_b$ energies of the decay, we calculate the Gegenbauer expansion and evolution of DAs. The corrections to the leading approximation will be done by making the twist-expansion and by taking the large $m_c$ mass correction in twist-3 and twist-4 DAs into account. The genuine $\mathcal{O}(4m_c^2/m_b \Lambda_{QCD})$  corrections are not included as we assume the collinearity of the wave functions.  Moreover, since we are aware of our model constraints in describing charmonia particles, we will show the stability of the model on the variation of parameters of the model, the consistency of our results with the 3ptSR calculation of the same form factors done with the same parameters used here and will also show consistency of the calculated form factors with the HQSS/NRQCD symmetry relations among form factors. 

The leading twist-2 DA of a $\eta_c$  meson is defined as follows \cite{Chernyak84}
\begin{eqnarray}
\langle 0 | \bar{c}(z) \gamma_{\mu} \gamma_5 [ z,-z] c(-z) |  \eta_c(p) \rangle = 
-i f_{\eta_c} p_\mu \int_{-1}^1 d \xi e^{i  \xi p z}  \phi(\xi,\mu) , 
\end{eqnarray}
%where $[z,-z]$ is a gauge factor
%where $[z,-z]= P\exp{ ig \int_{-z}^z dx^\mu A_{\mu}(x)}$ is a gauge integral, 
%\begin{eqnarray}
%[z,-z] = P\exp{ ig \int_{-z}^z dx^\mu A_{\mu}(x)}\,,
%\end{eqnarray}
while for the $J/\psi$ we have 
\begin{eqnarray}
\langle 0 | \bar{c}(z) \gamma_{\mu} [z,-z] c(-z) |J/\psi(p, \epsilon^{(\lambda=0)}) \rangle  =  f_{J/\psi} m_{J/\psi} p_{\mu}  
\int_{-1}^1 d \xi e^{i \xi p x}  \phi_{||}(\xi,\mu), 
\nonumber \\
\langle 0 | \bar{c}(z) \sigma_{\mu\nu}[z,-z] c(-z) |J/\psi(p, \epsilon^{(\lambda=\pm 1)}) \rangle  = i f_{J/\psi}^{\perp} (\epsilon_{\mu} p_{\nu} -\epsilon_{\nu} p_{\mu}  ) 
\int_{-1}^1 d u e^{i \xi p x} \phi_{\perp}(\xi,\mu), 
%\int_0^1 du \exp^{i u p_V x + i \bar{u} p_V y} \phi_{\perp}(u)
\end{eqnarray}
where $[z,-z]= P\exp{ ig \int_{-z}^z dx^\mu A_{\mu}(x)}$ is a gauge integral. 
In above $\xi = u - (1-u)$, $u$ is a fraction of a longitudinal momentum of a meson $M$ carried by a c-quark and $(1-u)$ is a fraction of  momentum carried by the c-antiquark. The DAs is defined at a scale $\mu$ at which the transverse momenta
are integrated up to and  all momenta below are
included in the nonperturbative DAs $\phi$. Other higher-twist amplitudes and higher-order corrections are defined similarly. For all details see, for example \cite{BharuchaStraubZwicky}.  
The vector and tensor  decay constants $f_{J/\psi}$ and $f_{J/\psi}^T$ are  defined as
\begin{eqnarray}
\langle 0|\bar c(0) \gamma_{\mu} c(0)| J/\psi(p,e^{(\lambda)})\rangle &=& f_{J/\psi} m_{J/\psi} e^{(\lambda)}_{\mu},
\nonumber \\
\langle 0|\bar c(0) \sigma_{\mu \nu} c(0)| J/\psi (p,e^{(\lambda)})\rangle &=& 
i f_{J/\psi}^T(\mu) (e_{\mu}^{(\lambda)}p_{\nu} - e_{\nu}^{(\lambda)}p_{\mu}) , 
\label{eq:decayC}
\end{eqnarray}
where $f_{J/\psi}^T$ is renormalization scale dependent: 
\begin{eqnarray}
f_{J/\psi}^T(\mu'^2) = \left( \alpha_s(\mu'^2)/\alpha_s(\mu^2) \right  )^{C_f/\beta_0}  f_{J/\psi}^T(\mu) 
\end{eqnarray}
and $\beta_0 = 11-2/3 n_f$, $n_f$ being the number of flavors involved.
The decay constant for $\eta_c$ is defined correspondingly as
\begin{eqnarray}
\langle 0 | \bar{c} \gamma_{\mu}\gamma_5 c |  \eta_c(p) \rangle = - i f_{\eta_c} p_{\mu}.
\end{eqnarray}
For the decay constants we will use the lattice results 
\begin{eqnarray}
\label{eq:latBc}
f_{B_c} &=& 0.427(6)(2)\, {\rm GeV}  \, ~\text{\cite{latticeBc,latticeBc1}}, 
\nonumber \\
f_{J/\psi} &=& 0.405(6)(2)\, \rm{GeV}  \, ~\text{\cite{Donald:2012ga}} ,\nonumber \\
f_{\eta_c} &=& 0.3947(24) \, {\rm GeV}  ~\text{\cite{latticeEtaC}}, 
\end{eqnarray}
while for $f_{J/\psi}^T$ we will use the value extracted from the ratio
\begin{eqnarray}
R^T_{J/\psi} =\frac{f_{J/\psi}^T(\mu= 2~{\rm GeV})}{f_{J/\psi}} = 0.975 \pm 0.010,
\end{eqnarray}
derived by considering combined QCDSR and lattice results \cite{BecirevicMelic}. The predictions for charmonia decay constants in \cite{BecirevicMelic} also nicely agree with the lattice results above.

\subsubsection{Distribution amplitudes for charmonia}
The leading twist-2 DAs 
%for the $J/\psi$ meson, $\phi_{\parallel,\perp}$, 
are expanded in terms of Gegenbauer polynomials as:
\begin{equation}
\label{Gegenbauer1}
\phi_P(u,\mu^2) = 6 u (1-u) \left( 1 + \sum\limits_{n=1}^\infty
  a_{n}^P(\mu^2) C_{n}^{3/2}(2u-1)\right).
\end{equation}
The leading term  is the asymptotic form  $\phi(u,\mu^2 \to  \infty) =6 u (1-u)$.  The Gegenbauer
coefficients $a_n$ are renormalized multiplicatively 
\begin{equation}
a_n^P(Q^2)  =  \left (   \alpha_s(Q^2)/\alpha_s(\mu^2)  \right )^{\gamma_n^P/(2\beta_0)}\, a_n^P(\mu^2),
\end{equation}
where the anomalous dimensions $\gamma^{\parallel,\perp}_n$ are given by
\begin{eqnarray}
\gamma^\parallel_n &=&  8C_F \left(\sum_{k=1}^{n+1} 1/k - \frac{3}{4} -
  \frac{2}{(n+1)(n+2)} \right),\\
\gamma^\perp_n &=&  8C_F \left(\sum_{k=1}^{n+1} 1/k - 1 \right).
\end{eqnarray}
  
Here $C_F = (N_c^2 -1)/(2 N_c)$ and $\beta_0 = 11/3 N_c - 2/3 N_f$, in which $N_c$ is the number of colors and $N_f$ the number of flavors. 
The coefficients $a_n^{||}$ appear in $\phi(\xi,\mu)$ and $\phi_{||}(\xi,\mu)$, while $a_n^{\perp}$ are coefficients in the expansion of the transversal twist-2 DA $\phi_{\perp}(\xi,\mu)$ of a $J/\psi$ meson.
%
%In the calculation of $B_c \to \eta_c$ form factors we use the following definitions:
%\begin{equation}
%\label{Gegenbauer2}
%   \phi_M(x,\mu) = 6x(1-x) \left[ 1 + \sum_{n=2}^\infty a_n^M(\mu)\,C_n^{(3/2)}(2x-1) \right] ,
%\end{equation}

%Both equations (\ref{Gegenbauer1}) and (\ref{Gegenbauer2}) can be inverted to give the coefficients of the confromal expansion 
The Eq.~(\ref{Gegenbauer1}) can be inverted to give the coefficients of the conformal expansion 
\begin{equation}
   a_n^P(\mu) = \frac{2(2n+3)}{3(n+1)(n+2)}\,\int_0^1\!du\,C_n^{(3/2)}(2u-1)\,\phi_P(u,\mu) \,,
\end{equation}
and with the help of these coefficients at some low-energy scale $\mu_0$,  the DA $\phi_P(\xi,\mu)$ can be reconstructed at any scale $\mu$. 

The distribution amplitudes can also be defined with the help of calculated moments of DAs at some scale $\mu$ as 
\begin{eqnarray}
\langle \xi^n \rangle_{\mu} = \int_{-1}^1 d\xi \xi^n \phi(\xi,\mu).
\end{eqnarray}
Charmonia particles are flavor-symmetric and therefore their DAs are symmetric around $u= 1/2$. The second moment is calculated in NRQCD \cite{Bodwin:1994jh,Braguta1}
\begin{equation}\label{mom2}
   \langle \xi^2 \rangle_{\mu_0} 
   = \frac{\langle v^2\rangle_M}{3} + {\cal O}(v^4) \,,
\end{equation}
where $\mu_0 \sim 1 {\rm GeV}$.

The values of the nonrelativistic speeds $v^2$ of quarks in $\eta_c$ and $J/\psi$ mesons are obtained in NRQCD by including the first-order $\alpha_s$ corrections and non-perturbative contributions proportional to $v^2$  in the analysis of  $\Gamma(\eta_c \to \gamma\gamma)$ and $\Gamma(J/\psi\to e^+ e^-)$ rates, respectively, and  $\langle v^2\rangle_{J/\psi}=0.225\,_{-0.088}^{+0.106}$ \cite{Bodwin:2007fz}, $\langle v^2\rangle_{\eta_c}=0.226\,_{-0.098}^{+0.123}$ \cite{Bodwin:2007fz}, $\langle v^2\rangle_{J/\psi} = \langle v^2\rangle_{\eta_c}=0.21\pm 0.02$ \cite{Braguta,Braguta1} have been extracted. As stated in \cite{NeubertKonigGrossman}, the two-loop \cite{Czarnecki:1997vz,Beneke:1997jm} and three-loop \cite{Beneke:2014qea} perturbative corrections to the NRQCD predictions for the $\Gamma(J/\psi\to e^+ e^-)$ decay rate is known to be large. In \cite{WangYang} and \cite{WangYang1} the ${\cal O}(v^2)$ and ${\cal O}(\alpha_s)$ corrections to twist-2 DAs of $\eta_c$ and $J/\psi$ have been calculated.  At leading order approximation in relative velocity $v$ there is no difference between $\eta_c$ and $J/\psi$ mesons and the results for the moments obtained are valid for both charmonia DAs. Based on the power-counting rules of NRQCD one would naively expect that $\langle v^2\rangle \sim 0.3$.  Taking all above into account, we will use the latest improved value \cite{Bodwin:2007fz,Chung:2010,Bodwin:2014} for both charmonia:
\begin{eqnarray}
\langle v^2 \rangle = 0.201 \pm 0.064 .
\end{eqnarray}

For the model of twist-2 DA at $\mu_0 = 1$ GeV we adopt the Gaussian model \cite{NeubertKonigGrossman}:
\begin{equation}\label{LCDAQQ}
   \phi(u,\mu_0) = N_\sigma\,\frac{4u(1-u)}{\sqrt{2\pi}\sigma}\,
    \exp\left[ - \frac{(u-\frac12)^2}{2\sigma^2} \right] ; \qquad
   \sigma^2 = \frac{\langle v^2\rangle_M}{12} \,,
\end{equation}
where $N_\sigma\approx 1$ is the normalization constant defined from 
\begin{eqnarray}
\int_{-1}^1 d \xi \phi(\xi,\mu) = 1.
\end{eqnarray}

We also use the Wandzura-Wilczek approximation where three-particle twist-3 DAs containing quarks and a gluon are neglected. In that case the twist-3 DAs of $\eta_c$ are fixed to their asymptotic forms including mass corrections \cite{BallBraunLenz}:
\begin{eqnarray}
\phi_p(u,\mu)\big|_{\rm WWA} &=& 1 + \rho_+(\mu) \phi_{p,+}(u,\mu), \nonumber \\
\phi_{\sigma}(u,\mu)\big|_{\rm WWA}&=& 6 u(1-u) + \rho_+ (\mu)\phi_{\sigma,+}(u,\mu),
\end{eqnarray}
where  $\rho_+(\mu) = 4 m_c^2(\mu)/m_{\eta_c}^2$ and
\begin{eqnarray}
\phi_{p,+}(u,\mu)&=&  \frac{1}{4} \left [  \int_0^u\!dv\,\frac{\phi^{'}(v,\mu)}{1-v}  - \int_u^1\!dv\,\frac{\phi^{'}(v,\mu)}{v}    \right ]  \,, \nonumber \\
\phi_{\sigma,+}(u,\mu) &= & - \frac{3}{2} u (1-u)\left [  \int_0^u\!dv\,\frac{\phi(v,\mu)}{(1-v)^2}  + \int_u^1\!dv\,\frac{\phi(v,\mu)}{v^2}    \right ]. 
\end{eqnarray}

For the $J/\psi$ meson the situation is somewhat more complicated. 
In the Wandzura-Wilczek approximation where three-particle DAs are neglected, by using equations of motion the twist-3 DAs can be expressed in terms of the leading twist-2 DAs $\phi_{\parallel,\perp}$  with the valence quark mass corrections 
\begin{eqnarray}
\delta_+(\mu) = \frac{2 m_c(\mu)}{m_{J/\psi}} \frac{1}{R_{J\psi}^T}\,, \quad 
\tilde{\delta}_+(\mu) = \frac{2 m_c(\mu)}{m_{J/\psi}} R_{J\psi}^T \,,
\end{eqnarray}
as \cite{ BallBraun1,BallBraun11,BallBraun2,BenekeFeldmann,BallBraunLenz1}:
\begin{eqnarray}
\tilde{h}_{||}^{(s)} = (1 - \delta_+(\mu)) h_{||}^{(s)}\, , \qquad
\tilde{g}_{\perp}^{(a)}  = (1 - \tilde{\delta}_+(\mu)) g_{\perp}^{(a)}\, , \nonumber
\end{eqnarray}
and 
\begin{eqnarray}
   g_\perp^{(v)}(x,\mu) \big|_{\rm WWA} 
   &=& \frac{1}{4} \left[ \int_0^u\!dv\,\frac{\Phi_{||}(y,\mu)}{1-v} 
    + \int_u^1\!dv\,\frac{\Phi_{||}(v,\mu)}{v} \right] + \tilde{\delta}_+(\mu)\phi_{\perp}(u,\mu), \\
 \tilde{g}_\perp^{(a)}(x,\mu) \big|_{\rm WWA} 
   &=&  (1-u)\int_0^u\!dv\,\frac{\Phi_{||}(v,\mu)}{1-v} 
    + u \int_u^1\!dv\,\frac{\Phi_{||}(v,\mu)}{v} , \\
h_{||}^{(t)}(u,\mu)\big|_{\rm WWA} &=&
   \frac{1}{2} \xi \left[ \int_0^u\!dv\,\frac{\Phi_{\perp}(y,\mu)}{1-v} 
    - \int_u^1\!dv\,\frac{\Phi_{\perp}(v,\mu)}{v} \right] + \delta_+(\mu) \phi_{\parallel}(u,\mu), \\
\tilde{h}_{||}^{(s)}(u,\mu)\big|_{\rm WWA} &=&
     (1-u) \int_0^u\!dv\,\frac{\Phi_{\perp}(v,\mu)}{1-v} 
    + u \int_u^1\!dv\,\frac{\Phi_{\perp}(v,\mu)}{v} ,
%h_{||}^{(t)}_{|WWA}(u,\mu) &=& (2 u-1) \Phi_v(u,\mu),\nonumber \\ 
%h_{||}^{'(s)}_{|WWA}(u,\mu) &=& -2 \Phi_v(u,\mu)\nonumber \\ 
%\int_0^u\!dv \left[ \phi_V^\perp(v,\mu) - h_{||}^{(t)}(v,\mu) \right]_{\rm WWA} &=& %x(1-x)\,\Phi_v(x,\mu) \,,
\end{eqnarray}
with 
\begin{eqnarray}
\Phi_{||}(u) &=& 2 \phi_{||}(u) + \tilde{\delta}_+ \xi \phi_{\perp}^{'}(u) \,, \nonumber \\
\Phi_{\perp}(u) &=& 2 \phi_{\perp}(u)  - \delta_+ \left (\phi_{\parallel}(u) -  \frac{\xi}{2} \phi_{\parallel}^{'}(u) \right ).
\end{eqnarray}
The $J/\psi$ twist-4 DAs will be taken in their asymptotic form:
\begin{eqnarray}
h_{\perp,3} &=& 6 u (1-u) \,, \quad g_{\parallel,3} = 6 u (1-u)\,, \nonumber \\
A_{||} &=& 24 u^2 (1-u)^2\,, \quad A_{\perp} = 12 u^2 (1-u)^2\,.
\end{eqnarray}

Some comments are in order. 
In the $B_c \to \eta_c$ decay we will retain only contributions up to twist-3 terms. It is well known that the standard twist expansion works very well for $B \to$ pseudoscalar form factors. It could be that in our case, for $B_c \to \eta_c$, twist-4 corrections are somewhat larger, due to the large and non-negligible $m_{c}$ mass, but since hadronic parameters for the twist-4 contribution for $\eta_c$ are not known we will not include them. 
In the decay $B_c \to J/\psi$ we keep all contributions up to twist-4, since their asymptotic form does not depend on the hadronic parameters. The $J/\psi$ DAs defined above do not correspond to matrix elements of operators with definite twist \cite{BallZwicky2004}: 
$\phi_{\perp,\parallel}$ are of twist-2,
$h_{\parallel}^{(s,t)}$ and $g_\perp^{(v,a)}$ contain a mixture of twist-2 and 3 contributions and ${\mathbb A}_{\perp,\parallel}$, $h_3$, $g_3$ are a mixture of twist-2, 3 and 4 contributions. Therefore it is usual to refer to  $g_\perp^{(v,a)},
h_\parallel^{(s,t)}$ as twist-3 LCDAs and to $h_3,g_3,{\mathbb
  A}_{\perp,\parallel}$ as twist-4 LCDAs. Also, as the mass of the vector particle in $B \to$ vector decays plays a significant role, in \cite{BallZwicky2004} it was proposed the following classification of relevance in the two-particle LCDA: 
$O(\delta^0):$ $\phi_\perp$;
$O(\delta^1):$ $\phi_\parallel,g_\perp^{(v,a)}$;
$O(\delta^2):$ $h_\parallel^{(s,t)},h_3,{\mathbb A}_\perp$;
$O(\delta^3):$ $g_3,{\mathbb A}_\parallel$,  
where now $\delta \sim m_{J/\psi}$ is treated as an expansion parameter. For more detailed discussion see \cite{Ball:2004ye,Ball:1998kk,BharuchaStraubZwicky}.   

\subsection{Parametrization of the form factors and the results}

The derivation of the sum rule expressions for the form factors proceeds in a standard way \cite{Ball:1998kk,Ball:2004ye,Altmannshofer:2008dz,BharuchaStraubZwicky}. 

The general expression for the calculation of the form factors is given by
\begin{eqnarray}
\label{eq:LCSR-ff}
F_{B_c \to M}(q^2) = \frac{m_b + m_c}{m_{B_c}^2 f_{B_c}} e^{m_b^2/\sigma^2} \int_{u_0^{B_c}}^1 \frac{d u}{u} \exp\left [ - \frac{m_b^2 - \bar{u}q^2 - u \bar{u} m_M^2}{u \sigma^2}\right ]  F(u,\mu, q^2),
\end{eqnarray}
where 
\begin{eqnarray}
u_0^{B_c} = \frac{1}{2 m_M^2} \sqrt{ (s_0^{B_c} - q^2 -m_M^2)^2 + 4 m_M^2 (m_b^2 - q^2)} -(s_0^{B_c} - q^2 -m_M^2), 
\end{eqnarray}
$\sigma$ is a Borel parameter and $\bar{u} = 1-u$. The functions $F(u,\mu,q^2)$ contain all twist contributions in terms of the  various twist DAs, and  the higher-twist contributions are suppressed  by the Borel parameter.  

The derived results at $q^2 =0$ are listed in Table~\ref{tab:ffactor}, together with the recent QCDSR result \cite{preparation} briefly discussed in Appendix A and earlier results found in the literature on the same form factors. The errors are obtained by varying all parameters in a given range and adding them in quadratures. %Note that we completely disagree with a previous LCSR calculation of the form factors in \cite{Huang:2007kb}.
\begin{table}[ht]
\centering
%\captionsetup{justification=centering,margin=1.4cm} \footnotemark
\addtolength{\tabcolsep}{-3pt}
\renewcommand{\arraystretch}{1.1}
 \begin{tabular}{||c || c|| c c c c c c c c || c ||}
 \hline
 \begin{tabular}{@{}c@{}}Form \\ Factor \end{tabular} & \begin{tabular}{@{}c@{}}this work \tablefootnote{The cited errors are obtained just by varying all parameters of the model and their smallness shows the stability of the sum rules used to obtain the predictions for the form factors. The errors do not include intrinsic uncertainties of the model itself which are hard to predict and could potentially increase the errors and lower the accuracy of the predictions.}
 \end{tabular} &  
 \begin{tabular}{@{}c@{}} QCDSR \\ \cite{preparation} \end{tabular} &  \begin{tabular}{@{}c@{}}QCDSR\\ ~\cite{Kiselev:2002vz}\end{tabular} &
 \begin{tabular}{@{}c@{}}SR \\ \cite{Huang:2007kb} \end{tabular} & \begin{tabular}{@{}c@{}}pQCD\\ ~\cite{Wen-Fei:2013uea}\end{tabular} & \begin{tabular}{@{}c@{}}CCQM\\ ~\cite{Tran:2018kuv,Issadykov:2017wlb}\end{tabular} & \begin{tabular}{@{}c@{}}RQM\\ ~\cite{Ebert:2003cn}\end{tabular} & \begin{tabular}{@{}c@{}}RQM\\ ~\cite{Nobes:2000pm}\end{tabular} & \begin{tabular}{@{}c@{}}LFQM\\ ~\cite{Wang:2008xt}\end{tabular} & \begin{tabular}{@{}c@{}}latt.\\ \cite{Colquhoun:2016osw}\end{tabular}\\ [0.5ex]
 \hline\hline
 $f^{\eta_c}_{+,0}(0)$ & $0.62\pm0.05$ &  $0.41\pm0.04$ & 0.66 & 0.87 & 0.48(7) & 0.75 & 0.47 & 0.54 & 0.61(5) & \textcolor{orange}{0.59}\\
 \hline
 $V^{J/\psi}(0)$ & $0.73\pm0.06$ &  $0.70\pm0.06$ & 1.03 & 1.69 & 0.42(2) & 0.78 & 0.49 & 0.73 & 0.74(4) & \textcolor{orange}{0.70}\\
 \hline
 $A_1^{J/\psi}(0)$ & $0.55\pm0.04$  & $0.50\pm0.05$ & 0.63 & 0.75 & 0.46(3) & 0.56 & 0.50 & 0.52 & 0.50(3) & \textcolor{orange}{0.48}\\
 \hline
 $A_2^{J/\psi}(0)$ & $0.35\pm0.03$  & $0.43\pm0.05$ & 0.69  & 1.69 & 0.64(3) & 0.55 & 0.73 & 0.51 & 0.44(5) & -\\
 \hline
 $A_0^{J/\psi}(0)$ & $0.54\pm0.04$  & $0.53\pm0.04$ & 0.60 & 0.27 & 0.59(3) & 0.56 & 0.40 & 0.53 & 0.53(3) & -\\
 \hline \hline
 $f^{\eta_c}_T(0)$ & $0.93\pm 0.07$ & - & - & - & - & 0.93 & - & - &  - & -\\
 \hline
 $T_{1,2}^{J/\psi}(0)$ & $0.47\pm 0.04$ & $0.48\pm0.03$ & - & - & - & 0.56 & - & - & - & -\\
 \hline
 $T_3^{J/\psi}(0)$ & $0.19\pm 0.01$ & $0.27\pm0.03$ & - & - & - & 0.20 & - & - & - & -\\
 \hline
\end{tabular}
\captionof{table}{Form factor predictions at $q^2 =0$. Recent relevant lattice results are given by the HPQCD collaboration \cite{Colquhoun:2016osw}, reported here in orange, without the systematical error.\label{tab:ffactor}} 
\end{table}
%\subsection{Parametrization of the form factors and the results}
%\footnotetext{The cited errors are obtained just by varying all parameters of the model and their smallness shows the stability of the sum rules used to obtain the predictions for the form factors. The errors do not include intrinsic uncertainties of the model itself which are hard to predict and could potentially increase the errors and lower the accuracy of the predictions.}

It is well know that the form factors extracted from the sum rules are valid in the low $q^2$ region. We use our results for the form factors and calculate them in the range $q^2 = \{-5,5\}$ GeV.  Then we extrapolate them from the low $q^2$ region to the $q_{\rm max}^2$ by using Bourrely-Caprini-Lellouch (BCL) parametrization \cite{BCL} of the form factor series expansion in powers of a conformal mapping variable, which satisfies unitarity, analyticity and perturbative QCD
scaling. The BCL parametrization is based on a rapidly converging series in
the parameter $z$ as 
\begin{eqnarray}
f(t) &=& \frac{1}{P(t)} \sum_{k=0} \alpha_k z^k( t,t_0), \nonumber \\
z( t,t_0)  &=&  \frac{ \sqrt{t_+ - t} - \sqrt{t_+ - t_0}}{\sqrt{t_+ - t} + \sqrt{t_+ - t_0}} ,
\label{z-exp}
\end{eqnarray}
weighted by a simple pole function $P(q^2) = 1- t/m_R^2$ which accounts for low-laying resonances present below the threshold production of real $B_c-M$ pairs at $t_+ = (m_{B_c} + m_M)^2$.  
The parameter $t_0$, $0 \leq t_0 \leq t_- = (m_{B_c} - m_M)^2$ is a free parameter that can be used to optimize the convergence of the series expansion.  For the truncation to only two terms in expansion Eq.~(\ref{z-exp}), it was shown that the optimized value of $t_0$ has the form \cite{BharuchaFeldmann2010}:
\begin{eqnarray}
t_{0{|\rm opt}} =  t_+  \left (1 - \sqrt{1 - \frac{t_-}{t_+}} \right ).
\end{eqnarray}
and that the other choices of $t_0$ do not make a visible change in the form factors parametrization. 

Masses of resonances appearing in the fits are determined by the properties of the form factors. 
The form factors $V$ and $T_1$ correspond to the vector components of the currents, and, as the $B_c$ meson is a pseudoscalar, they correspond to the axial vector components of the matrix elements. $A_{1,2}$, as well as $T_{2,3}$, correspond to the axial vector component of the $V-A$, while the form factor $A_0$ correspond to the pseudoscalar current and only contributes in the decays with the non-vanishing lepton masses (in the semileptonic $B_c$ decays with the $\tau$ lepton in our case). 
All relevant resonance masses are given in Table~\ref{tab:BCLfit}, together with the fitted parameters $\alpha_0, \alpha_1$ from Eq.~(\ref{z-exp}).

\begin{figure}[!htb]
%        \captionsetup{justification=centering,margin=1.4cm}
        \centering
        \includegraphics[width=0.3\textwidth]{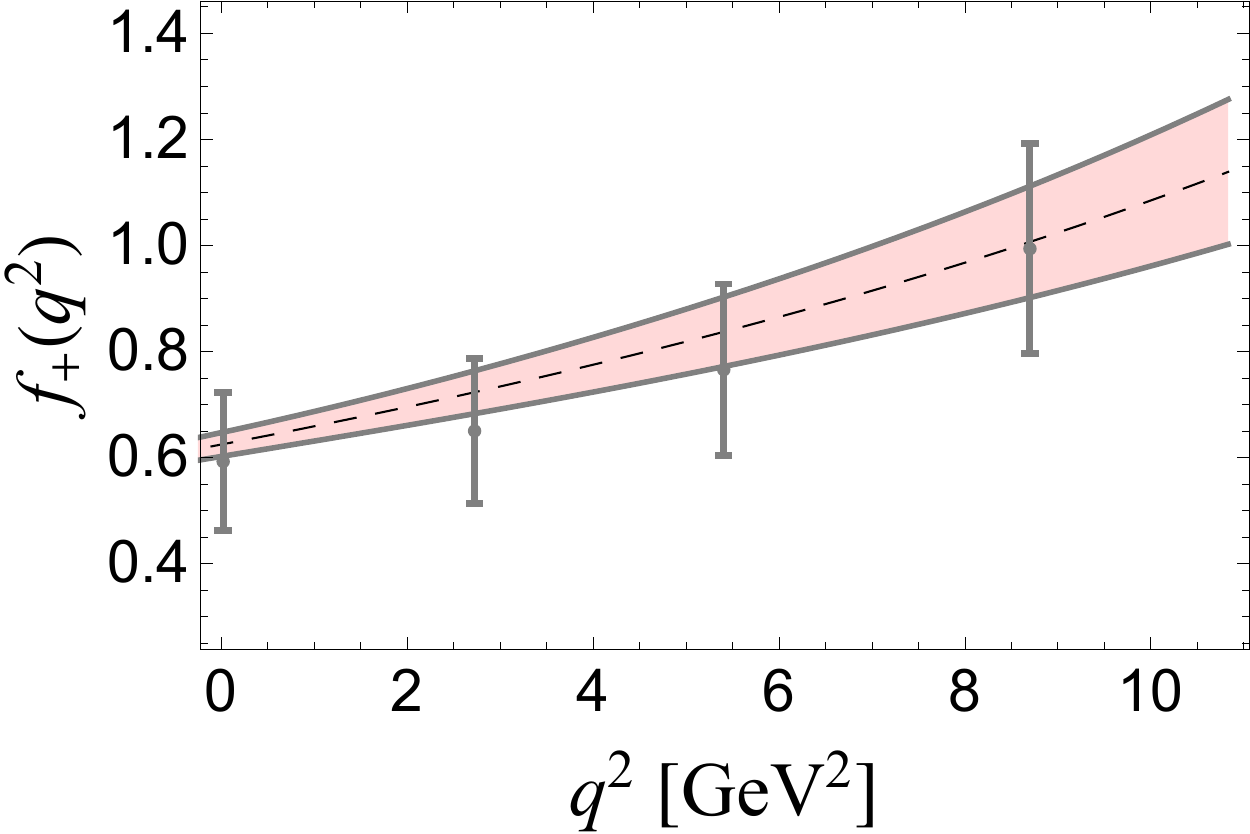}
        \includegraphics[width=0.3\textwidth]{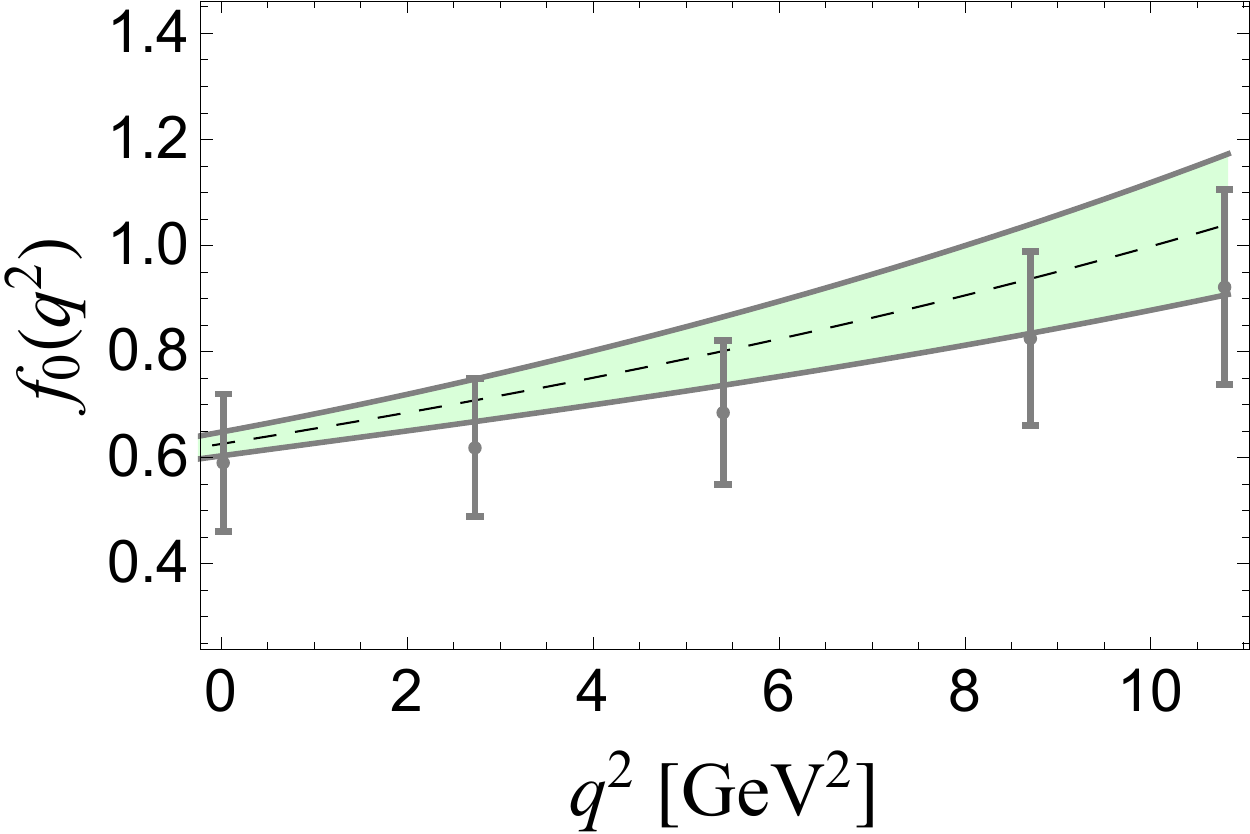}
        \includegraphics[width=0.3\textwidth]{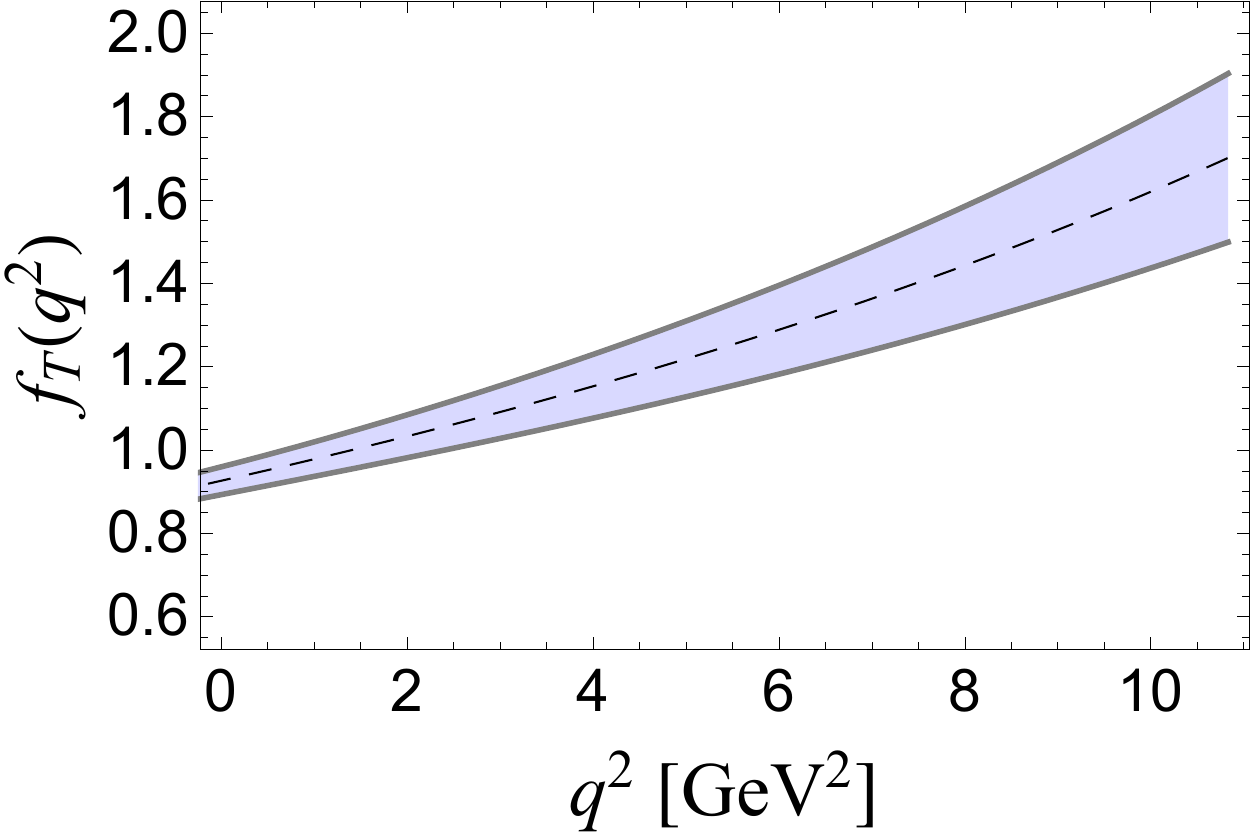}
         \caption{Pseudoscalar form factors for $B_c\rightarrow \eta_c$ calculated in this paper, including the lattice points from \cite{Colquhoun:2016osw} with added 20\% systematical error.}\label{fig:LCSR_EtaC}
    \end{figure}

\begin{figure}[!htb]
 %       \captionsetup{justification=centering,margin=1.4cm}
       \centering
        \includegraphics[width=0.45\textwidth]{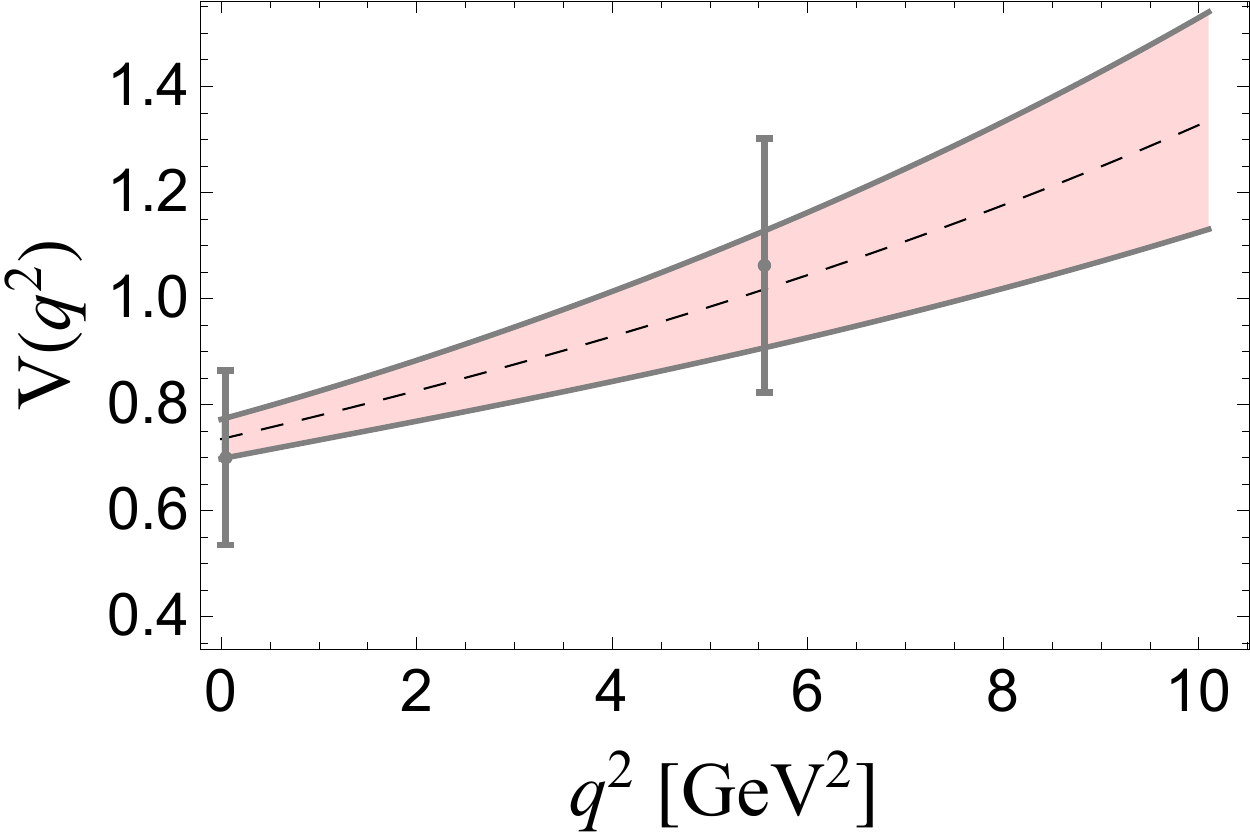}
        \includegraphics[width=0.45\textwidth]{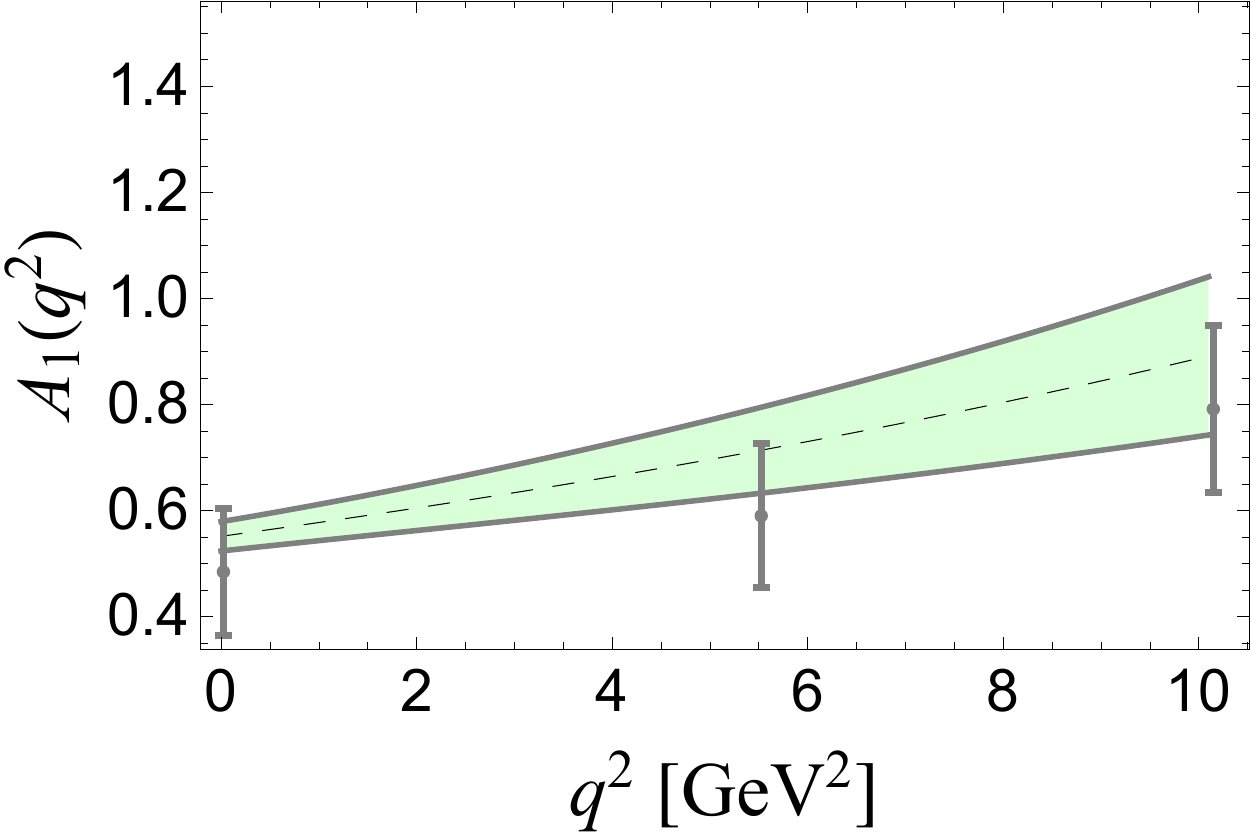}
        \includegraphics[width=0.45\textwidth]{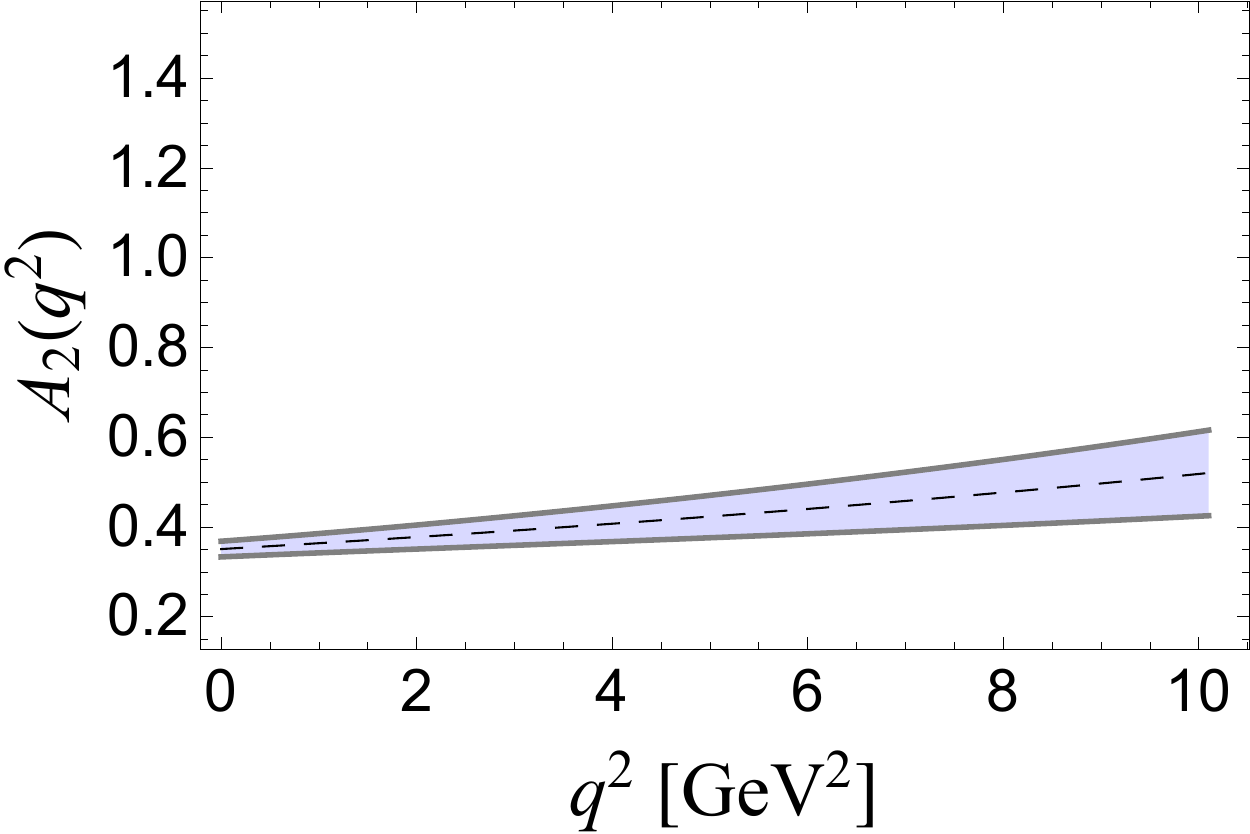}
        \includegraphics[width=0.45\textwidth]{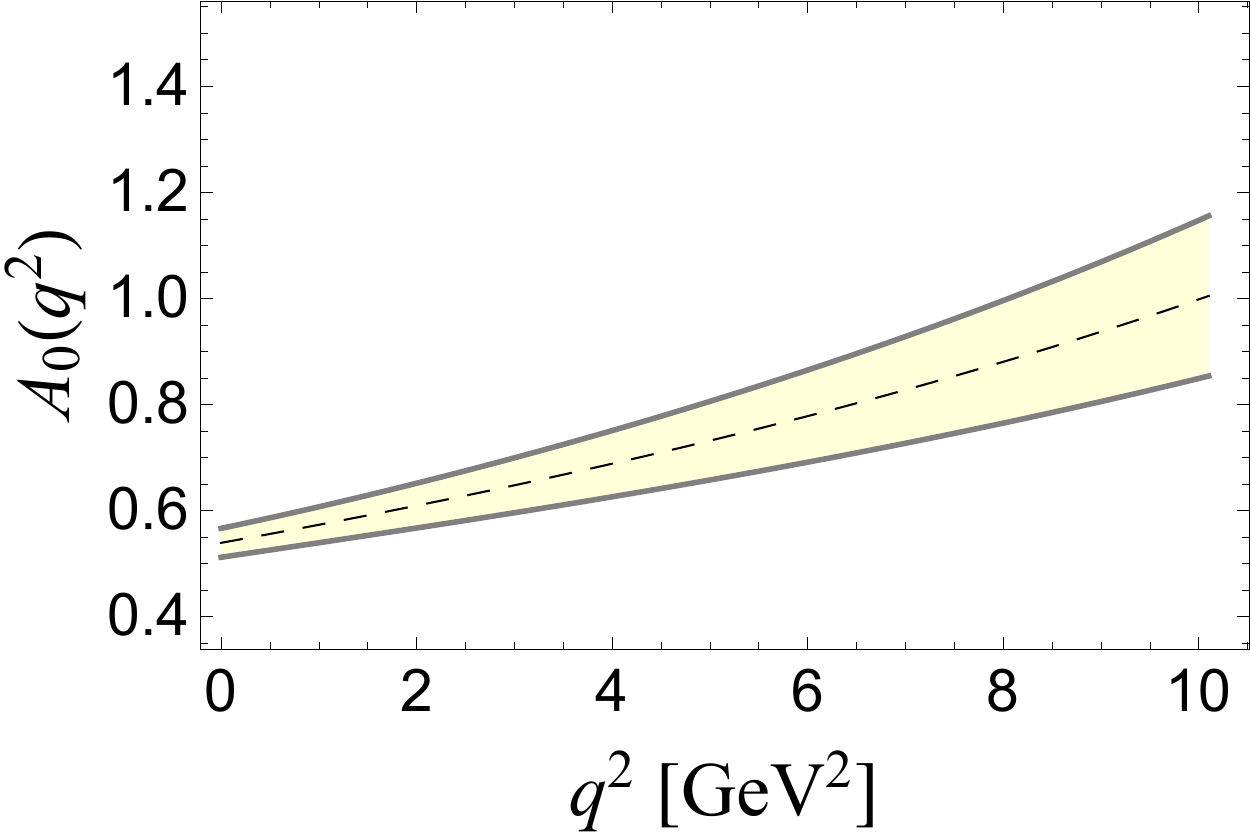}
         \caption{SM form factors for $B_c\rightarrow J/\psi$ calculated in this paper, including the lattice points from \cite{Colquhoun:2016osw} with added 20\% systematical error.} 
         \label{fig:LCSR_Jpsi1}
    \end{figure}

\begin{figure}[htb]
 %       \captionsetup{justification=centering}
       \centering
        \includegraphics[width=0.32\textwidth]{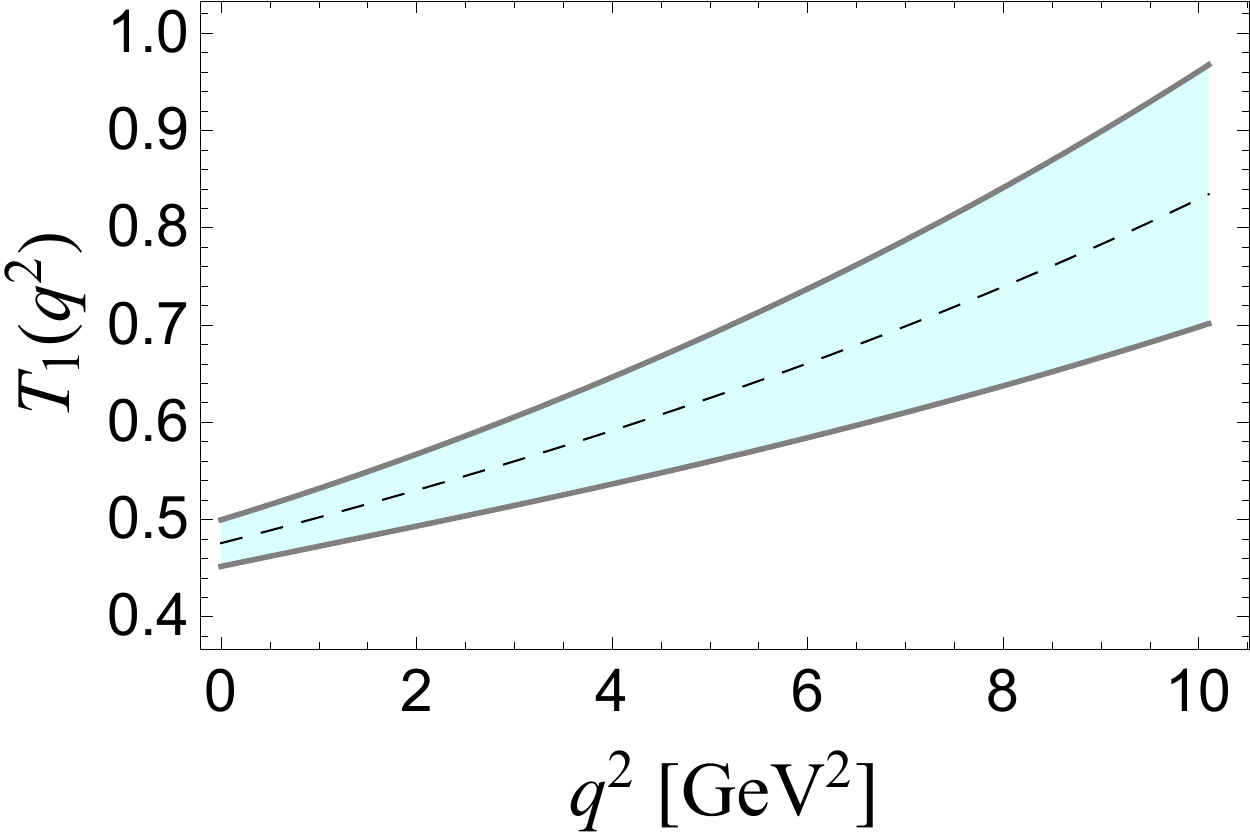}
        \includegraphics[width=0.32\textwidth]{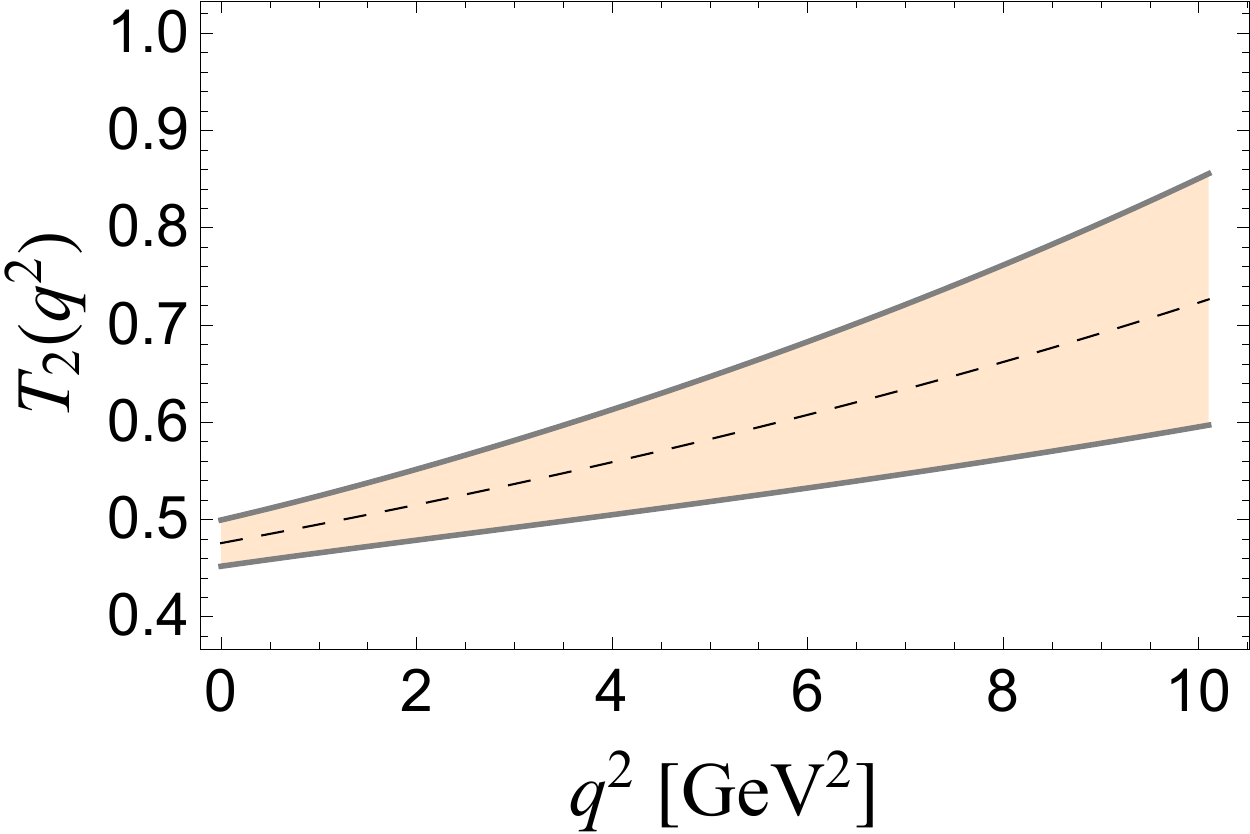}
        \includegraphics[width=0.32\textwidth]{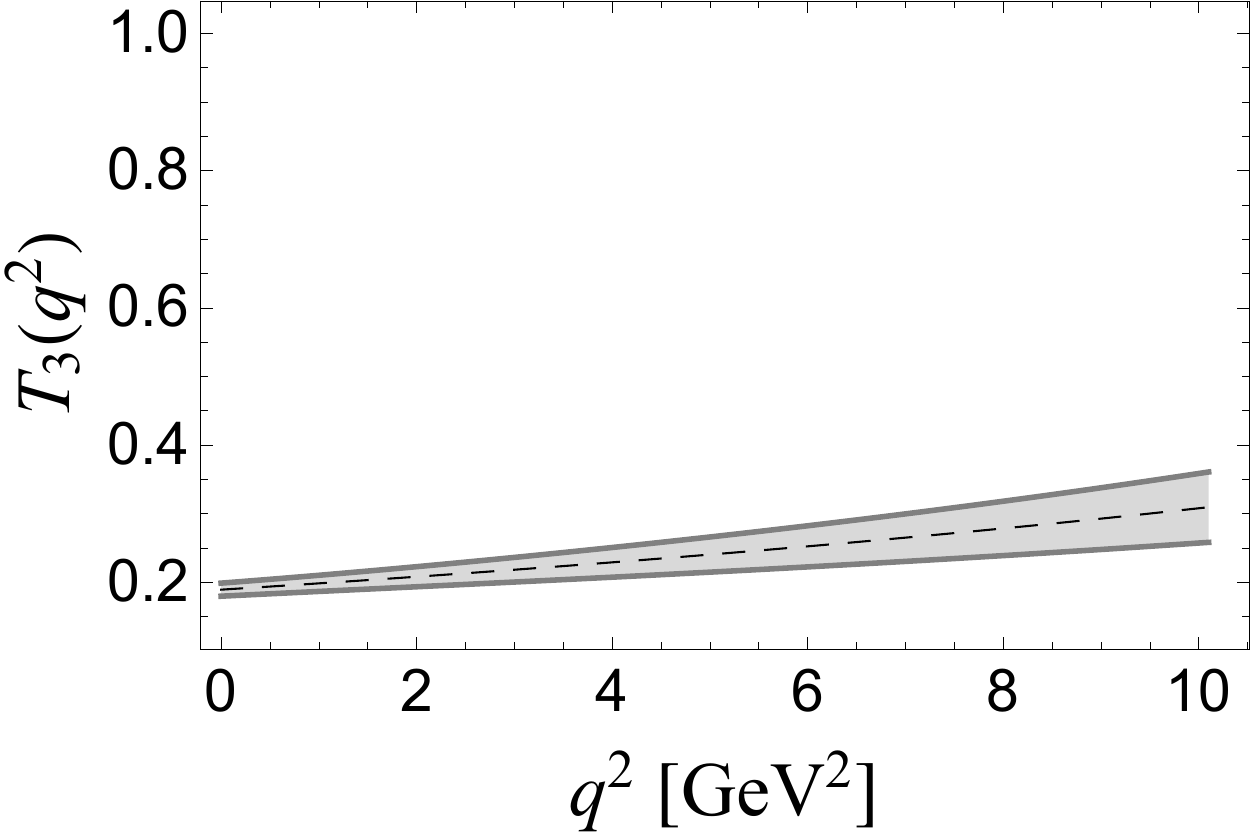}
               \caption{Tensor form factors for $B_c\rightarrow J/\psi$ calculated in this paper.}  \label{fig:LCSR_Jpsi2}
    \end{figure}

The predicted form factors in a full $q^2$ range are shown in Figs.~\ref{fig:LCSR_EtaC},~\ref{fig:LCSR_Jpsi1},~\ref{fig:LCSR_Jpsi2}.
%%%%%%%%%%%%%%%%%%%%%
\begin{table}[ht]
\centering
\addtolength{\tabcolsep}{-3pt}
\renewcommand{\arraystretch}{1.1}
 \begin{tabular}{||c | c c c c ||}
 \hline
Form factor &  $J^P$ & $m_R$  (GeV) & $\alpha_0$ & $\alpha_1$ \\ [0.5ex]
 \hline\hline
 $f_+$ & $1^-$& 6.34 & 0.62 & -6.13 \\
$f_0$ & $0^+$ & 6.71 & 0.63 & -4.86 \\
$f_T$ & $1^-$ & 6.34 & 0.93 & -9.36 \\
\hline
$V$ & $1^-$ & 6.34 & 0.74 & -8.66 \\
$A_1$ & $1^+$ & 6.75 & 0.55 & -4.67 \\
$A_2$ & $1^+$ & 6.75 & 0.35 & -1.78 \\
$A_0$ & $0^-$ & 6.28 & 0.54 & -6.80 \\
$T_1$ & $1^-$ & 6.34 & 0.48 & -4.88 \\
$T_2$ & $1^+$ & 6.75 & 0.48 & -2.93 \\
$T_3$ & $1^+$ & 6.75 & 0.19 & -1.69 \\
\hline
\end{tabular}
\captionof{table}{Summary of the BCL fit for $B_c \to \eta_c$  and $B_c \to J/\psi$ form factors. The masses of the low-laying $B_c$ resonances are taken  from \cite{Aad:2014laa,Aaij:2016qlz,Eichten:1994gt,Godfrey:2004ya}\label{tab:BCLfit}} 
\end{table}
%%%%%%%%%%%%%%%%%%%%%%%%%%%%%%%%%%%

%%%%%%%%%%% FIGURES - results
%%%%%%%%%%%%%%%%%%%%%%%%%%%%%%%%%%%%%%%%%%%%%%%%%%%%%
\subsubsection{HQSS/NRQCD symmetry relations among from factors at the zero recoil}
\label{sec:HQrelations}

It is interesting to check the HQSS and NRQCD limits of the form factors for $B_c \to \eta_c$ and $B_c \to J/\psi$ decays. These decays are specific since the decay proceeds through $b\to c$ quark decay and the  produced final state is a particle formed by two $c$ quarks. And although they look like as a heavy-to-heavy transitions and produce interesting symmetries in a heavy-quark limit \cite{Jenkins:1992nb}, the $c$-quark is significantly lighter than $b$ and the produced $c$-quark is quite energetic, which spoils exact heavy-flavor symmetries. On the other hand $c$-quark is heavy enough that such decays can be considered as  nonrelativistic so that the approximation of the zero-recoil point, i.e. the symmetry relations for a maximum momentum transfer $q^2_{\rm max} = (m_{B_c} - m_{J/\psi,\eta_c})^2$ still hold and the form factors can be related to a single function $\Delta$ \cite{Jenkins:1992nb, Kiselev:1999sc, Colangelo:1999zn}, with an unknown normalization. 
Following \cite{Jenkins:1992nb} we write for the form factors near zero recoil ($q' \ll m_c$): 
\begin{eqnarray}
	\bra{\eta_c(v,q^{\prime})}V_{\mu}(q^2)\ket{B_c(v)} & =& 2\sqrt{m_{B_c}m_{\eta_c}}\, \Delta(a_0q^{\prime})\,v_{\mu}, \\
	\bra{J/\psi(v,q^{\prime})}A_{\mu}(q^2)\ket{B_c(v)} & =& 2\sqrt{m_{B_c}m_{J/\psi}}\, \Delta(a_0q^{\prime})\,\epsilon_{\mu}^{\ast}, 
\end{eqnarray}
where $V_{\mu}= \bar{b}\gamma_{\mu}c$ and $A_{\mu} =\bar{b}\gamma_{\mu}\gamma_5 c $ and   $\epsilon_{\mu}$ is a polarization vector of $J/\psi$.  Here $v$ is the velocity of the $B_c$ meson, and $q^{\prime}$ is a small residual velocity carried by the final state meson (not to be confused by $q$, the momentum carried by the lepton pair system), so that
\begin{equation}
    \begin{split}
        p_{B_c\mu}=m_{B_c}v_{\mu}; \quad (p_{\eta_c,J/\psi})_{\mu}=m_{\eta_c,J/\psi}v_{\mu} + q^{\prime}_{\mu}.
    \end{split}
\end{equation}
The parameter $a_0$ is connected to the Bohr radius of the $B_c$ meson, it value is not important for the further discussion and will not be discussed here. 

We can now relate the $\Delta(a_0 q')$ function to the $B_c \to \eta_c$ form factor $f_+(q^2)$ at the zero recoil as 
\begin{equation}
     \Delta(a_0 q') \approx \sqrt{\frac{m_{B_c}}{m_{\eta_c}}}f_+(q_{\rm max}^2) ,
\end{equation}
which amounts, using the predicted $f_+(q_{\rm max}^2)$ from the calculation above, to 
\begin{equation}
    \Delta(a_0 q')_{\rm our} \approx 0.79\pm 0.09 .
\end{equation}
This value can be compared with the value obtained in the QCD relativistic potential model in  ~\cite{Colangelo:1999zn}.

In \cite{Kiselev:1999sc} it was shown that in the NRQCD approximation one can derive a generalized set of relations using the HQSS, so that the transition form factors of $B_c \to \eta_c$ and $B_c \to J/\psi$ decays can be given in terms of a single form factor, even for the case of non-equal four-velocities $v_1 \neq v_2$, of the initial and the final state heavy mesons. 
%In \cite{Kiselev:1999sc} it was shown that in the NRQCD approximation \textcolor{yellow}{(the heavy-quark spin-symmetry relations can be generalized to)} \textcolor{red}{one can derive} a \textcolor{red}{generalized} set of \textcolor{yellow}{(symmetry)} relations \textcolor{red}{using the HQSS}, so that the transition form factors of $B_c \to \eta_c$ and $B_c \to J/\psi$ decays can be given in terms of a single form factor, even for the case of non-equal four-velocities $v_1 \neq v_2$ \textcolor{yellow}{of the} \textcolor{red}{, respectively of initial and final state} heavy mesons \textcolor{yellow}{in the initial and final state}. 
If the following helicity basis for the form factors in $B_c \to J/\psi$ decay is defined
\begin{equation}
    \begin{split}
        g(q^2) & \equiv (H_{++}-H_{--})/\sqrt{\lambda(m_{B_c},m_{J/\psi},q^2)} = \frac{2}{m_{B_c}+m_{J/\psi}}V(q^2), \\
    f(q^2) & \equiv -(H_{++}+H_{--})/2 = (m_{B_c}+m_{J/\psi})A_1(q^2), \\
  \mathcal{F}_1(q^2) & \equiv -\sqrt{q^2}H_{00} \\
        & = \frac{1}{m_{J/\psi}}\bigg[-\frac{\lambda(m_{B_c},m_{J/\psi},q^2)}{2(m_{B_c}+m_{J/\psi})}A_2(q^2)\\
        & \qquad\,\,\,\,\,-\frac{1}{2}(q^2-m_{B_c}^2+m_{J/\psi}^2)(m_{B_c}+m_{J/\psi})A_1(q^2)\bigg], \\
 \mathcal{F}_2(q^2) & \equiv -2 \frac{\sqrt{q^2}}{\sqrt{\lambda(m_{B_c}^2,m_{J/\psi}^2,q^2)}} H_{t0} = 2 A_0(q^2),
    \end{split}
\end{equation}
and
\begin{equation}
    \lambda(m_{B_c},m_{J/\psi},q^2)=(q^2+m_{B_c}^2-m_{J/\psi}^2)^2-4m_{B_c}q^2. 
\end{equation}
the expressions from \cite{Kiselev:1999sc}, stemming from considering NRQCD and HQSS, and  relating different decay form factors at the point of zero recoil $q^2_{\text{max}}$ of $\bar{Q}q\rightarrow\bar{Q}^{'}q$ transitions can be expressed as \cite{Cohen:2018dgz}:
\begin{equation}
\label{eq:HQSSV}
    \begin{split}
        g(q^2_{\text{max}}) & = \frac{3+r_Q}{4m_{B_c}^2r_{J/\psi}}f(q^2_{\text{max}}), \\
    \mathcal{F}_1(q^2_{\text{max}}) & = m_{B_c}(1-r_{J/\psi})f(q^2_{\text{max}}), \\
    \mathcal{F}_2(q^2_{\text{max}}) & = \frac{2(1+r_{J/\psi})+(1-r_{J/\psi})(1-r_Q)}{4m_{B_c}r_{J/\psi}}f(q^2_{\text{max}}), 
    \end{split}
\end{equation}
for the $B_c\rightarrow J/\psi$ decay, and 
\begin{equation}
\label{eq:HQSSPS}
    f_0(q^2_{\text{max}})=\frac{1}{m^2_{B_c}-m^2_{\eta_c}}\frac{8m_{B_c}^2(1-r_{\eta_c})r_{\eta_c}}{2(1+r_{\eta_c})+(1-r_{\eta_c})(1-r_Q)}f_+(q^2_{\text{max}}), 
\end{equation}
for the $B_c\rightarrow \eta_c$ decay~\cite{Berns:2018vpl}, where some shorthand notation has been introduced:  $r_M=m_M/m_{B_c}$ (with $m_M=[m_{J/\psi},m_{\eta_c}]$), $r_Q=m_{Q^{'}}/m_Q=m_c/m_b$ and $r_q=m_q/m_Q=r_Q$. 
Additionally, the vector decay form factors can be related to the pseudoscalar ones as  \cite{Murphy:2018sqg}
\begin{equation}
\label{eq:HQSSVPS}
    \begin{split}
        f(q^2_{\text{max}}) & = \frac{8m_{B_c}r_{\eta_c}}{3+r_{\eta_c}-(1-r_{\eta_c})r_Q}f_+(q^2_{\text{max}}), \\
        g(q^2_{\text{max}}) & = \frac{1+r_Q}{m_{B_c}r_{J/\psi}}\frac{4r_{\eta_c}}{3+r_{\eta_c}-(1-r_{\eta_c})r_Q}f_+(q^2_{\text{max}}), \\
        \mathcal{F}_1(q^2_{\text{max}}) & = m_{B_c}^2(1-r_{J/\psi})\frac{8r_{\eta_c}}{3+r_{\eta_c}-(1-r_{\eta_c})r_Q}f_+(q^2_{\text{max}}), \\
        \mathcal{F}_2(q^2_{\text{max}}) & = \frac{1+r_{J/\psi}}{r_{J/\psi}}\frac{4r_{\eta_c}}{3+r_{\eta_c}-(1-r_{\eta_c})r_Q}f_+(q^2_{\text{max}}), 
    \end{split}
\end{equation}
where we have used that $r_q = r_Q$ for $B_c \to \eta_c,J/\psi$ and simplified the relations. 
It is expected that these relations are broken by terms of order $\mathcal{O}(m_c/m_b,\Lambda_{\text{QCD}}/m_c)\lesssim 30\%$. 

Here we check the consistency of these relations and our form factor results.  At the zero-recoil $\mathcal{F}_1(q^2)$ and $f(q^2)$ are the same up to a constant factor, Eq.~(\ref{eq:HQSSV}), which explicitly gives
\begin{equation}
 \frac{\mathcal{F}_1(q^2_{\text{max}})}{f(q^2_{\text{max}})}  = m_{B_c}(1-r_{J/\psi}) = 3.18 \,,
\end{equation}
whereas 
\begin{equation}
    \begin{split}
        \frac{g(q^2_{\text{max}})|_{}}{g(q^2_{\text{max}})|_{{\rm Eq.}(\ref{eq:HQSSV})}} & \approx 0.81, \\
        \frac{\mathcal{F}_2(q^2_{\text{max}})|_{}}{\mathcal{F}_2(q^2_{\text{max}})|_{{\rm Eq.}(\ref{eq:HQSSV})}} & \approx 0.89, \\
    \end{split}
\end{equation}
and
\begin{equation}
   \frac{f_0(q^2_{\text{max}})|_{}}{f_0(q^2_{\text{max}})|_{{\rm Eq.}(\ref{eq:HQSSPS})}}\approx 1.18.
\end{equation}
We see that the HQSS/NRQCD predictions are quite consistent with our sum rule  predictions for the form factors at the zero recoil and can be safely used in  model-independent bounds on $R$ ratios as it was done in \cite{Cohen:2018dgz,Berns:2018vpl,Murphy:2018sqg}, keeping in mind that their accuracy is limited to ${\cal O}(30\%)$.
Finally, using Eq.~(\ref{eq:HQSSVPS}) we obtain, 
\begin{equation}
    \begin{split}
        \frac{f(q^2_{\text{max}})|_{}}{f(q^2_{\text{max}})|_{{\rm Eq.}(\ref{eq:HQSSVPS})}} & \approx 1.02,\\
        \frac{g(q^2_{\text{max}})|_{}}{g(q^2_{\text{max}})|_{{\rm Eq.}(\ref{eq:HQSSVPS})}} & \approx 1.05,\\
        \frac{\mathcal{F}_1(q^2_{\text{max}})|_{}}{\mathcal{F}_1(q^2_{\text{max}})|_{{\rm Eq.}(\ref{eq:HQSSVPS})}} & \approx 1.02,\\
        \frac{\mathcal{F}_2(q^2_{\text{max}})|_{}}{\mathcal{F}_2(q^2_{\text{max}})|_{{\rm Eq.}(\ref{eq:HQSSVPS})}} & \approx 1.02, \\
    \end{split}
\end{equation}
an excellent agreement among the relations between $B_c\to \eta_c$ and $B_c \to J/\psi$ transition form factors derived from HQSS/NRQCD symmetry relations and our  exact results at the zero recoil. 

%%%%%%%%%%%%%%%%%%%%%%%%%%%%%%%%%%%%%%%%%%%%

\section{$R_{\eta_c}$, $R_{J/\psi}$ and decay distributions of  $B_c \rightarrow  \eta_c \ell \nu_\ell$ and $B_c \rightarrow  J/\psi \ell \nu_\ell$ }
\label{sec:diff_dist}

The general effective Lagrangian for the quark level transition $b\rightarrow c\ell\nu_\ell$ with $\ell = e,\mu,\tau$ is given by
\begin{eqnarray}\label{eq:Lag}
{\cal L} =\frac{G_F V_{cb}}{\sqrt{2}} \left [ ( 1+ V_L) {\cal O}_{V_L} +V_R {\cal O}_{V_R} +S_L {\cal O}_{S_L} +S_R {\cal O}_{S_R} + T_L {\cal O}_{T_L}\right ],
\end{eqnarray}
with the four-Fermi operators defined as
\begin{eqnarray}
\mathcal{O}_{V_L}&=& \left(\bar{c}\gamma^\mu (1-\gamma_5)b \right) \left(\bar{\ell}\gamma_\mu(1-\gamma_5)\nu_\ell \right),\quad \mathcal{O}_{V_R}= \left(\bar{c}\gamma^\mu (1+\gamma_5)b \right)\left(\bar{\ell}\gamma_\mu(1-\gamma_5)\nu_\ell \right), \nonumber \\
\mathcal{O}_{S_L}&=&\left(\bar{c} (1-\gamma_5)b\right) \left(\bar{\ell}(1-\gamma_5)\nu_\ell \right),\quad
\mathcal{O}_{S_R}=\left(\bar{c} (1+\gamma_5)b\right) \left(\bar{\ell}(1-\gamma_5)\nu_\ell \right), \nonumber \\
\mathcal{O}_{T_L}&=& \left(\bar{c}\sigma^{\mu\nu} (1-\gamma_5)b \right) \left(\bar{\ell}\sigma_{\mu\nu}(1-\gamma_5)\nu_\ell \right).
\end{eqnarray}
We use $\sigma_{\mu\nu}= i \left[\gamma_\mu,\gamma_\nu\right]/2$ and $V_{L,R},S_{L,R},T_L$ are the complex Wilson coefficients governing the NP contributions which are zero in the SM. Since we want to explain the possible lepton-flavour non-universality, we will assume that the NP effects contribute to the $\tau$ leptons only. The matrix element  of the semileptonic decays $B_c \rightarrow  J/\psi(\eta_c)\tau\nu_\tau$ then has the form:
\begin{eqnarray}
\mathcal{M}&=& \frac{G_F V_{cb}}{\sqrt{2}}\left[\left\lbrace (1+V_L+V_R)\bra{J/\psi(\eta_c)}\bar{c}\gamma^\mu b\ket{\bar{B}_c}+(V_R-V_L) \bra{J/\psi(\eta_c)}\bar{c}\gamma^\mu \gamma_5 b\ket{\bar{B}_c}\right\rbrace\bar{\ell}\gamma_\mu (1-\gamma_5)\nu_\ell  \right. \nonumber \\
&& \left. +(S_R+S_L) \bra{J/\psi(\eta_c)}\bar{c}b\ket{\bar{B}_c}\bar{\ell} (1-\gamma_5)\nu_\ell +(S_R-S_L) \bra{J/\psi(\eta_c)}\bar{c}\gamma_5 b\ket{\bar{B}_c}\bar{\ell}(1-\gamma_5)\nu_\ell \right. \nonumber \\
&& \left. +T_L \bra{J/\psi(\eta_c)}\bar{c}\sigma^{\mu\nu}(1-\gamma_5) b\ket{\bar{B}_c}\bar{\ell}\sigma^{\mu\nu}(1-\gamma_5)\nu_\ell   \right].
\end{eqnarray}
We note that the axial and the pseudoscalar hadronic currents do not contribute to the $B_c \rightarrow \eta_c$ decay, and therefore $V_R-V_L = 0, S_R-S_L = 0, \Rightarrow V_R = V_L, S_R = S_L$. The scalar hadronic current does not contribute to the $B_c \rightarrow J/\psi$ transition which leads to $S_L + S_R =0$.  We henceforth use the shorthand definition $S_R+S_L =S$ and $S_R-S_L = P$ in the text. 

 The constraints on the Wilson coefficients appearing in Eq.~(\ref{eq:Lag}) are obtained from the combined analysis of the BaBar, Belle and LHCb data for the branching fraction ratios $R_{D^{(\ast)}}$, the $\tau$ polarization asymmetry along the longitudinal directions of the $\tau$ lepton in $B \to D^\ast,$ as well as the longitudinal $D^\ast$ polarization in $B_c \rightarrow D^\ast \tau \nu_\tau$ decay \cite{Blanke:2018yud}. The leptonic branching fraction of the $B_c$ meson, $BR(B_c \to \tau \nu)$, is not yet measured, therefore the possible NP contributions come from precise experimental measurements of the $B_c$ lifetime, $\tau^{\mathrm{exp}}_{B_c} = (0.507\pm0.009)$ ps~\cite{Tanabashi:2018oca}.  The theoretical SM prediction of the $B_c$ lifetime still allows for up to 60\% contribution from NP~\cite{Beneke:1996xe, Blanke:2018yud} in the $B_c$ leptonic decay width. In particular, the best fit point for $S_R$ is dependent on the assumption of the  $B_c \to \tau \nu$ decay width. 
 
 We consider for our analysis the limit BR$(B_c \rightarrow \tau \bar{\nu})<$  30\% and the values of the Wilson coefficients from the combined analysis done in Ref.~\cite{Blanke:2018yud}.  They studied all one-dimensional scenarios with only one NP Wilson coefficient considered at a time  and the two-dimensional scenarios with  two NP Wilson coefficients considered simultaneously. The best fit points in the 1D scenarios and their $2\sigma$ ranges (given in square brackets below) at 1 TeV are given  in Table 1 of Ref.~\cite{Blanke:2018yud} and we list them below for completion: 
\begin{eqnarray}\label{eq:1Dfit}
&&V_L =  0.11~[0.06,~ 0.15],  \nonumber \\
&&S_R = 0.16~[0.08, ~0.23],  \quad S_L= 0.12~[0.01, ~0.20], \nonumber \\
&&S_L = 4 T_L = -0.07~[-0.15,~ 0.02].
\end{eqnarray}
 Only the real values of the coefficients were considered for the fit.  The possibility of allowing imaginary coefficients was examined in Ref.~\cite{Angelescu:2018tyl} and they obtained that the relation ${\rm Im}[S_L] = 4 \,{\rm Im}[T_L]$ is also permitted by the recent experiments. We therefore use the best fit value for $S_L = 4 \,T_L$ in Eq.~(\ref{eq:1Dfit}) for both, the real and the imaginary case.  The results of the fit for the NP Wilson coefficients in the 2D scenario at 1 TeV are taken from Table 2 of  Ref.~\cite{Blanke:2018yud}:
\begin{eqnarray}\label{eq:2Dfit}
&&(V_L,~ S_L)= -4 T_L  =  (0.08,~ 0.05),  \nonumber \\
&&(S_R,~S_L) = (-0.30,~-0.64), \nonumber \\
&& (V_L,~ S_R) = (0.09~0.06), \nonumber \\
&&  (\mathrm{Re}[S_L= 4 \, T_L],  \mathrm{Im}[S_L = 4 \,T_L]) = (-0.06, \pm 0.40).
\end{eqnarray}

All NP operators are generated by the addition of a single new particle to the SM. The relation $S_L = 4 \, T_L$  is generated in the $R_2$ leptoquark scenario with a scalar $SU(2)_L$ doublet~\cite{Dorsner:2016wpm, Becirevic:2016yqi} at the new physics scale.  The leptoquark model with an $SU(2)_L$ singlet scalar $S_1$ gives the relation $S_L= -4 \, T_L$ at the NP scale.  These relations are modified at the scale $m_b$ to:  $S_L(m_b) \simeq 8.1 \,T_L(m_b)$  for $R_2$, and  $S_L(m_b) \simeq -8.5 \, T_L(m_b)$, after including  one-loop electroweak corrections in addition to the three-loop QCD anomalous dimensions in the renormalization group running using the following relations~\cite{Gonzalez-Alonso:2017iyc}:
\begin{eqnarray}
V_L (m_b) &=& V_L (1~\mathrm{TeV}), \quad S_R (m_b) = 1.737 S_R (1~\mathrm{TeV}), \nonumber \\
\begin{pmatrix}
S_L (m_b) \\ T_L(m_b) 
\end{pmatrix}&=& \begin{pmatrix} 1.752 & -0.287 \\ -0.004 & 0.842\end{pmatrix} \begin{pmatrix}
S_L (1~\mathrm{TeV}) \\ T_L(1~\mathrm{TeV}) 
\end{pmatrix}.
\end{eqnarray}

We now discuss the differential decay rates  for the processes $B_c \rightarrow \eta_c \ell \nu_\ell$ and $B_c \rightarrow J/\psi \ell \nu_\ell$. The differential decay rate for these semi-leptonic processes depend on the angle $\theta_\ell$  which is the polar angle of the lepton $\ell$ (the angle between the lepton direction in the $W^*$ rest frame and the direction of the $W^*$ in the $B_c$ rest frame) and the momentum transfer $q^2$  ( $q =  p_{B_c} - p$) to the $\ell \nu_\ell$ pair. The differential $(q^2,\cos\theta_\ell)$ distribution can be calculated using the helicity techniques and is of the form
\begin{eqnarray}
\frac{d^2\Gamma}{dq^2d\cos\theta_\ell} &=& \frac{G_F^2|V_{cb}|^2|{\bf p_2}|v}{(2\pi)^3 64 m_{B_c}^2} H_{\mu\nu}L^{\mu\nu} (\theta_\ell),
\end{eqnarray}
where  $|{\bf p_2}| = \lambda^{1/2}(m_{B_c}^2,m_{\eta_c,J/\psi}^2,q^2)/2m_{B_c}$  is the momentum of $\eta_c(J/\psi)$ in the $B_c$ rest frame, $v= (1-m_\ell^2/q^2)$ is the lepton velocity in the $\ell^- \bar{\nu}_\ell$ center-of-mass frame
and $H_{\mu\nu}L^{\mu\nu}$ is the contraction of the hadronic and the leptonic tensors. The helicity techniques to calculate the angular distribution in the presence of new physics operators for the semi-leptonic decays considered here can be found in Ref.~\cite{Ivanov:2016qtw,Cohen:2018vhw}. 

The differential distribution for the $B_c\rightarrow \eta_c \tau\nu_\tau$ is written as
\begin{eqnarray}
\frac{d^2\Gamma(\eta_c)}{dq^2d\cos\theta_\ell}&=&\frac{G_F^2|V_{cb}|^2|{\bf p_2}|q^2 v^2}{(2\pi)^3 16m_{B_c}^2}
\Big\lbrace|1+V_L+V_R|^2\left[ |H_0|^2\sin^2\theta_\ell +2\delta_\ell |H_t-H_0\cos\theta_\ell|^2 \right]\nonumber \\
&+&|S|^2|H_P^S|^2
+16|T_L|^2\left[2\delta_\ell+( 1-2\delta_\ell)\cos^2\theta_\ell \right]|H_T|^2\nonumber \\
&+&2\sqrt{2\delta_\ell} \Big({\rm Re}S + S\, V_L \Big) H_P^S\left[ H_t-H_0\cos\theta_\ell \right]\nonumber \\
&+&8\sqrt{2\delta_\ell} \Big({\rm Re}T_L +T_LV_L\Big) \left[ H_0-H_t\cos\theta_\ell \right]H_T-8 H^S_P H_T \cos\theta_\ell \Big(T_LS \Big)
\Big\rbrace,
\label{eq:distretaC}
\end{eqnarray}
with the helicity flip-factor $\delta_\ell = m_\ell^2/2q^2$,  $T_LV_L = {\rm Re}T_L~{\rm Re}V_L  +  {\rm Im}T_L~{\rm Im}V_L $,  $T_LS =  {\rm Re}T_L~{\rm Re}S  + {\rm Im}T_L~{\rm Im}S$ and $S\,V_L = {\rm Re}S~{\rm Re}V_L  +  {\rm Im}S~{\rm Im}V_L $.  We consider the interference between the different NP operators as their effect can be similar to NP$^2$, if they are of the same value. The $H's$ in Eq.~(\ref{eq:distretaC}) are the hadronic helicity amplitudes written in terms of the invariant form factors defined in Eq.~(\ref{eq:etaC_ff})  and are of the form
\begin{equation}
H_t  =\frac{Pq}{\sqrt{q^2}} f_0 ,
\quad
H_0 = \frac{2m_{B_c}|{\bf p_2}|}{\sqrt{q^2}}f_+, \quad H_P^S =\frac{Pq}{m_b(\mu)-m_c(\mu)} f_0, \quad H_T = \frac{2m_{B_c}|{\bf p_2}|}{m_{B_c}+m_{\eta_c}}f_T,
\label{eq:hel_pp}
\end{equation}
with $P = p_{B_c}+p$ and $q = p_{B_c}-p$ ($p= p_{\eta_c}$ or $p= p_{J/\psi}$.  

Next, the differential distribution of the $\bar{B}_c\to J/\psi\ell^-\bar{\nu}_\ell$ decay is considered with  $V_RV_L = {\rm Re}V_R~ {\rm Re}V_L  +  {\rm Im}V_R~ {\rm Im}V_L $,  $T_LP =  {\rm Re}T_L~{\rm Re}P  + {\rm Im}T_L~ {\rm Im}P  $and $P\,V_L = {\rm Re}P~{\rm Re}V_L  +  {\rm Im}P~{\rm Im}V_L $, and is given by
\begingroup
\allowdisplaybreaks
\begin{eqnarray}
\frac{d^2\Gamma( J/\psi)}{dq^2d\cos\theta_\ell}
&=&\frac{G_F^2|V_{cb}|^2|{\bf p_2}|q^2v^2}{32 (2\pi)^3 m_{B_c}^2}\Big\lbrace
|1+V_L|^2
\Big[
(1-\cos\theta_\ell)^2|H_{++}|^2+(1+\cos\theta_\ell)^2|H_{--}|^2+2\sin^2\theta_\ell|H_{00}|^2\nonumber \\
&+&2\delta_\ell
\Big(
\sin^2\theta_\ell(|H_{++}|^2+|H_{--}|^2)+2|H_{t0}-H_{00}\cos\theta_\ell|^2
\Big)
\Big]\nonumber \\
&+&|V_R|^2
\Big[
(1-\cos\theta_\ell)^2|H_{--}|^2+(1+\cos\theta_\ell)^2|H_{++}|^2+2\sin^2\theta_\ell|H_{00}|^2\nonumber \\
&+&2\delta_\ell
\Big(
\sin^2\theta_\ell(|H_{++}|^2+|H_{--}|^2)+2|H_{t0}-H_{00}\cos\theta_\ell|^2
\Big)
\Big]-4 \Big({\rm Re}V_R +V_R V_L \Big)\nonumber \\
&&\Big[
(1+\cos^2\theta_\ell)H_{++}H_{--}+\sin^2\theta_\ell|H_{00}|^2+2\delta_\ell
\Big(
\sin^2\theta_\ell H_{++}H_{--}+|H_{t0}-H_{00}\cos\theta_\ell|^2
\Big)
\Big]\nonumber \\
&+&2|P|^2|H^S_V|^2+4 \sqrt{2\delta_\ell}    H^S_V(H_{t0}-H_{00}\cos\theta_\ell) \Big({\rm Re}P+P\,V_L\Big)+16 \cos\theta_\ell H^S_V H_T^0  T_LP\nonumber \\
&+&16|T_L|^2 \Big[
|H_T^0|^2\Big( 1+2\delta_\ell +(1-2\delta_\ell)\cos2\theta_\ell \Big)+2|H_T^+|^2\sin^2\frac{\theta_\ell}{2}\Big( 1+2\delta_\ell+(1-2\delta_\ell)\cos\theta_\ell \Big)\nonumber \\
&+&2|H_T^-|^2\cos^2\frac{\theta_\ell}{2}\Big( 1+2\delta_\ell-(1-2\delta_\ell)\cos\theta_\ell \Big)
\Big]-16\sqrt{2\delta_\ell} \Big({\rm Re}T_L +T_LV_L \Big) \nonumber \\
&& \Big[ H_{++}H_T^+ +H_{--}H_T^- +H_{00}H_T^0-\Big(H_{++}H_T^+ -H_{--}H_T^- +H_{t0}H_T^0\Big)\cos\theta_\ell \Big]\Big\rbrace.
\label{eq:distrJPsi}
\end{eqnarray}
\endgroup
The hadronic helicity amplitudes in terms of the form factors given in Eqs.~(\ref{eq:etaC_Jpsi1}, \ref{eq:etaC_Jpsi2}) are expressed as
\begin{eqnarray}
H_{\pm\pm} &= &\frac{-(m_{B_c}+m_{J/\psi})^2A_1\pm 2m_{B_c}|{\bf p_2}| V}{m_{B_c}+m_{J/\psi}}, \quad
H^S_V = \frac{2m_{B_c}}{m_b(\mu)+m_c(\mu)}|{\bf p_2}|A_0, \nonumber \\
H_{00} &= &  \frac{-(m_{B_c}^2 - m_{J/\psi}^2 - q^2)(m_{B_c}+m_{J/\psi})^2A_1 + 4m_{B_c}^2|{\bf p_2}|^2 A_2}{2m_{J/\psi}\sqrt{q^2}(m_{B_c}+m_{J/\psi})} 
,  H_{t0}  =  -\frac{2 m_{B_c} |{\bf p_2}| }{\sqrt{q^2}}A_0    \nonumber \\
%H_{t0} &= &  \frac{m_{B_c}|{\bf p_2}|\left((m_{B_c}-m_{J/\psi})A_2-(m_{B_c}+m_{J/\psi})A_1)+2 m_{J/\psi} (A_3-A_0)\right)}{m_{J/\psi}\sqrt{q^2}},
\nonumber \\
H_T^\pm &= &  -\frac{1}{\sqrt{q^2}}\left[
\pm\lambda^{1/2}[m_{B_c}^2,m_{J/\psi}^2,q^2]T_1+ (m_{B_c}^2-m_{J/\psi}^2) T_2
\right],\nonumber \\
H_T^0 &= &= -\frac{1}{2m_{J/\psi}}
\Big[
(m_{B_c}^2+3m_{J/\psi}^2-q^2)T_2 -\frac{\lambda[m_{B_c}^2,m_{J/\psi}^2,q^2]}{m_{B_c}^2-m_{J/\psi}^2}T_3
\Big].
\end{eqnarray}

\subsection{Results for the branching ratios and $R_{\eta_c}, R_{J/\psi} $ predictions}

%We list in the following the form factors at maximum recoil available from other authors

We first  give our predictions for branching fractions in the SM of both decays in Table~\ref{tab:brfr2}, where we put the branching fraction values updated using the latest value for the $B_c$ lifetime, $\tau_{B_c} = (0.507 \pm 0.009)$ ps ~\cite{Tanabashi:2018oca}, while in the brackets we cite the original published values of the $BR$s. If there are no brackets the branching fractions have already been calculated using the latest value for $\tau_{B_c}$. 
\begin{table}[ht]
\addtolength{\tabcolsep}{-4.8pt}
\renewcommand{\arraystretch}{1.}
 \begin{tabular}{||c || c||  c c c c c c c c c c ||}
 \hline
 Mode & \begin{tabular}{@{}c@{}}this work \end{tabular}  & \begin{tabular}{@{}c@{}}QCDSR\\ ~\cite{Kiselev:2002vz}\end{tabular} & \begin{tabular}{@{}c@{}}SR \\ \cite{Huang:2007kb} \end{tabular} & \begin{tabular}{@{}c@{}}pQCD\\ ~\cite{Wen-Fei:2013uea}\end{tabular} & \begin{tabular}{@{}c@{}}RCQM\\ ~\cite{Ivanov:2006ni}\end{tabular} & \begin{tabular}{@{}c@{}}CCQM\\ ~\cite{Tran:2018kuv,Issadykov:2017wlb}\end{tabular} & \begin{tabular}{@{}c@{}}RQM\\ ~\cite{Ebert:2003cn}\end{tabular} &\begin{tabular}{@{}c@{}}RQM\\ ~\cite{Scora:1995ty}\end{tabular} & \begin{tabular}{@{}c@{}}RQM\\ ~\cite{Nobes:2000pm}\end{tabular} & \begin{tabular}{@{}c@{}}RQM\\ ~\cite{AbdElHady:1999xh}\end{tabular} & \begin{tabular}{@{}c@{}}LFQM\\ ~\cite{Wang:2008xt}\end{tabular}\\ [0.5ex]
 \hline\hline
 $B_c\rightarrow \eta_c l \bar{\nu}_{l}$ & $0.82^{+0.12}_{-0.11}$ &  \begin{tabular}{@{}c@{}}0.85 \\ (0.75) \end{tabular} & \begin{tabular}{@{}c@{}}1.85 \\ (1.64) \end{tabular} & \begin{tabular}{@{}c@{}}0.50 \\ (0.44) \end{tabular} & \begin{tabular}{@{}c@{}}0.91 \\ (0.81) \end{tabular} & 0.95 & \begin{tabular}{@{}c@{}}0.47 \\ (0.42) \end{tabular} & 0.89 & \begin{tabular}{@{}c@{}}0.52 \\ (0.52) \end{tabular} & 0.85 & \begin{tabular}{@{}c@{}}0.74 \\ (0.67) \end{tabular}\\
 \hline
 $B_c\rightarrow \eta_c \tau \bar{\nu}_{\tau}$ & $0.26^{+0.06}_{-0.05}$ &  \begin{tabular}{@{}c@{}}0.25 \\ (0.23) \end{tabular} & \begin{tabular}{@{}c@{}}0.55 \\ (0.49) \end{tabular} & \begin{tabular}{@{}c@{}}0.15 \\ (0.14) \end{tabular} & \begin{tabular}{@{}c@{}}0.25 \\ (0.22) \end{tabular} & 0.24 & - & - & - & - & \begin{tabular}{@{}c@{}}0.21 \\ (0.19) \end{tabular}\\
 \hline\hline
 $B_c\rightarrow J/\psi l \bar{\nu}_{l}$ & $2.24^{+0.57}_{-0.49}$ &  \begin{tabular}{@{}c@{}}2.16 \\ (1.9) \end{tabular} & \begin{tabular}{@{}c@{}}2.67 \\ (2.37) \end{tabular} & \begin{tabular}{@{}c@{}}1.13 \\ (1.00) \end{tabular} & \begin{tabular}{@{}c@{}}2.33 \\ (2.07) \end{tabular} & 1.67 & \begin{tabular}{@{}c@{}}1.39 \\ (1.23) \end{tabular} & 1.42 & \begin{tabular}{@{}c@{}}1.49 \\ (1.47) \end{tabular} & 2.33 & \begin{tabular}{@{}c@{}}1.64 \\ (1.49) \end{tabular}\\
 \hline
 $B_c\rightarrow J/\psi \tau \bar{\nu}_{\tau}$ & $0.53^{+0.16}_{-0.14}$ & \begin{tabular}{@{}c@{}}0.54 \\ (0.48) \end{tabular} & \begin{tabular}{@{}c@{}}0.73 \\ (0.65) \end{tabular} & \begin{tabular}{@{}c@{}}0.33 \\ (0.29) \end{tabular} & \begin{tabular}{@{}c@{}}0.55 \\ (0.49) \end{tabular} & 0.40 & -  & - & - & - & \begin{tabular}{@{}c@{}}0.41 \\ (0.37) \end{tabular}\\
 \hline
\end{tabular}
\captionof{table}{Branching fractions of $B_c\rightarrow J/\psi,\eta_c$ decays calculated in different models and given in \%, with $l$ denoting a light lepton, $e$ or $\mu$.\label{tab:brfr2}} 
\end{table}

The ratios of semileptonic branching fractions using our calculated form factors from Eq.~(\ref{eq:LCSR-ff}) are  
\begin{eqnarray}
R_{\eta_c}|_{\rm SM}& \equiv\frac{\Gamma(B_c\rightarrow \eta_c \tau \bar{\nu}_{\tau})}{\Gamma(B_c\rightarrow \eta_c \mu \bar{\nu}_{\mu})} = 0.32 \pm 0.02\,,
 \\
        %\begin{cases}
 %                                           0.322^{+0.019}_{-0.021} & \text{LCSR}.\\
 %                                           0.301 & \text{QCDSR}.
 %                                           \end{cases}\\
R_{J/\psi}|_{\rm SM}& \equiv\frac{\Gamma(B_c\rightarrow J/\psi \tau \bar{\nu}_{\tau})}{\Gamma(B_c\rightarrow J/\psi \mu \bar{\nu}_{\mu})} = 0.23 \pm 0.01 .  %0.236^{+0.011}_{-0.013}\quad \text{LCSR}\\ 
%    \end{split}
\end{eqnarray}

We see that above results agree with the recent model-independent analysis of $R_{J/\psi}$~\cite{Cohen:2018dgz,Murphy:2018sqg} and $R_{\eta_c}$~\cite{Murphy:2018sqg,Berns:2018vpl}. See also the discussion in Sec.\ref{sec:HQrelations}. 
% $R(J/\psi)$  [CohenLammLebed1807.02730,MurphySoni1808.05932v3] and 
%$R(\eta_c)$ [MurphySoni1808.05932v3,BernsLamm1808.07360].
%%%%%%%%%%%%%%%%%%%%%%%%%%%%%%%%%%%%%%%%%%%%%%%%%%%%%%%%%%%%%%%%%%%%%

\begin{figure}[htb]
 \includegraphics[width=\textwidth]{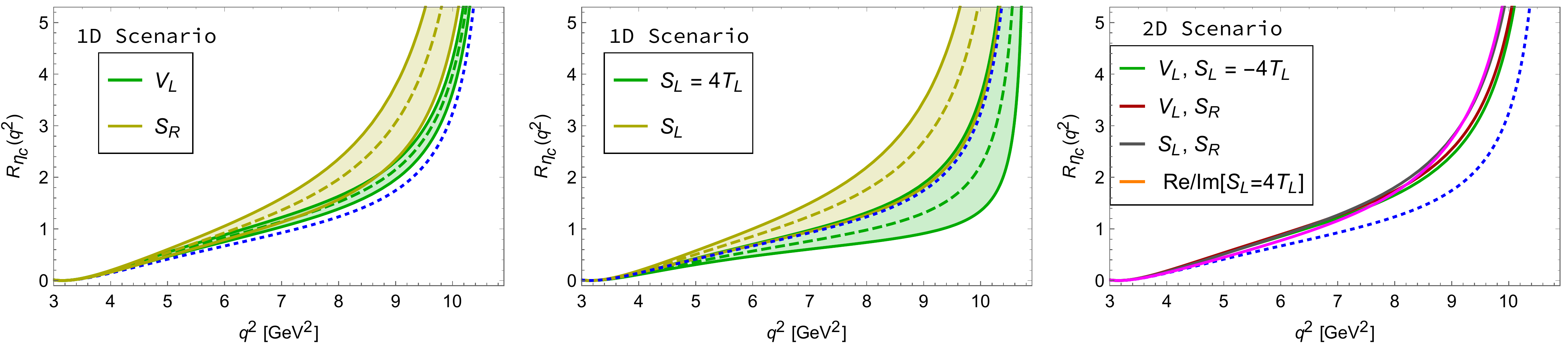}
        \includegraphics[width=\textwidth]{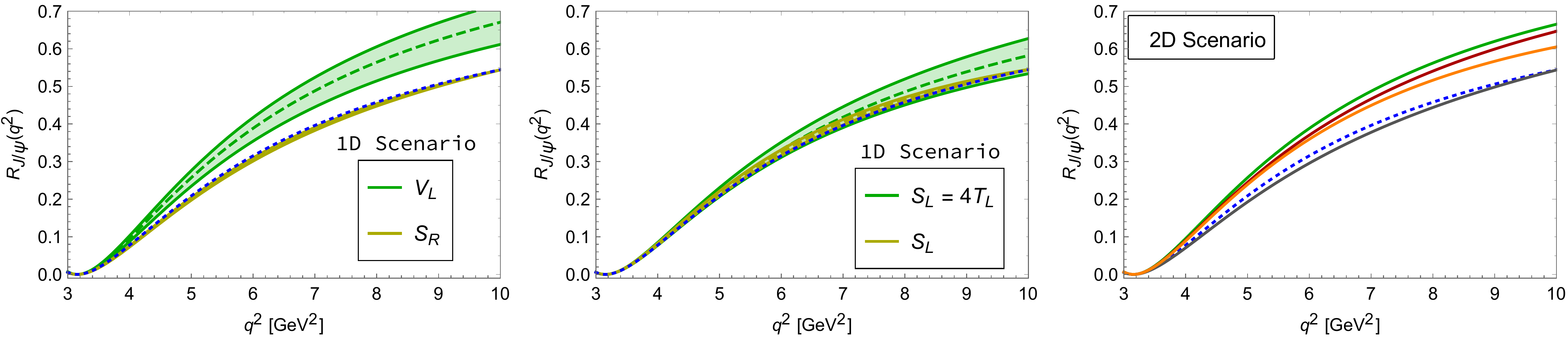}
        \caption{Ratios of branching fractions $R_{\eta_c}(q^2)$ (upper panel),  $R_{J/\psi}(q^2)$ (lower panel) as a function of $q^2$. The blue dotted lines are the SM prediction, the  green dashed line is for the best fit values of the NP couplings in the 1D scenario as discussed in the text. The green band represents the NP effects from the 2$\sigma$ allowed regions in the 1D scenarios. The third figure in both panels is the result for the best fit points in the 2D scenarios.}
        \label{fig:Rq2}
    \end{figure}

The ratios of the branching fractions $R_{J/\psi,\eta_c}$ are computed next in the context of different NP scenarios using the form factors calculated in Sec.~\ref{sec:LCSR_ff}. The values of the NP operators' effective couplings considered for our analysis are discussed before and are given by Eqs.~(\ref{eq:1Dfit},~\ref{eq:2Dfit}). In Fig.~\ref{fig:Rq2} we show the the $q^2$ dependence of the ratios $R_{\eta_c}$ and $R_{J/\psi}$ in the presence of only one NP operator (first two figures of both panels). The third figure in both panels shows the ratio in presence of two NP operators. The SM value is always shown by the blue dotted line. We see that the ratio increases for most of NP contributions for both $J/\psi$ and $\eta_c$.  The $S_L =4 \,T_L$ case with the coupling being pure real or imaginary results in a decrease in the ratio $R_{\eta_c}$. This is due to the negative interference between  $S_L$ and $T_L$ , Eq.~(\ref{eq:distretaC}). The shaded region shows the $2\sigma$ allowed region for $V_L, S_L=4T_L, S_{L,R}$ parameters in the 1D fit, with the central value shown by a dashed line.  In case of the 2D scenarios the results are  presented for the best fit point.  As expected, the ratio $R_{\eta_c}$ is more sensitive to the scalar and the tensor operators, whereas $R_{J/\psi}$ is more sensitive to $V_L$.
The values of $R_{J/\psi}$ and $R_{\eta_c}$ in the presence of different NP scenarios are listed in Table~\ref{tab:avg_R}. The results are presented for the best fit points, as well as for the 2$\sigma$ allowed regions in the 1D scenario. 

 Note that any of the considered NP scenarios derived from the recent global fit analysis on available experimental data on semileptonic $B \to (D,D^{\ast}) \ell \nu_\ell$
decays \cite{Blanke:2018yud} cannot explain the 2$\sigma$ tension with the experiment Eq.~\ref{eq:expRJpsi} of $R_{J/\psi}$  ratio.

\begin{center}
\addtolength{\tabcolsep}{-5pt}
%\captionsetup{justification=centering,margin=1.4cm}
\renewcommand{\arraystretch}{1.2}
 \begin{tabular}{||c ||c|| c |c|c|c|c|c|c|c|}
 \hline
 &SM&$V_L$ &$S_L$ & $S_R$ &$S_L = 4 T_L$ & $ (V_L,S_L=-4T_L)$&$(S_R,S_L)$& $(V_L,S_R)$&Re,Im$[S_L=4T_L]$   \\ [0.5ex]
 \hline\hline
$R_{\eta_c}$&0.32  &$0.39_{0.36}^{0.42}$ &$0.44_{0.33}^{0.55}$&$0.49_{0.40}^{0.59}$&$0.26_{0.20}^{0.34}$&0.42&0.45&0.44&0.43  \\ \hline \hline
$R_{J/\psi}$  &0.23  &$0.29_{0.26}^{0.31}$ &$0.24_{0.23}^{0.24}$&$0.23_{0.23}^{0.22}$&$0.25^{0.26}_{0.23}$&0.29&0.22&0.27&0.26 \\
  \hline
\end{tabular}
\captionof{table}{The values of $R_{\eta_c}$ and $R_{J/\psi}$  in the presence of different NP scenarios. The subscript and the superscript are the values for the $2\sigma$ range of the NP couplings.} \label{tab:avg_R} 
\end{center}

\subsection{Forward-backward asymmetry, convexity parameter and the $\tau$ polarization}

\begin{figure}[htb]
 \includegraphics[width=\textwidth]{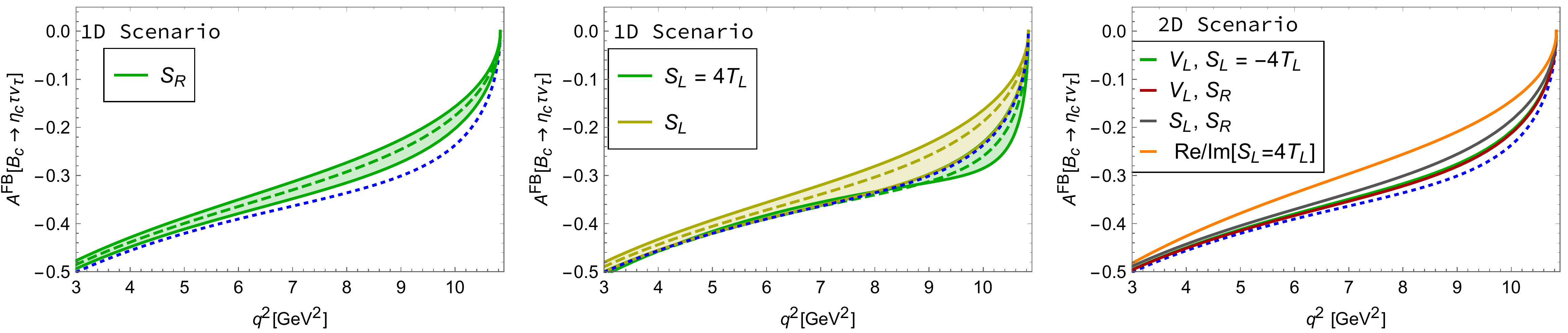}
        \includegraphics[width=\textwidth]{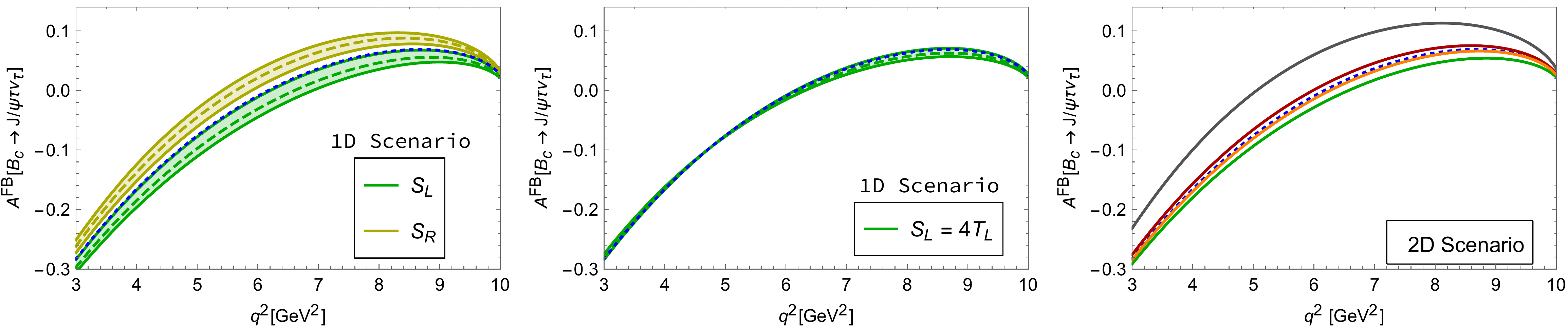}
        \caption{Forward-backward asymmetry $A^{FB}(q^2)$ for $\eta_c$ (upper panel),  and $J/\psi$ (lower panel) as a function of $q^2$. The blue dotted lines are the SM prediction, the  green dashed line is for the best fit values of the NP couplings in the 1D scenario as discussed in the text. The green band represents the NP effects from the 2$\sigma$ allowed regions. The third figure in both panels is the result for the best fit points in the 2D scenarios.}
        \label{fig:FBA}
    \end{figure}

\begin{figure}[htb]
 \includegraphics[width=\textwidth]{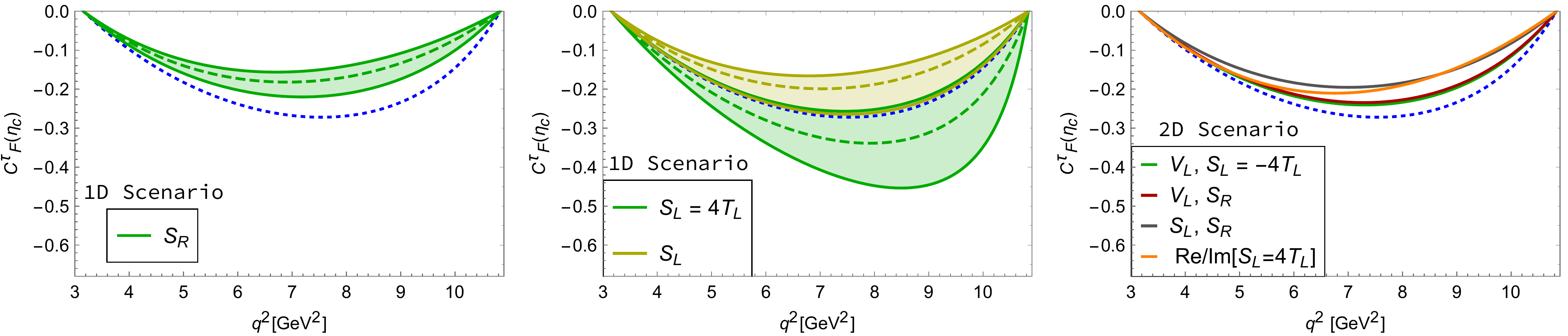}
        \includegraphics[width=\textwidth]{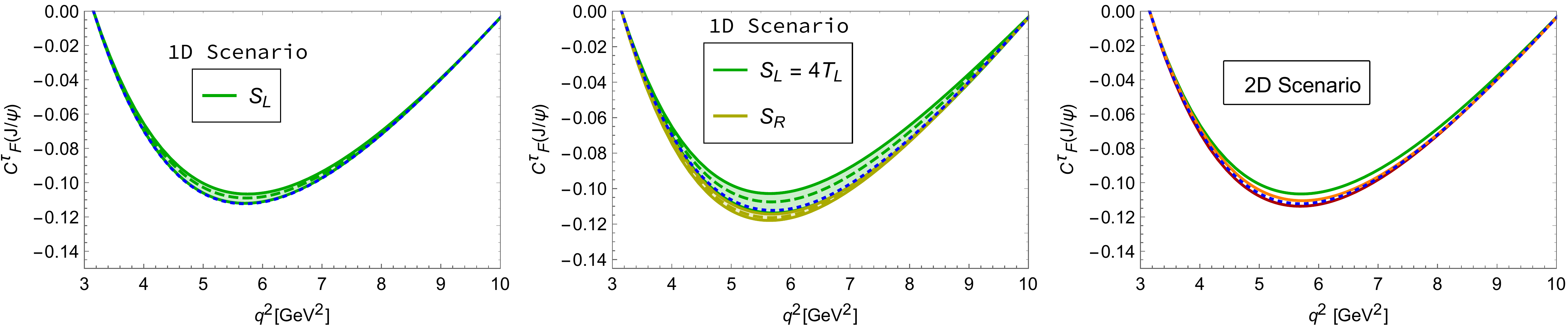}
         \caption{Convexity parameter $C^\tau_F(q^2)$  for $\eta_c$ (upper panel),  and $J/\psi$ (lower panel) as a function of $q^2$. The blue dotted lines are the SM prediction, the  green dashed line is for the best fit values of the NP couplings in the 1D scenario as discussed in the text. The green band represents the NP effects from the 2$\sigma$ allowed regions. The third figure in both panels is the result for the best fit points in the 2D scenarios.}
     \label{fig:C2}     
    \end{figure}

The differential distributions defined in Eqs.~(\ref{eq:distretaC},~\ref{eq:distrJPsi}) can be written in a simple form as a function of $\cos \theta_\ell$ as 
\begin{eqnarray}
\frac{d\Gamma}{dq^2 d\cos\theta_\ell}&=& \frac{G_F^2|V_{cb}|^2|{\bf p_2}|q^2v^2}{32 (2\pi)^3 m_{B_c}^2 }(\mathcal{A}(q^2) + \mathcal{B}(q^2) \cos\theta_\ell + \mathcal{C}(q^2) \cos^2\theta_\ell).
\end{eqnarray}
Observables depending on the polar angle distribution of the emitted leptons such as the forward-backward lepton asymmetry and the convexity parameter are considered first. They are defined by  
\begin{eqnarray}
A^{FB}(q^2)&=& \frac{\Big(\int_0^1-\int_{-1}^0\Big)d \cos\theta_\ell\frac{d^2\Gamma}{dq^2d\cos\theta_\ell}}{\Big(\int_0^1+\int_1^0\Big)d \cos\theta_\ell\frac{d^2\Gamma}{dq^2d\cos\theta_\ell}} = \frac{\mathcal{B}(q^2)}{2\big(\mathcal{A}(q^2)+ \mathcal{C}(q^2) /3\big)}, \nonumber \\
C^\tau_F(q^2)&=& \frac{1}{d\Gamma/dq^2}\frac{d^2(d\Gamma/dq^2)}{d(\cos\theta_\ell)^2 } = \frac{\mathcal{C}(q^2)}{\big(\mathcal{A}(q^2)+ \mathcal{C}(q^2) /3\big)} .
\end{eqnarray}
The $\mathcal{A}(q^2), \mathcal{B}(q^2)$ and $\mathcal{C}(q^2)$ functions can be easily obtained from Eqs.~(\ref{eq:distretaC},~\ref{eq:distrJPsi}). We present in Figs.~\ref{fig:FBA},~\ref{fig:C2} the $q^2$ dependence of the forward-backward asymmetry $A_{FB}(q^2)$ and the convexity parameter $C^{\tau}_F(q^2)$. These observables are not sensitive to the case where $V_L$ is the only NP contribution. The $B_c \rightarrow \eta_c$ transition appears to be more sensitive to the new physics operators as compared to the $B_c \to J/\psi$ transition. In case of the $J/\psi$ decay mode, the presence of the $S_L, S_R$ coefficients in the 2D scenario leads to a significant deviation from $A^{FB}(q^2)$ prediction in the SM. The present allowed values of the coupling have a very small effect on $C^\tau_F(q^2)$ in case of $J/\psi$, whereas in case of $\eta_c$ the $S_L=4 \,T_L$ case enhances $C^\tau_F(q^2)$  only at large values of $q^2$.

\begin{figure}[htb]
 \includegraphics[width=\textwidth]{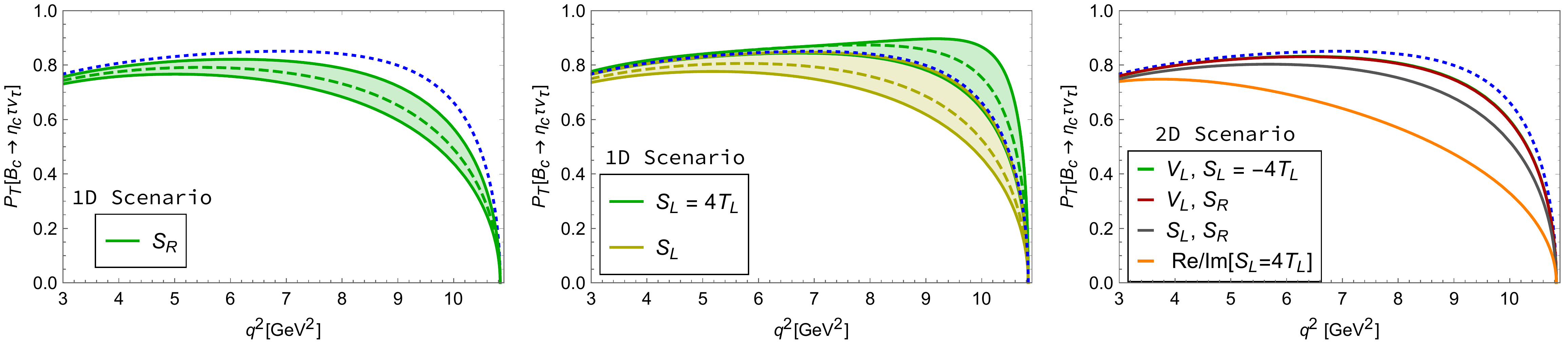}
  \includegraphics[width=\textwidth]{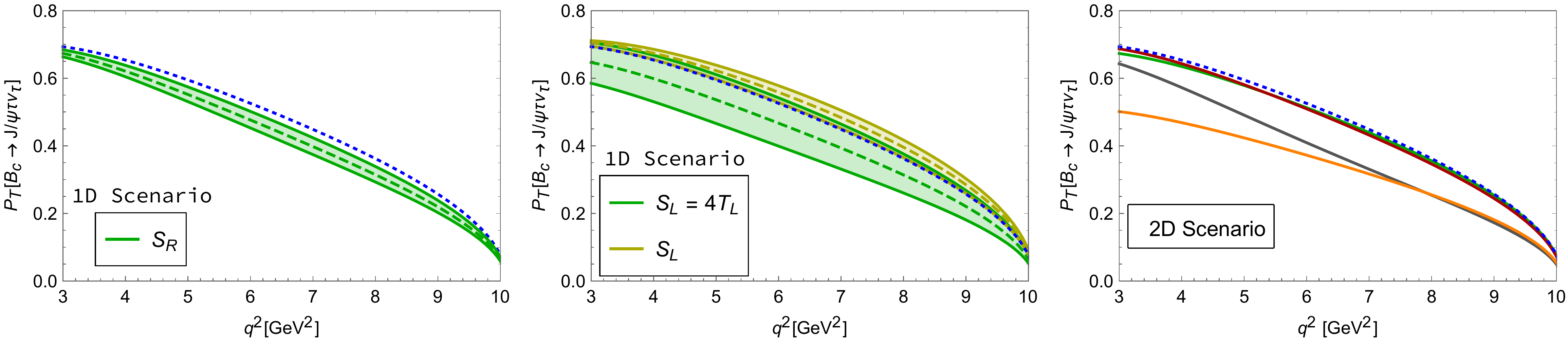}
 \caption{Transverse polarization of $\tau$  ($P_T^{\eta_c,J/\psi}$) in the decay for $B_c \to \eta_c\tau \nu$ (upper panel),  and $B_c \to J/\psi \tau \nu_\tau$ (lower panel) as a function of $q^2$. The blue dotted lines are the SM prediction, the  green dashed line is for the best fit values of the NP couplings in the 1D scenario as discussed in the text. The green band represents the NP effects from the 2$\sigma$ allowed regions. The third figure in both panels is the result for the best fit points in the 2D scenarios.}
 \label{fig:tau_tp}
 \end{figure}

Now we discuss the effect on the polarization of the emitted $\tau$ in the $W^-$ rest frame in the presence of the NP operators. The differential decay rate for a given spin projection in a given direction can be easily obtained with the inclusion of the spin projection operators $(1+ \gamma_5 \slashed{s}_i)/2$ for $\tau$ in the calculation. The longitudinal and the transverse polarization components of the $\tau$ are then defined as:
\begin{eqnarray}
P_{L,T}(q^2)&=& \frac{d\Gamma(s_i^\mu)/dq^2-d\Gamma(-s_i^\mu)/dq^2}{d\Gamma(s_i^\mu)/dq^2+d\Gamma(-s_i^\mu)/dq^2} = \frac{\mathcal{P}_{L,T}(q^2)}{2(\mathcal{A}(q^2)+ \mathcal{C}(q^2) /3)}, \quad i = L,T,
\end{eqnarray}
where $s^\mu_L$ and $s^\mu_T$ are the longitudinal and the transverse polarization four-vectors of $\tau^-$ in the $W^-$ rest frame and are given by~\cite{Faessler:2002ut, Gutsche:2013oea, Gutsche:2015mxa}
\begin{eqnarray}
s_L^\mu&=&\frac{1}{m_\tau}(|\vec{p}_\tau|,E_\tau\sin\theta_\tau,0,E_\tau\cos\theta_\tau),\quad
s_T^\mu=(0,\cos\theta_\tau,0,-\sin\theta_\tau).
\end{eqnarray}
 The longitudinal and transverse polarizations in the $B_c \to \eta, J/\psi \tau\nu_{\tau}$ decays are  given as :
\begin{eqnarray}
\label{eq:PL}
\mathcal{P}_L^{\eta_c}(q^2)&=&
\Big\lbrace
|1+V_L+V_R|^2\big[-|H_0|^2+\delta_\tau(|H_0|^2+3|H_t|^2)\big]+3\sqrt{2\delta_\tau}H_P^S H_t \Big({\rm Re}S+S\, V_L \Big) \nonumber \\
&+&
\frac{3}{2}|S|^2|H_P^S|^2
+8|T_L|^2(1-4\delta_\tau)|H_T|^2 -4\sqrt{2\delta_\tau}\Big({\rm Re}T_L+T_LV_L\Big) H_0 H_T
\Big\rbrace, \\
\mathcal{P}_L^{J/\psi}(q^2)&=&\Big\lbrace
(|1+V_L|^2+|V_R|^2)\big[-\sum\limits_{n=\pm,0}|H_{nn}|^2+\delta_\tau \Big(\sum\limits_{n=\pm,0}|H_{nn}|^2+3|H_{t0}|^2 \Big)\big]+2{\rm Re}V_R \nonumber \\
&&\big[(1-\delta_\tau)(|H_{00}|^2+2H_{++}H_{--})+3\delta_\tau |H_{t0}|^2\big]-3\sqrt{2\delta_\tau}\Big({\rm Re}P +P\,V_L\Big) H_V^S H_{t0}\nonumber \\
&+&\frac{3}{2}|P|^2|H_V^S|^2+8|T_L|^2(1-4\delta_\tau)\sum\limits_{n}|H_T^n|^2+4\sqrt{2\delta_\tau}\Big({\rm Re}T_L +T_LV_L\Big)\sum\limits_{n=\pm,0}H_{nn}H_T^n
\Big\rbrace, \nonumber
\end{eqnarray}
%and 
\vspace{-0.75cm}
\begingroup
\allowdisplaybreaks
\begin{eqnarray}\label{eq:tau_TP}
\mathcal{P}_T^{\eta_c}(q^2)&=&\frac{3\pi\sqrt{\delta_\tau}}{2\sqrt{2}}\Big\lbrace
|1+V_L+V_R|^2 H_0 H_t
+\frac{1}{\sqrt{2\delta_\tau}} \Big({\rm Re}S + S\,V_L\Big)H_P^S H_0  \nonumber \\
&+& 4\sqrt{2\delta_\tau}\Big({\rm Re}T_L+T_LV_L\Big) H_t H_T + 4 H_P^S H_T T_L S
\Big\rbrace  \label{eq:PT}
, \\
\mathcal{P}_T^{J/\psi}(q^2)&=&\frac{3\pi\sqrt{\delta_\tau}}{4\sqrt{2}}\Big\lbrace
(|1+V_L|^2-|V_R|^2)(|H_{--}|^2-|H_{++}|^2)+2(|1+V_L|^2+|V_R|^2)H_{t0}H_{00}\nonumber \\
&-&4{\rm Re}V_RH_{t0}H_{00}-\frac{2} {\sqrt{2\delta\tau}} \Big({\rm Re}P +P\,V_L \Big)H_V^SH_{00}+16|T_L|^2(|H_T^-|^2-|H_T^+|^2)\nonumber \\
&+&4 \Big({\rm Re}T_L+T_LV_L\Big)\Big[\frac{1+2\delta_\tau}{\sqrt{2\delta_\tau}}(H_{++}H_T^+-H_{--}H_T^-)-2\sqrt{2\delta_\tau}H_{t0}H_T^0\Big]+ 8 H_S^V H_T^0   T_LP
\Big\rbrace. \nonumber 
\end{eqnarray}
\endgroup
The transverse polarization of $\tau$ as can be seen from Eq.~(\ref{eq:PT}) has an overall factor of $\sqrt{\delta_\tau}$ and therefore vanishes in the limit of zero lepton mass and the emitted lepton is then fully longitudinally polarized. Therefore,  the $\tau$ lepton can be largely transversely polarized as compared to the muons or the electrons.  
\begin{figure}[htb]
 \includegraphics[width=\textwidth]{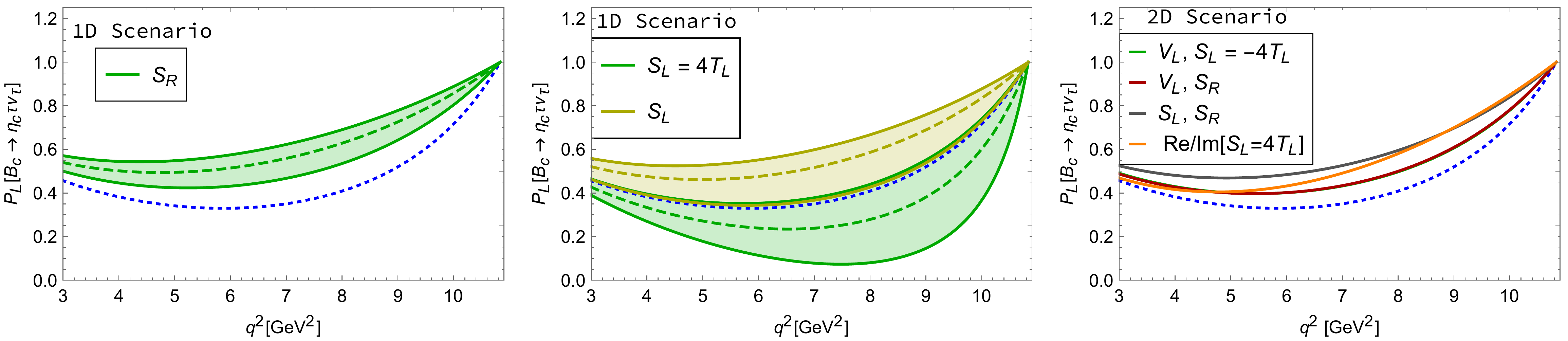}
        \includegraphics[width=\textwidth]{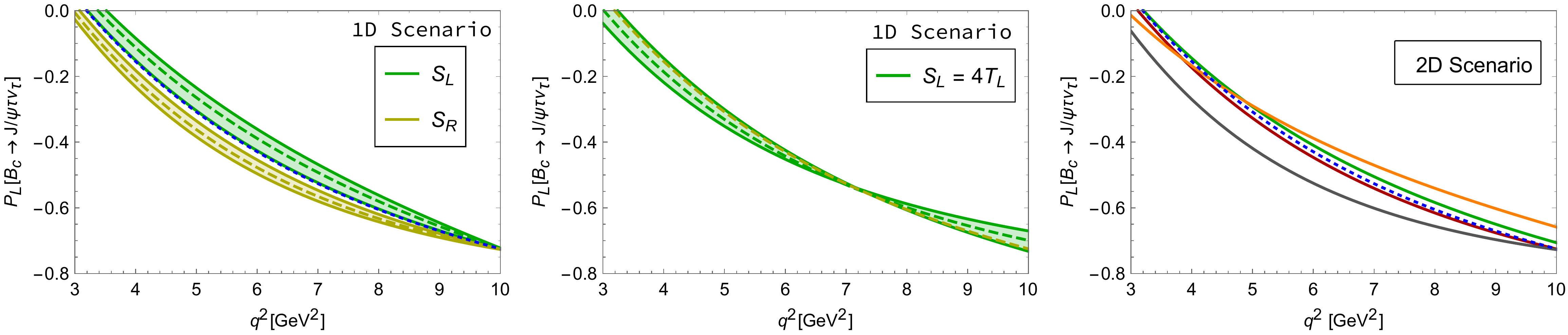}
  \caption{Longitudinal polarization of $\tau$  ($P_L^{\eta_c,J/\psi}$) in the decay of $B_c \to \eta_c\tau \nu$ (upper panel),  and $B_c \to J/\psi \tau \nu_\tau$ (lower panel) as a function of $q^2$. The blue dotted lines are the SM prediction, the  green dashed line is for the best fit values of the NP couplings in the 1D scenario as discussed in the text. The green band represents the NP effects from the 2$\sigma$ allowed regions. The third figure in both panels is the result for the best fit points in the 2D scenarios.}
  \label{fig:tau_lp}
 \end{figure}
 The $q^2$ dependence of the $\tau$ polarization in presence of different NP operators is shown in Figs.~\ref{fig:tau_lp},~\ref{fig:tau_tp}. The following observations can be made from the figures. The longitudinal and transverse polarizations of $\tau$ in the $\eta_c$  decay mode are more sensitive to the NP operators compared to the $J/\psi$ decay mode. The tau transverse polarization in the $J/\psi$ decay mode is again mostly affected by the NP operator $S_L = 4 \, T_L$ at low values of $q^2$, whereas the $S_L, S_R$ parameters in the 2D scenario lead to a deviation from the SM prediction for both the longitudinal and the transverse $\tau$ polarization.  The predictions for the mean forward-backward asymmetry, the convexity parameter and the tau polarization in the presence of different NP operators are summarised  in Table~\ref{tab:avg_obs}.%, whereas the respected SM values are given in Table~\ref{tab:avg_obs_sm}.
\begin{center}
%\captionsetup{justification=centering,margin=1.4cm}
\addtolength{\tabcolsep}{-5pt}
\renewcommand{\arraystretch}{1.1}
 \begin{tabular}{||c ||c|| c|c|c|c|c|c|c|c|}
 \hline
 &SM &$S_L$ & $S_R$ &$S_L = 4 T_L$ & $ (V_L,S_L=-4T_L)$&$(S_R,S_L)$& $(V_L,S_R)$& Re,Im[$S_L=4T_L$]    \\ [0.5ex]
 \hline\hline
 $A_{FB}^{\eta_c}$  &$-0.35$  &$-0.31^{-0.29}_{-0.34}$&$-0.30^{-0.28}_{-0.32}$&$-0.36^{-0.34}_{-0.36}$&$-0.33$ &$- 0.31$ &$-0.33$ &$-0.27$ \\ \hline
 $C_{F}^{\tau,\eta_c}$ &$-0.22$ &$-0.16^{-0.13}_{-0.21}$&$-0.14^{-0.12}_{-0.17}$&$-0.27^{-0.21}_{-0.35}$&$-0.19$ &$-0.15$ &$-0.19$ &$-0.16$  \\ \hline
$P_{L}^{\eta_c}$  & 0.42 &$0.58_{0.43}^{0.66}$&$0.62_{0.53}^{0.68}$&$0.31_{0.14}^{0.45}$&0.50 & 0.59& 0.50&0.57 \\ \hline
 $P_{T}^{\eta_c}$ &0.81 &$0.73_{0.67}^{0.80}$&$0.70_{0.66}^{0.76}$&$0.84_{0.80}^{0.86}$&0.77 &0.72 &0.77 &0.43  \\ \hline \hline
  $A_{FB}^{J/\psi}$  & 0.02  &$0.005_{-0.01}^{0.02}$&$0.04_{0.03}^{0.05}$&$0.02_{0.01}^{0.02}$&0.006 &0.07 &0.03 &0.02 \\
  \hline 
 $C_{F}^{\tau,J/\psi}$ &  $-0.07$ &$-0.07^{-0.07}_{-0.07}$&$-0.07^{-0.07}_{-0.07}$&$-0.07^{-0.06}_{-0.07}$&$-0.07$ & $-0.08$&$-0.08$ &$-0.08$ \\
  \hline 
 $P_{L}^{J/\psi}$    &$-0.53$  &$-0.50^{-0.48}_{-0.53}$&$-0.57^{-0.55}_{-0.58}$&$-0.53^{-0.53}_{-0.53}$&$-0.51$ &$-0.60$ &$-0.54$ &$-0.48$\\
  \hline   
$P_{T}^{J/\psi}$    &0.40 &$0.43_{0.40}^{0.45}$&$0.35_{0.33}^{0.38}$&$0.35_{0.29}^{0.41}$&0.39 & 0.29&0.38 &0.28 \\
  \hline \hline
\end{tabular}
\captionof{table}{The integrated values of the forward-backward asymmetry, the convexity parameter and the longitudinal and transverse polarization of $\tau$ in the whole $q^2$ region, in case of different NP scenarios discussed in the text. The subscript and the superscript are the values for the $2\sigma$ range of the NP couplings.} \label{tab:avg_obs} 
\end{center}

 \section{Decay distribution of  $B_c \rightarrow  J/\psi \, ( J/\psi\to \mu^+\mu^- ) \, \ell \nu_\ell$ decay}
\label{sec:diff_dist_decay}
 
\begin{figure}
   \centering
 \includegraphics[width=8cm, height= 5cm]{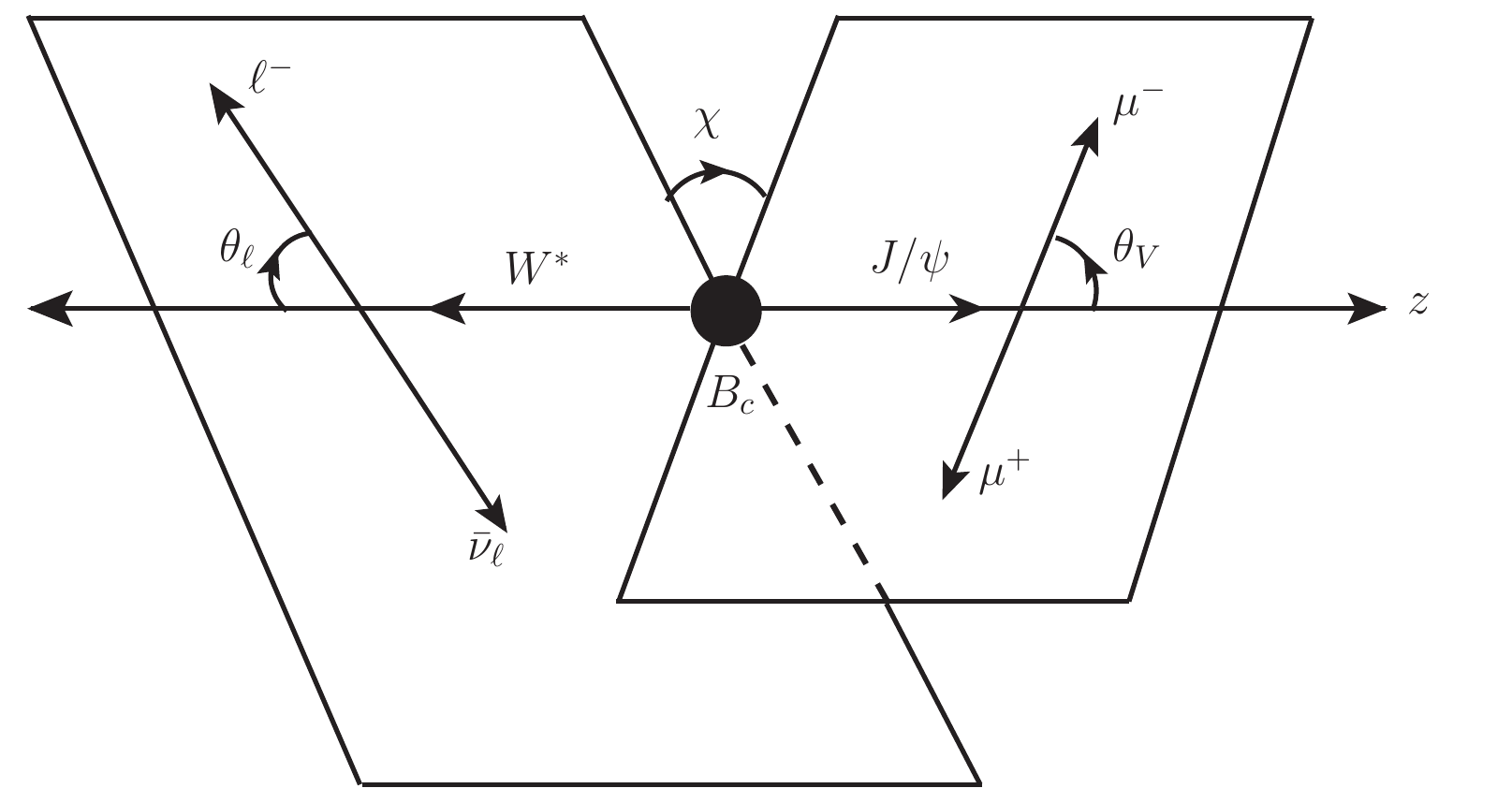}
 \caption{Angular conventions for the $B_c\to J/\psi \ell\nu_\ell, J/\psi \to \mu^+\mu^-$ decay.}
 \label{fig:4body}
 \end{figure}

 We consider in this section the process $B_c \rightarrow J/\psi \, (J/\psi \rightarrow \mu^+ \mu^-) \,\ell \nu_\ell$,  with the 4-fold differential decay rate being dependent on three angles $\theta_V, \theta_\ell, \chi$  and the momentum transfer $q^2$. The angle $\theta_\ell$ is same as defined before, $\theta_V$ is the polar angle between the direction of the emitted $\mu^-$ in the $J/\psi$ rest frame and the parent $J/\psi$ in the $B_c$ rest frame, and $\chi$ is the azimuthal angle between the $W^* \ell \nu$ plane and the $J/\psi \mu^+ \mu^-$ plane. The angles are shown in Fig.~\ref{fig:4body} and are defined as usually being taken in the literature.  
  The $J/\psi$ is too light to decay to $\tau^+\tau^-$, therefore the outgoing leptons can be either a pair of $\mu$ or of $e$. We ignore the mass $m_\mu, m_{e}$ from the $J/\psi$ decays but the mass of lepton from $W^*$ decay is retained. The total differential decay rate for the $\mu^-_L\mu^+_R$ ($\sigma \sim \lambda_\ell^--\lambda_\ell^+ = -1$)  final state is given by Eq.~(\ref{eq:tot_dr}) below.  The corresponding expressions for $\mu^-_R\mu^+_L$ final state can be obtained by setting $\theta_V \rightarrow \theta_V+\pi$ in Eq.~\ref{eq:tot_dr}.
 \begin{eqnarray}\label{eq:tot_dr}
 \frac{d\Gamma(B_c\to J/\psi\ell \nu_\ell, J/\psi \to \mu^+_R \mu^-_L)}{dq^2 d\cos\theta_\ell d\cos\theta_V d\chi } &=&  \frac{3G_F^2|V_{cb}|^2|{\bf p_2}|q^2v^2}{8 (4\pi)^4 m_{B_c}^2 } {\rm BR}(J/\psi \to \mu^-_L\mu^+_R) 
 \Big[ |1+V_L|^2 \mathcal{T}_{V_L} + |V_R|^2   \mathcal{T}_{|V_R|^2}    \nonumber \\
&+ &   \mathcal{T}_{V_R^{int}} +  2 |P|^2 (H_S^V)^2\sin^2\theta_V+  \mathcal{T}_{P^{int}}  +  |T_L|^2   \mathcal{T}_{|T_L|^2}   +  \mathcal{T}_{T_L^{int}} \Big], 
 \end{eqnarray}
 with
  \begin{eqnarray}
\mathcal{T}_{V_L}&=& \sin^2\theta_V\Big(2H_{00}^2 (\sin^2\theta_\ell+ 2\delta_\tau \cos^2\theta_\ell )+4 \delta_\tau H_{t0}^2 \Big)+8 H_{++}^2 \sin^2\frac{\theta_\ell}{2}\sin^4\frac{\theta_V}{2}\Big( 2 \delta_\tau \cos^2\frac{\theta_\ell}{2} + \sin^2\frac{\theta_\ell}{2}  \Big) \nonumber \\
&+&8 H_{--}^2 \cos^2\frac{\theta_\ell}{2}\cos^4\frac{\theta_V}{2}\Big( 2 \delta_\tau \sin^2\frac{\theta_\ell}{2} + \cos^2\frac{\theta_\ell}{2}  \Big)  + H_{++} H_{--} \sin^2\theta_\ell\sin^2\theta_V \cos2\chi (1-2\delta_\tau) \nonumber \\
&-& 8\sin\theta_\ell \sin\theta_V\cos\chi H_{00}\Big(H_{++} \sin^2\frac{\theta_V}{2}(\sin^2\frac{\theta_\ell}{2}+ \delta_\tau \cos\theta_\ell)+H_{--}\cos^2\frac{\theta_V}{2} (\cos^2\frac{\theta_\ell}{2}- \delta_\tau \cos\theta_\ell) \Big) \nonumber \\
&+& 8\delta_\tau\sin\theta_\ell \sin\theta_V  \cos\chi H_{t0}\Big(H_{++} \sin^2\frac{\theta_V}{2}-H_{--}\cos^2\frac{\theta_V}{2}\Big)-8\sin\theta_V^2\cos\theta_\ell \delta_\tau H_{t0}H_{00}\nonumber,
\end{eqnarray}
\begin{eqnarray}
 \mathcal{T}_{|V_R|^2}&=&\sin^2\theta_V \Big(2H_{00}^2 (\sin^2\theta_\ell+ 2\delta_\tau \cos^2\theta_\ell )+4 \delta_\tau H_{t0}^2 \Big)+8 H_{--}^2 \sin^2\frac{\theta_\ell}{2}\sin^4\frac{\theta_V}{2}\Big( 2 \delta_\tau \cos^2\frac{\theta_\ell}{2} + \sin^2\frac{\theta_\ell}{2}  \Big)  \nonumber \\
&+&8 H_{++}^2 \cos^2\frac{\theta_\ell}{2}\cos^4\frac{\theta_V}{2}\Big( 2 \delta_\tau \sin^2\frac{\theta_\ell}{2} + \cos^2\frac{\theta_\ell}{2}  \Big)  + H_{++} H_{--} \sin^2\theta_\ell\sin^2\theta_V \cos2\chi (1-2\delta_\tau)  \nonumber \\
&-& 8\sin\theta_\ell \sin\theta_V\cos\chi H_{00}\Big(H_{--} \sin^2\frac{\theta_V}{2}(\sin^2\frac{\theta_\ell}{2}+ \delta_\tau \cos\theta_\ell)+H_{++}\cos^2\frac{\theta_V}{2} (\cos^2\frac{\theta_\ell}{2}- \delta_\tau \cos\theta_\ell) \Big) \nonumber \\
&+&8\delta_\tau\sin\theta_\ell \sin\theta_V  \cos\chi H_{t0}\Big(H_{--} \sin^2\frac{\theta_V}{2}-H_{++}\cos^2\frac{\theta_V}{2}\Big)-8\sin\theta_V^2\cos\theta_\ell \delta_\tau H_{t0}H_{00}\nonumber,
\end{eqnarray}
\begin{eqnarray}
   \mathcal{T}_{V_R^{int}}&=&-2{\rm Re}V_R \sin^2\theta_V \Big(2H_{00}^2 (\sin^2\theta_\ell+ 2\delta_\tau \cos^2\theta_\ell )+4 \delta_\tau H_{t0}^2 \Big)-
  \sin^2\theta_\ell\sin^2\theta_V  (1-2\delta_\tau)  \nonumber \\
 && 
  \Big[ H_{++}^2({\rm Re}V_R  \cos2\chi + {\rm Im}V_R  \sin2\chi )+ H_{--}^2({\rm Re}V_R  \cos2\chi - {\rm Im}V_R  \sin2\chi )\Big]  \nonumber \\
&-& 2{\rm Re}V_R H_{++} H_{--} \left[(1+\cos^2\theta_V)\Big(1+\cos^2\theta_\ell+2\delta_\tau\sin^2\theta_\ell\Big)+4\cos\theta_V\cos\theta_\ell\right]
 \nonumber \\
&+& 16 {\rm Re}V_R\cos\theta_\ell\sin^2\theta_V H_{00}H_{t0} +\Big(4\sin\theta_V\cos\theta_\ell-\sin2\theta_V \sin2\theta_\ell (2\delta_\tau-1)\Big)H_{00}  \nonumber \\
&&\Big(H_{++}({\rm Re}V_R  \cos\chi + {\rm Im}V_R  \sin\chi )+H_{--}({\rm Re}V_R  \cos\chi - {\rm Im}V_R  \sin\chi )\Big)  \nonumber \\
&+&4 \delta_\tau  \sin\theta_\ell\sin2\theta_VH_{t0}\Big[H_{++}({\rm Re}V_R  \cos\chi + {\rm Im}V_R  \sin\chi )+H_{--}({\rm Re}V_R  \cos\chi - {\rm Im}V_R  \sin\chi )\Big]\nonumber,
\end{eqnarray} 
 \begin{eqnarray} 
\mathcal{T}_{P^{int}} &=& 4\sqrt{2 \delta_\tau} H_S^V \Big[{\rm Re} P \sin^2\theta_V \Big(H_{t0}-H_{00}\cos\theta_\ell\Big) +\sin\theta_V  \sin\theta_\ell \left\lbrace H_{++}\sin^2\frac{\theta_V}{2} \right. \nonumber \\
&&\left.  \Big({\rm Re} P \cos\chi+ {\rm Im}P)\sin\chi\Big)-H_{--}\cos^2\frac{\theta_V}{2} \Big({\rm Re} P \cos\chi- {\rm Im}  P\sin\chi\Big)\right\rbrace\Big],\nonumber \\
 \mathcal{T}_{|T_L|^2} &=& 16 |T_L|^2 \left[2 |H_{T}^0|^2 \sin^2\theta_V (\cos^2\theta_\ell+ 2\delta_\tau \sin^2\theta_\ell )+8|H_{T}^+|^2 \sin^2\frac{\theta_\ell}{2}\sin^4\frac{\theta_V}{2}\Big( 2 \delta_\tau \sin^2\frac{\theta_\ell}{2} + \cos^2\frac{\theta_\ell}{2}  \Big) \right. \nonumber \\
&+& \left.8 |H_{T}^-|^2 \cos^2\frac{\theta_\ell}{2}\cos^4\frac{\theta_V}{2}\Big( 2 \delta_\tau \cos^2\frac{\theta_\ell}{2} + \sin^2\frac{\theta_\ell}{2}  \Big)  - H_{T}^+ H_{T}^- \sin^2\theta_\ell\sin^2\theta_V \cos2\chi (1-2\delta_\tau) \right. \nonumber \\
&-& \left. 4\sin\theta_\ell\sin\theta_V\cos\chi H_T^0\left\lbrace H_T^+ \sin^2\frac{\theta_V}{2} \Big(2\delta_\tau \sin^2\frac{\theta_\ell}{2}+\cos\theta_\ell\Big)+H_T^- \cos^2\frac{\theta_V}{2} \Big(2\delta_\tau \cos^2\frac{\theta_\ell}{2}-\cos\theta_\ell\Big)\right\rbrace
\right]\nonumber,
\end{eqnarray}
\begin{eqnarray}
\mathcal{T}_{T_L^{int}} &=&16\sqrt{2\delta_\tau}\left[H_T^0 \left\lbrace {\rm Re}T_L \sin^2\theta_V \Big( H_ {t0} \cos\theta_\ell - H_ {00}\Big)  + \sin\theta_\ell \sin\theta_V \Big( H_ {++} \sin^2 \frac{\theta_V}{2} 
 ({\rm Re} T_L \cos\chi + {\rm Im} T_L \sin \chi) \right. \right.  \nonumber \\
 && \left. \left. + H_ {--} \cos^2\frac{\theta_V}{2} ({\rm Re} T_L \cos\chi - {\rm Im} T_L \sin \chi)\Big) 
   \right\rbrace  +H_T^+ \left\lbrace \sin\theta_\ell\sin\theta_V\Big({\rm Re} T_L \cos\chi -{\rm Im} T_L \sin \chi\Big)\right. \right.  \nonumber \\
 && \left. \left.  (H_{00}-H_{t0})\sin^2\frac{\theta_V}{2}-4{\rm Re} T_L \sin^2\frac{\theta_\ell}{2} \sin^4 \frac{\theta_V}{2}  H_{++}\right\rbrace+
 H_T^- \left\lbrace \sin\theta_\ell\sin\theta_V\Big({\rm Re} T_L \cos\chi +{\rm Im} T_L \sin \chi\Big)\right. \right.  \nonumber \\
 && \left. \left.  (H_{00}+H_{t0})\cos^2\frac{\theta_V}{2}-4{\rm Re}T_L \cos^2\frac{\theta_\ell}{2} \cos^4 \frac{\theta_V}{2}  H_{--}\right\rbrace \right]\nonumber.
    \end{eqnarray} 
We only list the interference terms of NP with SM and do not show the NP-NP interference terms, but they are included in our calculations. The expressions above are now more involved for the 4-fold differential distribution and contain various combinations with $\theta_\ell$, $\theta_V$ and $\chi$ angles, with the imaginary couplings being proportional to $\sin\chi$. The constraints on the NP coefficients ($V_L$, $S_L$  and $S_R$) in the 1D scenario are obtained using the condition that they are purely real. The global fit results which we consider here, do not include the vector operator $V_R$, as this vector operator with right-handed coupling to the quarks does not arise at the dimension-six level in the $SU(2)_L$-invariant effective theory. The relation $S_L = 4 \,T_L$ in the pure imaginary case is in more agreement with the SM compared to the case with the real Wilson coefficients. 
However, the effects of the real and the imaginary components of these NP coefficients can be isolated by constructing different angular asymmetries. 

We first consider the forward-backward asymmetry in $\theta_V$ and both $\theta_V, \theta_\ell$ with the angle $\chi$ fully integrated over: 
 \begin{eqnarray}
 \label{eq:fb1}
A_{FB}^{J/\psi}(\theta_V)&=&\frac{1}{\Gamma}\int dq^2\int_0^{2\pi}d\chi \int_{-1}^{1}d\cos\theta_{\ell} \left(\int_{0}^{1}-\int_{-1}^{0}\right) d\cos\theta_V~ \mathcal{G}[q^2,\theta_\ell,\theta_V,\chi] \nonumber \\
&&=\frac{8\pi}{3\Gamma}\left[   |1+V_L|^2 (1+\delta_\ell)\Big(H_{--}^2-H_{++}^2\Big) + 8 |T_L|^2 (1+4 \delta_\ell) (|H_T^-|^2-|H_T^+|^2) \right.   \nonumber \\                        && \left. -12\sqrt{2\delta_\ell}    \Big({\rm Re}T_L + T_LV_L  \Big)  \Big(H_{--}H_T^- - H_{++}H_T^+  \Big)   \right], \nonumber \\
A_{FB}^{J/\psi}(\theta_V,\theta_\ell)&=&\frac{1}{\Gamma}\int dq^2\int_0^{2\pi}d\chi  \left(\int_{0}^{1}-\int_{-1}^{0}\right) d\cos\theta_{V} \left(\int_{0}^{1}-\int_{-1}^{0}\right) d\cos\theta_\ell ~ \mathcal{G}[q^2,\theta_\ell,\theta_V,\chi] \nonumber \\
&&=\frac{2\pi}{\Gamma}\left [|1+V_L|^2\Big(H_{--}^2+H_{++}^2\Big)+ 32 |T_L|^2 \delta_\ell (|H_T^-|^2+|H_T^+|^2)   \right.   \nonumber \\     
&& \left. -8\sqrt{2\delta_\ell}    \Big({\rm Re}T_L +T_L V_L  \Big)  \Big(H_{++}H_T^+ + H_{--}H_T^-  \Big) \right],
\end{eqnarray}
where $\mathcal{G} = \left(\frac{d^4\Gamma}{dq^2\, d\cos\theta_\ell\, d\cos\theta_{V}\, d\chi} \right) $ and $\Gamma$ in the denominator is the decay width of $B_c \to \mu^+\mu^- \ell \nu_\ell$, obtained by integrating Eq.~(\ref{eq:tot_dr}) and is given by
\begin{eqnarray}\label{eq:tDW}
\Gamma(B_c \to \mu^+\mu^- \ell \nu_\ell) &=& \frac{16\pi}{9}\Big[2|1+V_L|^2 \left\lbrace (1+\delta_\tau)\Big(H_{00}^2+H_{++}^2+H_{--}^2 \Big)+ 3 \delta_\tau H_{t0}^2 \right\rbrace + 3 |H_S^V|^2  |P|^2  \nonumber \\
 &+&  6\sqrt{2 \delta_\tau}H_S^V H_{t0} \Big({\rm Re}P+P\,V_L\Big) + 16 |T_L|^2 (1+4\delta_\tau)\Big(|H_T^0|^2+|H_T^+|^2+|H_T^-|^2\Big) \nonumber \\
&-&24\sqrt{2 \delta_\tau} \Big({\rm Re}T_L+T_LV_L\Big)\Big(H_{00}H_T^0+H_{++}H_T^++H_{--}H_T^-\Big)\Big].
\end{eqnarray}

\begin{figure}[htb]
\includegraphics[width=\textwidth]{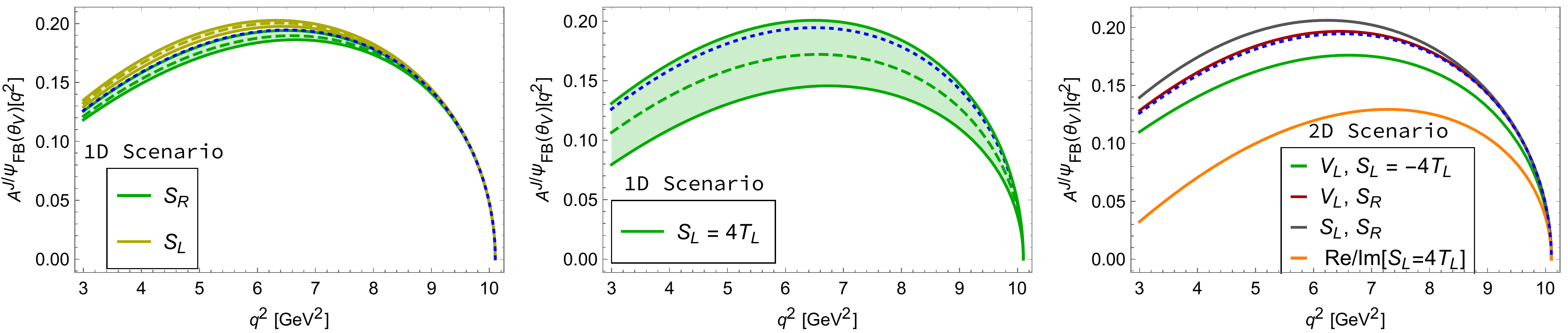}
\includegraphics[width=\textwidth]{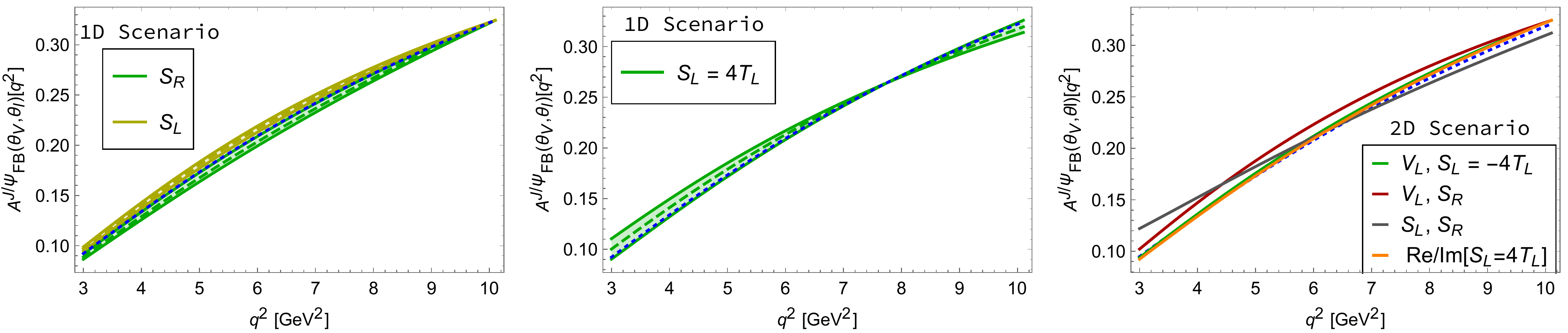}
\caption{Forward-backward asymmetry $A^{FB} (\theta_V)$ [upper-panel]  and   $A^{FB} (\theta_V, \theta_\ell)$ [lower-panel]  as a function of $q^2$. The blue dotted lines are the SM prediction, the green dashed line is for the best fit values of the NP couplings in the 1D scenario as discussed in the text. The green band represents the NP effects from the 2$\sigma$ allowed regions. The third figure in both panels is the result for the best fit points in the 2D scenarios.}
\label{fig:asym_theta_VL}
\end{figure}

It can be seen from Eq.~(\ref{eq:fb1})  that the numerator is not sensitive to the scalar type NP operators. Therefore the sensitivity to the scalar NP comes only from the total decay width in the denominator, Eq.~(\ref{eq:tDW}). In Fig.~\ref{fig:asym_theta_VL} we show   $A_{FB}^{J/\psi}(\theta_V)$ and $A_{FB}^{J/\psi}(\theta_V,\theta_\ell)$ as a function of $q^2$ with the values of the new physics couplings as given in Eqs.~(\ref{eq:1Dfit}, \ref{eq:2Dfit}).  The current bound on the NP couplings makes the observable $A_{FB}^{J/\psi}(\theta_V)$ sensitive to  $S_L = 4 \,T_L$ in the 1D scenario and to the same combination with both the real and the imaginary components present in case of 2D scenario. As for the asymmetry $A_{FB}^{J/\psi}(\theta_V)$, the deviation from the SM in case of 2D scenario for the combination Re[$S_L = 4 \,T_L$], Im[$S_L =4 \,T_L$] can be as large as 50-70\% in the region of small $q^2$. However, the other observable with the asymmetry in both $\theta_V$ and $\theta_\ell$, $A_{FB}^{J/\psi}(\theta_V,\theta_\ell)$  is not a good observable to look for NP scenarios in the current situation. 

One can build additional asymmetries in the angle $\chi$ along with $\theta_V$ and $\theta_\ell$. These asymmetries are proportional to both $\cos\chi$ and $\sin \chi$, and their corresponding expressions are given below: 
 \begin{eqnarray}
A_{FB}^{J/\psi}(\chi,\theta_V)&=&\frac{1}{\Gamma}\int dq^2 \left(\int_{-\pi/2}^{\pi/2}-\int_{\pi/2}^{3\pi/2}\right)d\chi \int_{-1}^{1}d\cos\theta_{\ell} \left(\int_{0}^{1}-\int_{-1}^{0}\right) d\cos\theta_V~ \mathcal{G}[q^2,\theta_\ell,\theta_V,\chi] \nonumber \\
&=&\frac{-4\pi}{3\Gamma}\left[|1+V_L|^2 \left\lbrace H_{00} \Big(H_{--}-H_{++}\Big)+  2\delta_\ell  H_{t0}  \Big(H_{--}+H_{++} \Big)   \right\rbrace-2 H_S^V H_T^+ T_LP   \right. \nonumber \\     
& +& \left.  \sqrt{2\delta_\ell}  \Big(H_{--}+H_{++} \Big)    H_S^V\Big({\rm Re}P +P\,V_L \Big) + 32 \delta_\ell H_T^0 \Big(H_T^--H_T^+\Big)|T_L|^2 \right.  \nonumber \\     
&-& \left. 4\sqrt{2\delta_\ell}\Big({\rm Re}T_L +T_L V_L\Big) \Big(H_{00} (H_T^--H_T^+)+H_T^0(H_{--}-H_{++})+H_{t0}(H_T^-+H_T^+)\Big)
 \right], \nonumber \\
A_{FB}^{J/\psi}(\chi,\theta_V,\theta_\ell)&=&\frac{1}{\Gamma}\int dq^2 \left(\int_{-\pi/2}^{\pi/2}-\int_{\pi/2}^{3\pi/2}\right)d\chi\left(\int_{0}^{1}-\int_{-1}^{0}\right) d\cos\theta_{\ell} \left(\int_{0}^{1}-\int_{-1}^{0}\right) d\cos\theta_V ~ \mathcal{G}[q^2,\theta_\ell,\theta_V,\chi] \nonumber \\
&=&\frac{16}{9\Gamma} (2\delta_\ell-1) \left[ |1+V_L|^2     H_{00} \Big(H_{--}+H_{++} \Big)- 16 |T_L|^2  H_T^0 \Big(H_T^-+H_T^+\Big) \right].
 \end{eqnarray}
\begin{figure}[htb]
\includegraphics[width=\textwidth]{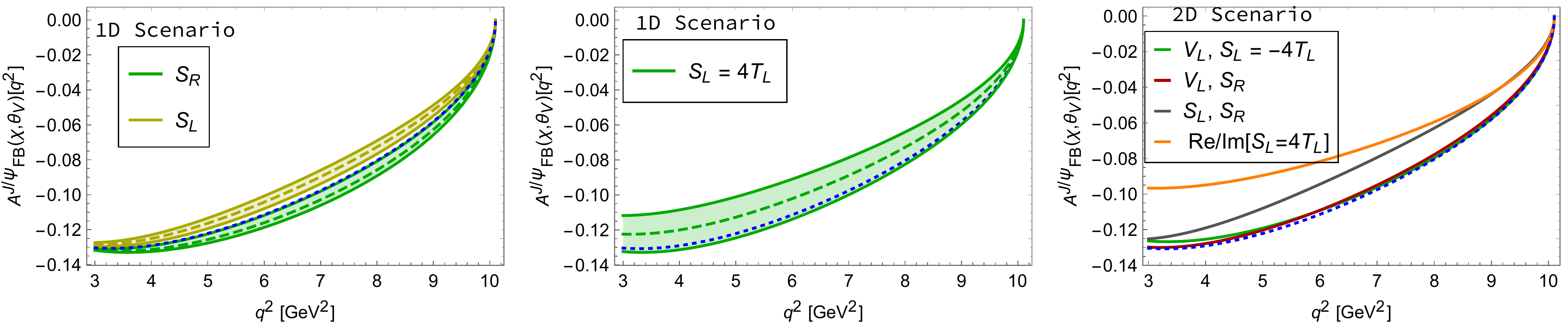}
\includegraphics[width=\textwidth]{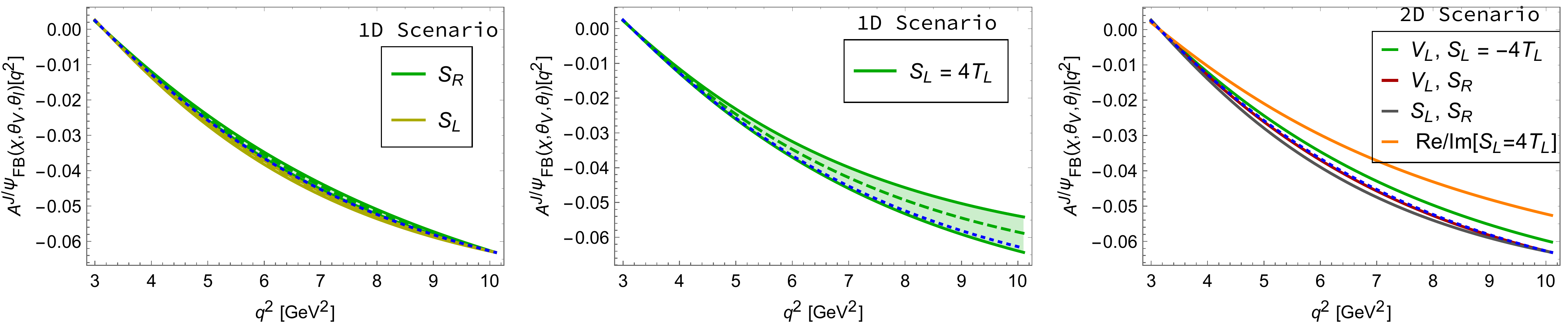}
\caption{Asymmetries $A^{FB} (\chi, \theta_V)$ [upper-panel]  and   $A^{FB} (\chi,\theta_V, \theta_\ell)$ [lower-panel]  as a function of $q^2$. The blue dotted lines are the SM prediction, the green dashed line is for the best fit values of the NP couplings in the 1D scenario as discussed in the text. The green band represents the NP effects from the 2$\sigma$ allowed regions. The third figure in both panels is the result for the best fit points in the 2D scenarios.}
\label{fig:asym_theta_chiVL}
\end{figure}
We show in Fig.~\ref{fig:asym_theta_chiVL}, the asymmetries $A_{FB}^{J/\psi}(\chi,\theta_V)$ [upper-panel] and   $A^{FB} (\chi,\theta_V, \theta_\ell)$ [lower-panel]  as a function of $q^2$.  They behave similar to the asymmetries ($A_{FB}^{J/\psi}(\theta_V), A_{FB}^{J/\psi}(\theta_V,\theta_\ell)$) discussed above, where $\chi$ was integrated over the whole range. These observables do not provide any additional information compared to $A_{FB} (\theta_V)$ and   $A_{FB} (\theta_V, \theta_\ell)$ discussed before. The 2D scenario in case of $A_{FB}^{J/\psi}(\chi,\theta_V)$  with Re[$S_L = 4 \,T_L$], Im[$S_L =4 \,T_L$] results in about 10-20\% deviation from the SM value at low values of $q^2$. 

\begin{figure}[htb]
\includegraphics[width=15cm,height=5cm]{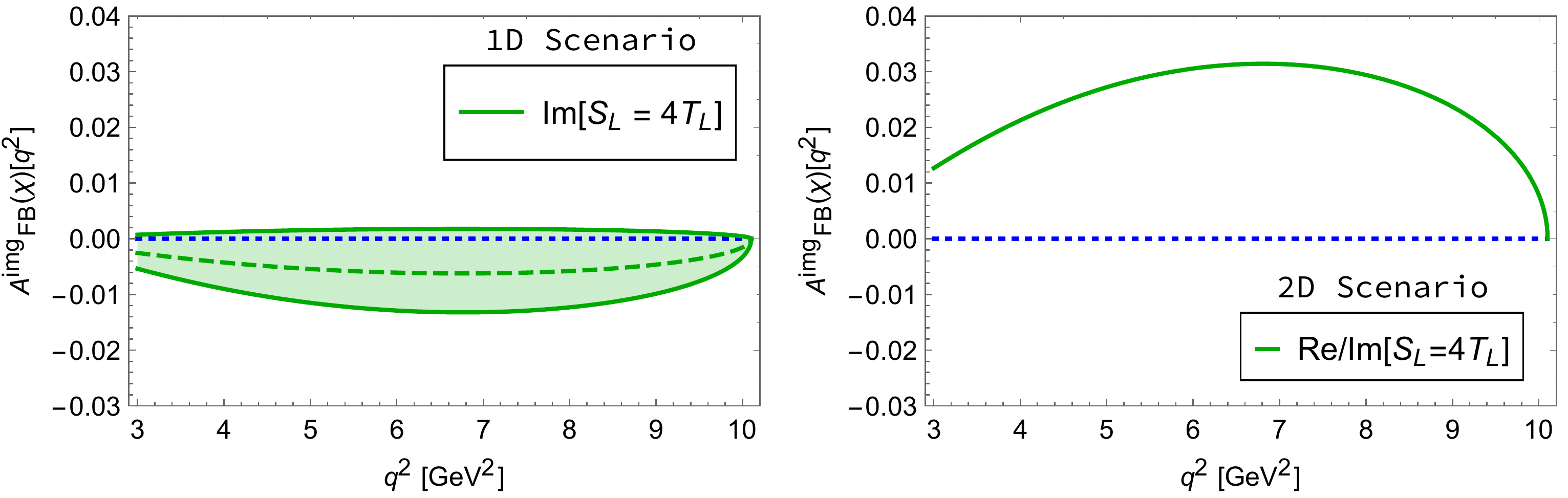}
\includegraphics[width=15cm,height=5cm]{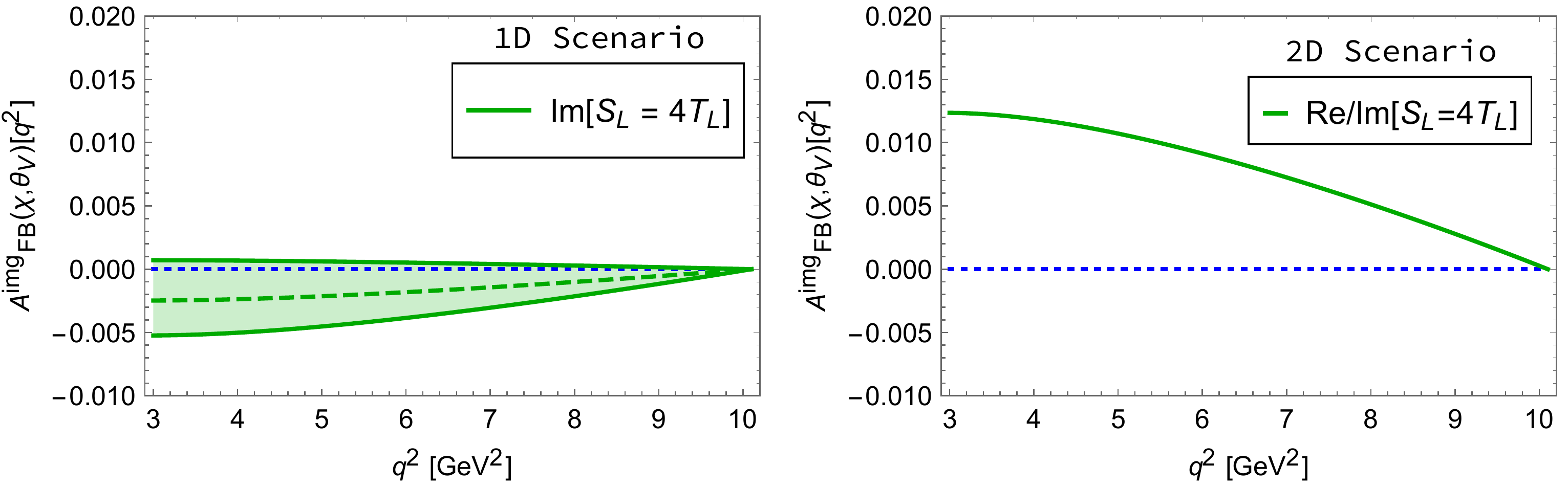}
\caption{Asymmetries $A^{img}_{FB}(\chi)$ [upper-panel]  and   $A^{img}_{FB}(\chi,\theta_V)$ [lower-panel]  as a function of $q^2$. The SM value being zero is shown by a blue dotted line, the green dashed line is for the best fit values of the NP couplings in the 1D scenario as discussed in the text. The green band represents the NP effects from the 2$\sigma$ allowed regions.  The second figure is for the relevant 2D scenario.}
\label{fig:asym_img}
\end{figure}

Finally, we consider the observables which will be sensitive to only the imaginary component of the NP operators. These asymmetries are zero within the SM. There are three possible combinations, ($a$) asymmetry depending only on $\chi$, ($b$) asymmetry depending on $\chi$ and $\theta_V$, and ($c$) asymmetry depending on $\chi$, $\theta_V$ and  $\theta_\ell$. The relevant expressions are: 
 \begin{eqnarray}
A^{img}_{FB}(\chi)&=&\frac{1}{\Gamma}\int dq^2 \left(\int_{0}^{\pi}-\int_{\pi}^{2\pi}\right)d\chi \int_{-1}^{1}d\cos\theta_{\ell}  \int_{-1}^{1} d\cos\theta_V~ \mathcal{G}[q^2,\theta_\ell,\theta_V,\chi] \nonumber \\
&=&\frac{\pi^2}{\Gamma}\left[\sqrt{2\delta_\ell}  \Big(4\left\lbrace {\rm Im}T_L + (T_L V_L)^\ast \right\rbrace \left\lbrace H_{00} (H_T^--H_T^+)+H_T^0(H_{++}-H_{--}) \right.\right. \nonumber \\
&+&\left. \left. H_{t0}(H_T^++H_T^-)\right\rbrace +H_S^V  \left\lbrace{\rm Im} P+(P\,V_L)^\ast \right\rbrace\Big(H_{--}+H_{++}\Big)\Big) + 4 H_S^V H_T^+ (P\,T_L)^\ast \right], \nonumber \\
 \end{eqnarray} 
where  $(T_L V_L)^\ast = {\rm Im}T_L {\rm Re} V_L- {\rm Im} V_L{\rm Re} T_L,~(P\,V_L)^\ast= {\rm Im}P\, {\rm Re} V_L- {\rm Im} V_L{\rm Re}P,~ (P\,T_L)^\ast= {\rm Im}P\, {\rm Re} T_L- {\rm Im} T_L{\rm Re}P$, 
\begin{align}
A^{img}_{FB}(\chi,\theta_V)&=\frac{1}{\Gamma}\int dq^2 \left(\int_{0}^{\pi}-\int_{\pi}^{2\pi}\right)d\chi \int_{-1}^{1}d\cos\theta_{\ell} \left(\int_{0}^{1}-\int_{-1}^{0}\right) d\cos\theta_V ~ \mathcal{G}[q^2,\theta_\ell,\theta_V,\chi]\nonumber \\
&=\frac{4\pi}{3\Gamma}\sqrt{\delta_\ell}\left[4\sqrt{2}\left\lbrace {\rm Im} T_L -(T_LV_L)^\ast\right\rbrace \left\lbrace H_{00} (H_T^-+H_T^+)+H_{t0} (H_T^--H_T^+)-H_T^0 (H_{++}+H_{--})\right\rbrace \right. \nonumber \\
&\left. +\sqrt{2}H_S^V  \Big(H_{--}-H_{++}\Big)\left\lbrace {\rm Im} P- (P\,V_L)^\ast \right\rbrace  +\frac{4}{\sqrt{\delta_\ell}} H_S^V H_T^+ (P\,T_L)^\ast    \right],
 \end{align}
 \begin{align}
A^{img}_{FB}(\chi,\theta_V,\theta_\ell)&=\frac{1}{\Gamma}\int dq^2 \left(\int_{0}^{\pi}-\int_{\pi}^{2\pi}\right)d\chi\left(\int_{0}^{1}-\int_{-1}^{0}\right) d\cos\theta_{\ell} \left(\int_{0}^{1}-\int_{-1}^{0}\right) d\cos\theta_V~ \mathcal{G}[q^2,\theta_\ell,\theta_V,\chi] \nonumber \\
&=\frac{16}{9\Gamma}(2\delta_\ell-1) \left[ H_{00}\Big(H_{--}-H_{++}\Big)  \Big(2 {\rm Im} V_R  +  {\rm Im} V_R {\rm Re} V_L -   {\rm Re} V_R  {\rm Im} V_L    \Big)\right].
 \end{align}
The asymmetry $A^{img}_{FB}(\chi,\theta_V,\theta_\ell)$ in only sensitive to the NP operator $V_R$ and is therefore not relevant for our case since these $V_R$ coefficients  are not considered in the global fits as discussed before.  We show in Fig.~\ref{fig:asym_img} $A^{img}_{FB}(\chi)$ [upper-panel]  and   $A^{img}_{FB}(\chi,\theta_V)$ [lower-panel]  as a function of $q^2$. These observables are only shown for  $ {\rm Im} S_L= 4  \,{\rm Im}T_L$ in the 1D scenario  and  $ {\rm Re} S_L= 4 \, {\rm Re}T_L$, $ {\rm Im} S_L= 4  \,{\rm Im}T_L$ in the 2D scenarios as these were the only cases considered in the global fit in Ref.~\cite{Blanke:2018yud}. The forward-backward asymmetry depending  only on $\chi$  in the light of results from the current global fit shows about 1\% deviation from the SM in the 1D scenario and up to 3\% deviation in the 2D scenario, in the mid-range of $q^2 = 5-9$ GeV$^2$.  The asymmetry $A^{img}_{FB}(\chi,\theta_V)$  in case of the 2D scenario will have around only 1\% deviation in low $q^2$ region, whereas it is insensitive to NP in the 1D scenario.
The predictions for the integrated forward-backward asymmetries in the presence of different NP operators is summarised  in Table~\ref{tab:avg_obs_img}. 
\begin{center}
\addtolength{\tabcolsep}{-4pt}
%\captionsetup{justification=centering,margin=1.4cm}
\renewcommand{\arraystretch}{1.1}
\begin{tabular}{||c || c ||c |c|c|c|c|c|c|c|}
\hline
& SM  &$S_L$ & $S_R$ &$S_L = 4 \,T_L$ & $ (V_L,S_L=-4 \,T_L)$&$(S_R,S_L)$& $(V_L,S_R)$& Re,Im[$S_L=4 \,T_L$]    \\ [0.5ex]
\hline\hline
$A_{FB}^{J/\psi}(\theta_V)$  &$0.16$  &$0.16_{0.15}^{0.16}$&$0.17_{0.16}^{0.17}$&$0.14_{0.12}^{0.17}$&$0.15$ &$0.17$ &$0.16$ &$0.09$ \\ \hline
$A_{FB}^{J/\psi}(\theta_V,\theta_\ell)$  &$ 0.21$ &$0.20_{0.20}^{0.21}$&$0.21_{0.21}^{0.22}$&$0.21_{0.21}^{0.22}$&$0.21$ &$0.22$ &$0.21$ &$0.21$  \\ \hline
$-A_{FB}^{J/\psi}(\chi,\theta_V)$    &$0.09$   &$0.10_{0.09}^{0.10}$&$0.09_{0.09}^{0.09}$&$0.09_{0.08}^{0.10}$&$0.10$ & $0.08$&$0.10$ &$0.07$ \\ \hline 
$-A_{FB}^{J/\psi}(\chi,\theta_V,\theta_\ell)$  &0.03   &$0.03_{0.03}^{0.03}$&$0.03_{0.03}^{0.03}$&$0.03_{0.03}^{0.03}$&0.03 & 0.03& 0.03&0.02 \\ \hline
$A_{img}^{FB}(\chi)$    &$0.0$   &$0.0$&$0.0$&$-0.004_{-0.01}^{0.001}$&$0.0$ &$0.0$ &$0.0$ &$0.02$\\  \hline 
 $A_{img}^{FB}(\chi,\theta_V)$  &$0.0$   &$0.0$&$0.0$&$-0.002_{-0.003}^{0.0}$&$0.0$ &$0.0$ &$0.0$ &$-0.001$  \\ \hline
\end{tabular}
\captionof{table}{The integrated values of the forward-backward asymmetries in the whole $q^2$ region, in case of different NP scenarios discussed in the text. The subscript and the superscript are the values for the $2\sigma$ range of the NP couplings.} \label{tab:avg_obs_img} 
\end{center}

\section{Conclusions}\label{sec:conclusion}

Experimental measurements of semileptonic decays of the $B$ mesons have lead to intriguing experimental tensions with the SM in the last years. The LHCb measurement of $B_c \to J/\psi l \nu_l$ decays has lead to the speculation whether the observed potential lepton flavour universality (LFU) violation in $B$ decays can be also seen in the semileptonic $B_c$ channels. 

However, the SM prediction for the $B_c$ decays require a knowledge of the transition form factors of $B_c \rightarrow \eta_c, J/\psi$ and the ignorance of the form factor theoretical errors yields a degree of uncertainty in the prediction. Preliminary results for these form factors exist at couple of $q^2$ values from the lattice QCD, but they do not cover the entire allowed range of the momentum transfer and are still given without systematical errors. We have calculated the form factors in the sum rule approach and have given the results in the full $q^2$ region. Our results are in good agreement with the existing lattice points. The SM branching ratios of the $B_c$ meson to $J/\psi$ and $\eta_c$ are  calculated and compared with the results from other approaches. Our predictions for the semileptonic ratios $R_{J/\psi}|_{\rm SM} = 0.23 \pm 0.01$ and $R_{\eta_c}|_{\rm SM}= 0.32 \pm 0.02$ are in agreement with other derivations and support the existing tension at 2$\sigma$ level with the experiment on $R_{J/\psi}$, Eq.~(\ref{eq:expRJpsi}).
With more data on $B_c$ decays from HL-LHC all observables in $B_c \to \eta_c, J/\psi$ semileptonic decays will be within the reach of LHCb and tested in the near future.
%With the extended experimental LHCb program, future studies with more data will boost or disapprove this evidence of LFU violation in $B_c$ decays. 

The possible NP effects in the semileptonic decays of $B_c$ to $\eta_c$ and $J/\psi$ is also studied based on the effective Hamiltonian approach consisting of all possible four-fermi operators. The constraints on these NP operators can be obtained from the experimental data on  $R_{D(\ast)}$, the $\tau$ and $D^\ast$ longitudinal polarization from $B \to D^{\ast}$ decay and the leptonic $B_c \to \tau \mu$ branching ratio. We take into account the latest constraints from Ref.~\cite{Blanke:2018yud} and analyse the effects of the NP operators on various observables. The ratio $R_{J/\psi}$ is sensitive to $V_L$ in the high $q^2$ range whereas $R_{\eta_c}$ is more sensitive to the scalar and the tensor operators, as expected. 

The sensitivity of all the considered observables in this work to the different NP operators is summarized in Table~\ref{tab:sensitivity}. We find that most of the observables in the $\eta_c$ decay mode are sensitive to the NP coupling $S_R$. The transverse polarization of $\tau$ is mostly affected by the current best fit point of the combination of coefficients Re,Im[$S_L =4 \,T_L$] in the 2D NP scenario. The 2D NP scenario with the presence of both $S_R,S_L$ leads to the largest deviation from the SM predictions for most of the observables in the case of $J/\psi$, apart from $R_{J/\psi}$.
\begin{center}
\addtolength{\tabcolsep}{-4pt}
%\captionsetup{justification=centering,margin=1.4cm}
%\renewcommand{\arraystretch}{1.5}
\begin{tabular}{||c | c |c |c|c|c|c|c|c|c|}
\hline
& $V_L$  &$S_L$ & $S_R$ &$S_L = 4 \,T_L$ & $ (V_L,S_L=-4 \,T_L)$&$(S_R,S_L)$& $(V_L,S_R)$& Re,Im[$S_L=4 \,T_L$]    \\ [0.5ex]
\hline\hline
$R_{\eta_c}$ & & & $\checkmark$ & & & & &  \\ \hline
$A^{\eta_c}_{FB}$& &$\checkmark^\ast$ &$\checkmark^\ast$ & & & & &$\checkmark$  \\ \hline
$C^{\tau,\eta_c}_{F}$& &$\checkmark$ &$\checkmark$ &$\checkmark^\ast$ & & & &  \\ \hline
$P_L^{\eta_c}$& & &$\checkmark$ &$\checkmark^\ast$ & & & &  \\ \hline
$P_T^{\eta_c}$& & & & & & & &$\checkmark$  \\ \hline \hline
$R_{J/\psi}$ &$\checkmark$ & & & &$\checkmark$ & & &  \\ \hline
$A^{J/\psi}_{FB}$& & & & & &$\checkmark$ & &  \\ \hline
$P_L^{J/\psi}$& & & & & &$\checkmark$ & &  \\ \hline
$P_T^{J/\psi}$& & & &$\checkmark^\ast$ & &$\checkmark$ & &$\checkmark$  \\ \hline 
% \hline
% $A_{FB}^{J/\psi}(\theta_V)$ & & & & & & & &$\checkmark$ \\ \hline
% $-A_{FB}^{J/\psi}(\chi,\theta_V)$   & & & & & & & &$\checkmark$ \\ \hline 
% $A_{img}^{FB}(\chi)$    & & & & & & & &$\checkmark$\\  \hline 
%  $A_{img}^{FB}(\chi,\theta_V)$  & & & &$\checkmark$ & & & &$\checkmark$ \\ \hline
\end{tabular}
\captionof{table}{Summary of the sensitivity of the observables to the NP couplings. The best fit value of the NP coupling which is most sensitive to the observable is marked with $\checkmark$. The boxes with $\checkmark^\ast$ are the ones where $2\sigma$ ranges of NP parameters give the largest deviation from the SM value.} \label{tab:sensitivity} 
\end{center}
In addition, the full 4-fold differential distribution of the decay rate $B_c\rightarrow J/\psi \ell \nu_\ell$, with $J/\psi$ decaying to a pair of leptons of opposite helicity is considered for the first time in the presence of new physics operators. We find that the asymmetry in the angle $\theta_V$ $(A_{FB}^{J/\psi}(\theta_V))$ is mostly sensitive to the NP couplings Re,Im[$S_L =4T_L$], in the 2D NP scenarios. The asymmetries in the angle $\chi$, which are zero in the SM and are sensitive to the imaginary part of the NP coupling, are also considered and found to be sensitive to $S_L = 4\,T_L$ combination of parameters. Therefore,  with the current allowed parameter space for the $S_L = 4\,T_L$ NP parameters obtained from the global fit to experimental data on semileptonic $B \to D,D^{\ast}$  decays, the asymmetries constructed with $\theta_V$, $\chi$ and $(\theta_V,\chi)$ angles lead to significant deviation from the SM prediction.

However,  it is important to stress that none of the NP scenarios derived from the recent global fit analysis of the available experimental data on semileptonic $B \to (D,D^{\ast}) \ell \nu_\ell$
decays ~\cite{Blanke:2018yud} can also simultaneously explain the current 2$\sigma$ tension with the experimental  $R_{J/\psi}$ ratio. 
With the extended experimental LHCb program, future studies with more data will be needed to boost or disapprove this evidence of LFU violation in $B_c$ decays. 
%With more data on $B_c$ decays from HL-LHC all observables in $B_c \to \eta_c, J/\psi$ semileptonic decays will be within the reach of LHCb and tested in the near future.

\section*{Acknowledgments}
This work is partly supported by the EU grant RBI-T-WINNING (grant EU H2020 CSA-2015 number 692194) and by the European Union through the European Regional Development Fund - the Competitiveness and Cohesion Operational Programme (KK.01.1.1.06). M.P. acknowledge support of the Slovenian Research Agency through research core funding No. P1-0035.

%%%%%%%%%%%%%%%%%%%%%%%%%%%%%%%%%%%%%%%%%%%%%%%%%%%%%%%%%%%%%%
\appendix

\section{Form factors calculated in the three-point QCDSR model \cite{preparation}}\label{sec:app1}

We have previously calculated in \cite{preparation} the same form factors using a more traditional, albeit somewhat modified approach of three-point QCD sum rules (3ptQCDSR) and we present the corresponding results in Table~\ref{tab:ffactor}. 
%\ref{tab:brfr2}. 
Here we just briefly discuss the method of our calculation and the main difference to the LCSR-inspired approach used in the paper. In 3ptQCDSR mesonic states are interpolated by the currents as
	\begin{equation}
		\begin{split}
			j_{B_c}(x) & = \bar{c}(x)\mathrm{i}\gamma_5 b(x),\\
			j_{J/\psi}^{\nu}(x) & = \bar{c}(x)\gamma^{\nu}c(x), 
		\end{split}
	\end{equation}
taken at large virtualities. 
By inserting a set of hadronic states in the correlation function defined as
\begin{equation}
\Pi^{\mu\nu}(p_{B_c},p_{J/\psi})\equiv\mathrm{i}^2
\int\!\!\!\int\mathrm{d}^4x\,\mathrm{d}^4y\,\mathrm{e}^{-i(p_{B_c}x-p_{J/\psi}y)}\expval{T\big\{j_{J/\psi}^{\nu}(y)j^{\mu}_{\mathrm{V-A}}(0)j^{\dagger\vphantom{\mu}}_{B_c}(x)\big\}}{0},
\end{equation}
one can extract the form factors by calculating the perturbative part of the correlator and the nonperturbative contributions given in terms of 
universal vacuum condensates built from the quark and gluon operators of increasing dimension (here we have calculated only the leading nonperturbative contribution coming from the gluon condensate) and matching the QCD result via dispersion relation to a sum over hadronic states. At the end, the expressions are Borel transformed in order to improve the convergence.

Since it is known that in the sum rule calculation of  heavy meson decay constants higher orders of perturbation series can contribute as much as 40-50\%, whereas the 3-point function is calculated at LO, in order to reduce the uncertainties we have performed the following procedure: in the form factors calculation we have taken for the $s_0$ threshold parameters the same values as those that reproduce the corresponding charmonia decay constants obtained from lattice QCD when the decay constants are calculated in the sum rules by taking into account only the LO perturbative part and the gluon-condensate contribution, i.e. with the same approximations as for the form factors, whereas the Borel mass parameter is taken in the region where stability is achieved (we aim at the $\sim 5\%$ stability in the Borel masses in the given Borel window).  Furthermore in order to reduce the uncertainties even more, we do not vary the decay constants and thresholds independently, but rather in the 3ptQCDSR calculation we always use the decay constants (varied inside the range allowed by lattice) together with the corresponding thresholds fixed by the decay constants calculation. The hope is that all the higher order/higher dimension operator contributions are then simulated through the appropriate threshold modification in the 3-point QCDSR calculation. The parameters obtained that way are given in Table~\ref{tab:lists} below. One can notice that in contrast to the LCSR-inspired calculation used in the main text, here the pole mass of the $b$ quark is used, together with the $c$-quark mass derived from the ratio of masses extracted from the lattice calculations. 
 \begin{center}
 \addtolength{\tabcolsep}{-4pt}
%\captionsetup{justification=centering,margin=1.4cm}
 \renewcommand{\arraystretch}{1.1}
 		\begin{tabular}{ |c|c| }
 			\hline
 			$m_b = 4.6^{+0.1}_{-0.1}$ GeV & $s_{B_c}=52-54$ GeV$^2$ \\ [0.75ex] 
 			$m_c = Zm_b$, $\forall Z\approx 0.29^{+0.1}_{-0.1}$ & $s_{J/\psi} = 15.5-16.5$ GeV$^2$ \\[0.75ex] 
 			\hline\hline 
 			$\expval{\frac{\alpha_s}{\pi}GG}=0.012^{+0.006}_{-0.010}$ GeV$^4$ & $M^2_{B_c}=60-80$ GeV$^2$\\ [0.75ex]
 			 & $M^2_{J/\psi}=20-25$ GeV$^2$\\ [0.75ex]
 			\hline
 		\end{tabular}
 \captionof{table}{Parameters used in the 3ptQCDSR calculation \cite{preparation}.}
 \label{tab:lists}
   \end{center}
%%%%%%%%%%%%%%%%%%%%%%%%%%%
%%%%%%%%%%%%%%%%%%%%%%%%%%%%%%%%%%%%%%%%%%%%%%%%%%%%%%%%%%%%%%

 \end{document}